\documentclass[12pt]{article}
\usepackage{amsmath,mathrsfs,bm,amssymb,color,theorem,hiroshima}
\usepackage{latexsym}
\newtheorem{theorem}{Theorem}[section]
\newtheorem{proposition}[theorem]{Proposition}
\newtheorem{lemma}[theorem]{Lemma}

\newtheorem{remark}[theorem]{Remark}

\makeatletter
\@addtoreset{equation}{section}
\makeatother

\title
{\sc  Spectral zeta function  and  ground state of quantum Rabi model\footnote{
We dedicate this paper in honor of Professor Takashi Hara on his 60th+ birthday.}}
\author{Fumio Hiroshima\footnote{Kyushu university} and Tomoyuki Shirai\footnote{Kyushu university}}
\date{\today}
\begin{document}
\maketitle
\begin{abstract}
The  spectral zeta function 
$\zeta_g(s;g^2+\tau)=\sum_{n=0}^\infty \frac{1}{(E_g(n)+g^2+\tau)^s}$
of the quantum Rabi Hamiltonian is considered, 
where $\tau>0$ and $E_n(g)$ denotes $n$th eigenvalue of the quantum Rabi Hamiltonian $H$. 
Let $\zeta(s;\tau)=\sum_{n=0}^\infty\frac{1}{(n+\tau)^s}$ be the Hurwitz zeta function.
It is shown that 
$\lim_{|g|\to\infty} \zeta_g(s;g^2+\tau)=2\zeta (s;\tau)$. 
Moreover the path measure ${\Pi_\infty}$ 
associated with the ground state of $H$
 is constructed on a discontinuous path space, and several applications are shown. 
\end{abstract}

\section{Spectral zeta function of quantum Rabi model}
The quantum Rabi model describes a two-level atom coupled to a single mode photon by the dipole interaction term. 
The single photon is represented by the $1$D harmonic oscillator. 
This model is initially introduced by I.I.Rabi at 1937 \cite{rab36} in a semiclassical region and then 
the quantized version is introduced by E.T. Jaynes and F.W. Cumming \cite{JC63}.  
Suppose that 
the eigenvalues of the two-level atom is $\{-\triangle,\triangle\}$. 
Here $\triangle>0$ is a constant. 
Let $\s_x,\s_y$ and $\s_z$ be the $2\times 2$ Pauli matrices:
\begin{align*}
 \s_x= \begin{pmatrix} 0 & 1 \\ 1 & 0 \end{pmatrix},\quad
 \s_y = \begin{pmatrix} 0 & -i \\ i & 0 \end{pmatrix},\quad
 \s_z = \begin{pmatrix} 1 & 0 \\ 0 & -1 \end{pmatrix}.
\end{align*}
Then 
the Hamiltonian of the two-level atom is represented 
by 
$\triangle\s_z$. 
On the other hand let 
$a$ and $\add$ be the annihilation operator 
and the creation operator in $\LR$. 
They are given by 
\begin{align*}
 a=\frac{1}{\sqrt 2} \left(\frac{\rd}{\rd x}+ x\right),\quad 
 \add=\frac{1}{\sqrt2} \left(- \frac{\rd}{\rd x}+ x\right).
 \end{align*}
They satisfy the canonical commutation relation 
$[a,\add]=\one$, and $a^\ast=\add$, where $a^\ast$ denotes the adjoint of $a$.  
The harmonic oscillator is given by $\add a$, i.e., 
\[\add a=-\half\frac{\rd^2}{\rd x^2}+\half x^2-\half.\]
The harmonic oscillator $\add a$ is self-adjoint on $D(\frac{\rd^2}{\rd x^2})\cap D(x^2)$ and the spectrum of $\add a$ is 
$\spec(\add a)={\NN}\cup\{0\}$. 
The quantum Rabi Hamiltonian is defined 
as a self-adjoint operator on 
the tensor product Hilbert space 
${\CC}^2\otimes\LR$ by 
\begin{align*}
K=\triangle\s_z\otimes \one+\one\otimes \add a+ 
g\s_x\otimes (a+\add). 
\end{align*}
Here 
\[
\s_x\otimes (a+\add)=\sqrt 2 \begin{pmatrix} 0 & 1 \\ 1 & 0 \end{pmatrix}\otimes x\]
is the interaction term and 
$g\in\RR$ stands for a coupling constant. 
It can be seen that $K$ has the parity symmetry: 
\[[K, \s_z\otimes (-\one)^{\add a}]=0.\]
The parity symmetry sometimes referred to $\ZZ_2$-symmetry. 
Putting $g=0$ in $K$ we see that 
$\spec(\triangle \s_z\otimes\one+\one\otimes \add a)=\{n\pm \triangle\}_{n=0}^\infty$. 
It can be also seen that the spectrum of $K$ is purely discrete. 
We set 
\[\spec(K)=\{E_n(g)\}_{n=0}^\infty,\]
 where 
$E_n(g)\leq E_{n+1}(g)$ for $n\geq0$. 
Then 
$E_{2n}(0)=n-\triangle $ and 
$E_{2n+1}(0)=n+\triangle $. 
The Hurwitz type spectral zeta function of the quantum Rabi model is defined by 
\begin{align}
\label{hur}
\zeta_g(s;\tau)=\sum_{n=0}^\infty \frac{1}{(E_n(g)+\tau)^s},\quad \tau>0,\ \Re(s)>1.
\end{align}
As is seen in Lemma \ref{lowerbound} below, 
since $K+g^2\geq -\triangle$, 
instead of $\zeta_g(s;\tau)$, 
we are interested in investigating the asymptotic behaviour of 
\[\zeta_g(s;g^2+\tau)=\sum_{n=0}^\infty \frac{1}{(E_n(g)+g^2+\tau)^s},\quad \tau>0,\ \Re(s)>1\]
as $|g|\to\infty$. 

We review the results on spectra of the quantum Rabi Hamiltonian over the past decade. The quantum Rabi model has many applications in numerous fields ranging from not only physics but also pure mathematics. 
Although the spectrum of the quantum Rabi Hamiltonian is easy to obtain numerically, the exact investigation on the spectrum is difficult. 
In \cite{kus85} the energy level crossing of spectral curves 
$g\mapsto E_n(g)$ is investigated. 
In particular it is pointed out in \cite{kus85} that 
for $m\geq1$, 
there exist $m$ 
distinct solutions $g$ 
satisfying 
$E_{2m}(g)=E_{2m+1}(g)$ and
\[E_{2m}(g)+g^2=E_{2m+1}(g)+g^2=m.\] 
Thus $E_{2m}(g)(=E_{2m}(g))$ is a 
degenerate eigenvalue. 
There may exist however an eigenvalue of the form $m-g^2$ with $m\in\ZZ$ 
but not degenerate, which is observed in \cite{MPS14} only numerically.
See also \cite{WY14}, where a relationship between 
Lie algebra ${\mathfrak sl}_2$ and degenerate eigenvalues is given. 
It is shown in \cite{bra11,bra13a,bra13b} that 
non-degenerate eigenvalues are given by zeros of the so-called $G$-function
constructed by using $\ZZ_2$-symmetry of the quantum Rabi Hamiltonian. 
On the other hand {\it asymmetric} quantum Rabi model is defined by adding $\epsilon \s_x$ to $K$. It breaks $\ZZ_2$-symmetry. 
In \cite{LB15} the energy level crossing of spectral curves of 
asymmetric quantum Rabi model with $\epsilon=1/2$ is investigated, 
and then it is studied in \cite{RW22a,RBW21,KRW21} from mathematics point of view. 
In \cite{BZ21} the oscillation of map $n\mapsto E_n(g)$ 
is explicitly given for fixed $g$. While we are interested in 
the asymptotic behavior of $E_n(g)+g^2$ as $|g|\to\infty$. 
Numerically it is expected as 
\begin{align}\label{asy}
\lim_{|g|\to\infty}E_{2m}(g)+g^2=
\lim_{|g|\to\infty}E_{2m+1}(g)+g^2=m.
\end{align}
The meromorphic continuation of \eqref{hur} 
is proven by \cite{sug18} in the same way as 
\cite{IW05b} where that of non-commutative harmonic oscillators is discussed. 
It is also proven in \cite{rey23,RW21} by using a contour integral expression like the Riemann zeta function. 
See also \cite{IW05,och08,KW23} for 
investigating special values of the spectral zeta function of non-commutative harmonic oscillators.

The  spectral zeta function can be represented as 
\begin{align}
\label{tr}
\zeta_g(s;g^2+\tau)=
{\rm Tr}(K+g^2+\tau)^{-1}=
\frac{1}{\Gamma(s)}\int_0^\infty t^{s-1}
{\rm Tr} \left(e^{-t(K+g^2+\tau)}\right) \rd t,
\end{align}
where $\Gamma$ denotes the gamma function.
Hence it is worth while investigating 
the asymptotic behavior of the semigroup 
$e^{-t(K+g^2+\tau)}$ as $|g|\to \infty$. 
The integral kernel of $e^{-tK}$ are studied in 
\cite{HH14,RW21,RW22b,rey23}. 
Let 
\[\zeta(s;\tau)=\sum_{n=0}^\infty \frac{1}{(n+\tau)^s},\quad \Re (s)>1\]
be the Hurwiz zeta function. 
In this paper it is shown that 
\begin{align}
\label{mainthem}
\lim_{|g|\to\infty}
 \zeta_g(s;g^2+\tau)=
 2\zeta(s;\tau),\quad \Re (s)>1
 \end{align}
 in Theorem \ref{main2} 
 by the Feynman-Kac formula of 
$e^{-t(K+g^2+\tau)}$ in \eqref{tr}. 
A similar result is obtained in the recent paper 
\cite{RW23}, but the method is different from ours. 
Although both of \cite{RW23} and our paper investigate the semigroup generated by 
the quantum Rabi Hamiltonian,  
our method is an application of the Feynman-Kac formula established in \cite{HH14}. 
Moreover \eqref{asy} can be immediately proven in Corollary \ref{conv} as the 
byproduct of \eqref{mainthem}. 
The Feynman-Kac formula of the semigroup generated by 
Schr\"odinger operators with spin $1/2$ is established in 
\cite{HIL12}. We also refer to \cite{ALS83,ARS91,ALS98,GMM17,GV81}. 
We emphasize that Schr\"odinger operators with spin $1/2$ contain matrix coefficients but 
the integrand in the Feynman-Kac formula derived in \cite{HIL12} is {\it scalar}. Asymptotic behaviors of quantum Rabi model as $|g|\to\infty$ are also studied in e.g., \cite{HMS17}. 
In \cite{hir15,cai22} a relationship between SUSY and asymptotic behavior as $|g|\to\infty$ is discussed. 

In addition to the analysis of the  spectral zeta function, 
we discuss measures associated with the ground state of the quantum Rabi Hamiltonian. 
The quantum Rabi model can be regarded as the one mode version of the spin-boson model in quantum field theory. 
In \cite{HHL14} the path measure associated with the ground state of the spin-boson model is discussed. 
In this paper we also show the existence of the measure ${\Pi_\infty}$ associated with the ground state $\grr$ of the quantum Rabi Hamiltonian. 
Then 
under some condition we can see that 
\[(\grr, \cO \grr)=\EE_{{\Pi_\infty}}[f_{\cO}]\]
for some observable $\cO$ with a function $f_{\cO}$.

This paper is organized as follows. 
 In Section \ref{s2} we give Feynman-Kac formulas and show the first main theorem 
 as Theorem \ref{main1}. 
 In Section \ref{s3} asymptotic behavior of the  spectral zeta function is discussed 
 and the second main theorem is given as Theorem \ref{main2}. 
 In Section \ref{s5} we show the existence of the path measure associated with the ground state of the quantum Rabi Hamiltonian in Theorem \ref{main3}, 
and several applications are shown. 
In Section~\ref{shirai} some random process derived from the pair interaction is investigated.

 \section{Feynman-Kac formulas}
\label{s2}
\subsection{Unitary transformations}
In this section we define self-adjoint operators $H$ and $L$. 
Both are unitary equivalent to $K$. 
Let $\s=(\s_x,\s_y,\s_z)$. 
The rotation group in $\RR^3$ 
has an adjoint representation on $SU(2)$.
Let $n\in \RR^3$ be a unit vector and $\theta\in [0,2\pi)$. 
Thus 
$e^{(i/2)\theta n\cdot \s}(x\cdot \s) e^{-(i/2)\theta n\cdot \s}=Rx\cdot\s$, where 
$R$ denotes the $3\times 3$ matrix representing 
the rotation around $n$ with angle $\theta$.
In particular for $n=(0,1,0)$ and $\theta=\pi/2$, 
we have 
$U\s_x U^{-1}=\s_z$ and 
$U\s_z U^{-1}=-\s_x$, 
where 
\begin{align}\label{U}
U=e^{i\frac{\pi}{4}\s_y}.
\end{align}
Let us define the unitary operator $\cS_g$. 
Let $p=-i\frac{\rd }{\rd x}$ and $F$ denotes the Fourier transform on 
$\LR$. Then 
$\cS_g$ is defined by 
\begin{align}\label{S}
\cS_g=
\begin{pmatrix}F& 0\\
0&F
\end{pmatrix}
\begin{pmatrix}0& e^{i\sqrt 2 gp}\\
 e^{-i\sqrt 2 gp}&0
\end{pmatrix}.
\end{align}
Let $\gr$
be the normalized ground state of $\add a$, i.e., 
$\add a \gr=0$ and it is explicitly given by 
\[\gr(x)=\pi^{-1/4}e^{-|x|^2/2}.\]
Since $\gr$ is strictly positive, we can define the unitary operator 
$\cU_{\gr}:\LR\to L^2(\RR, \gr^2 \rd x)$ by 
\begin{align}\label{Ug}
\cU_{\gr} f=\gr^{-1} f.
\end{align}
We set the probability measure $\gr^2(x) \rd x$ on $\RR$ by $\rd \mu$, i.e.,
\[\rd \mu(x)=\frac{1}{\sqrt\pi}e^{-|x|^2}\rd x.\]
Define 
\[\cH=\CC^2\otimes L^2(\RR,\rd \mu). \]
Thus the composition of unitary operators 
\eqref{U}, \eqref{S} and \eqref{Ug} is denoted by 
${\cal U}_g=\cU_{\gr}\cS_gU$ and 
$K$ is transformed to 
the operator: 
\begin{align}
\label{H}
H={\cal U}_g K {\cal U}_g^{-1}=
\begin{pmatrix}-\half \frac{\rd^2}{\rd x^2}+x\frac{\rd}{\rd x} &0\\
0& -\half \frac{\rd^2}{\rd x^2}+x\frac{\rd}{\rd x}
\end{pmatrix}
-g^2-\triangle 
\begin{pmatrix}0&e^{i2\sqrt 2gx}\\
e^{-i2\sqrt 2gx}&0
\end{pmatrix}
\end{align}
in $\cH$. 
From this when $\triangle=0$, we see that 
$K\cong \one \otimes \add a-g^2$. 
Next we define self-adjoint operator $L$. 
Let $\cU=\cU_{\gr}U $. 
$L$ is defined by 
\begin{align}
L&=\cU K \cU^{-1}\nonumber=-\triangle\s_x\otimes\one+g\s_z\otimes(\bdd+b)+\one\otimes\bdd b\nonumber \\
&\label{L}
=
\begin{pmatrix}-\half \frac{\rd^2}{\rd x^2}+x\frac{\rd}{\rd x} &0\\
0& -\half \frac{\rd^2}{\rd x^2}+x\frac{\rd}{\rd x}
\end{pmatrix}-
\begin{pmatrix}-\sqrt 2 gx & \triangle\\
\triangle & \sqrt 2 gx
\end{pmatrix}.
\end{align}
Here 
$b$ and $\bdd$ are the annihilation operator and the creation operator in $L^2(\RR,\rd \mu)$, which are defined by 
$\gr^{-1}\ass \gr=\bss$. 
It is actually given by 
\[b=a+\frac{x}{\sqrt2},\quad \bdd=\add-\frac
{x}{\sqrt2}.\] 
They satisfy that 
\[\bdd+b=\sqrt 2x,\quad \bdd b=-\half \frac{\rd^2}{\rd x^2}+x\frac{\rd}{\rd x},\]
$[b,\bdd]=1$ and $b^\ast =\bdd$ in $L^2(\RR,\rd \mu)$. 
We shall use $H$ for studying the asymptotic behaviour of the  
spectral zeta function $\zeta_g(s;g^2+\tau)$, 
since the  term $g^2$ appears explicitly.
 On the other hand $L$ shall be used to construct the path measure associated with the ground state, 
 since the off diagonal part of $L$ is of the simple form: 
 $-\triangle\begin{pmatrix} 0 & 1 \\ 1 & 0 \end{pmatrix}$.

Although the lemma below is shown in \cite{sug18}, we show a proof of it since it is immediate. 
\bl{lowerbound}
We have
$\inf\spec(H)\geq -g^2-\triangle$. 
\el 
\proof
Since
$S=-\triangle 
\begin{pmatrix}0&e^{i2\sqrt 2gx}\\
e^{-i2\sqrt 2gx}&0
\end{pmatrix}$ is a bounded operator with 
the operator norm 
$\|S\|=\triangle$ and 
$\inf\spec(-\half \frac{\rd^2}{\rd x^2}+x\frac{\rd}{\rd x})=0$, 
the lemma follows from \eqref{H}. 
\qed

Setting $\triangle=0=g$ in $H$, we set 
\[H_0=\one\otimes\bdd b=
\begin{pmatrix}-\half \frac{\rd^2}{\rd x^2}+x\frac{\rd}{\rd x} &0\\
0& -\half \frac{\rd^2}{\rd x^2}+x\frac{\rd}{\rd x}
\end{pmatrix}.\]
\bl{weakzero}
We see that 
\begin{align*}
&(1) \ w-\lim_{|g|\to\infty} \cU_g=0,\\
&(2) \ \lim_{|g|\to\infty} (\phi, (H+g^2)\psi)=(\phi, H_0\psi), \quad \phi,\psi\in D(H).
\end{align*}
\el 
\proof
This follows from Riemann-Lebesgue Lemma. 
\qed

\subsection{Path integral representations}
In this section we shall construct Feynman-Kac formulas of 
$e^{-tH}$ and $e^{-tL}$. 
\subsubsection{Ornstein-Uhrenbeck process}
Let 
\[h=\bdd b=-\half \frac{\rd^2}{\rd x^2}+x\frac{\rd}{\rd x}.\]
Let $\pro X$ be the Ornstein-Uhrenbeck process on a probability space 
\[(\cX, \cB_\cX, \rP^x).\]
We see that 
${\rP^x}(X_0=x)=1$ and 
\begin{align*}
\int_{\RR} \EE_{\rP}^x \left[X_t \right] \rd \mu(x)=0,\quad 
\int _{\RR} \EE_{\rP}^x \left[X_t X_s\right] \rd \mu(x) =
\half {e^{-|t-s|}}.
\end{align*}
Here $\EE_{\rP}^x\left[\cdots\right] $ denotes the expectation with respect to 
the probability measure $\rP^x$. 
The generator of $X_t$ is given by $-h$ and 
\begin{align}\label{F1}
(\phi, e^{-th}\psi)_{L^2(\RR,\rd \mu)}=\int_{\RR} \EE_{\rP}^x\left[\overline{\phi(X_0)} \psi(X_t)\right] \rd \mu(x).
\end{align}
It is well known that the Ornstein-Uhrenbeck process can be represented by $1$D-Brownian motion. 
Let $(B_t)_{t\geq0}$ be $1$D-Brownian motion starting from $x$ at $t=0$ 
on a probability space 
$(\cX, \cB_\cX, \cW^0)$. 
The distributions of 
$X_s$ under $\rP^x$ 
and 
$e^{-s}\left(x+\frac{1}{\sqrt{2}}B_{e^{2s}-1}\right)$ under $\cW^0$ are identical. 
We denote this as 
\begin{align}
\label{X}
X_s\ \stackrel{d}{=}\ e^{-s}\left(x+\frac{1}{\sqrt{2}}B_{e^{2s}-1}\right)\quad s\geq0.
\end{align}
We shall often times use \eqref{X} in this paper. 
Then we can compute the density function $\kappa_t$ 
of $X_t$ as 
\begin{align*}
\EE^x_\rP[f(X_t)]=
\EE^0_{\cW}\left[f\left(e^{-t}\left(x+\frac{1}{\sqrt{2}}B_{e^{2t}-1}\right)\right)\right]=
\int_{\RR} f(y) \kappa_t(y,x) \rd y,
\end{align*}
where 
\begin{align}\label{kt}
\kappa_t(y,x)=\frac{1}{\sqrt{\pi(1-e^{-2t})}}
\exp\left(-\frac{|y-e^{-t}x|^2}{1-e^{-2t}}\right).
\end{align}
 The Mehler kernel $M_t$ is defined by 
{ \[M_t(x,y)=\frac{\gr(x)}{\gr(y)}\kappa_t(y,x)
=\frac{1}{\sqrt{\pi(1-e^{-2t})}}\exp\left(-\half\frac{(1+e^{-2t})(x^2+y^2)-4xy e^{-t}}{1-e^{-2t}}\right).
\]}
For the later use we extend 
the Ornstein-Uhrenbeck process 
$\pro X$ to 
the Ornstein-Uhrenbeck process 
$\proo {\hat X}$ on the whole real line on the probability space 
$(\bar \cX, \cB_{\bar \cX},\bar\rP^x)$. 
Here 
$\bar \cX=\cX\times \cX$, 
$\cB_{\bar \cX}=\cB_{\cX}\times \cB_{\cX}$ and 
$\bar\rP^x=
\rP^x\otimes \rP^x$. 
Define for $w=(w_1,w_2)\in \cX\times\cX$ 
\begin{align}\label{XX}
\hat X_t(w)=\left\{\begin{array}{ll} X_t(w_1),& t\geq0,\\
X_{-t}(w_2),&t<0.
\end{array}
\right.
\end{align}
Then $\hat X_t$ and $\hat X_{-s}$ for any $s,t>0$ are independent.
We also see that 
\begin{align}\label{F12}
(\phi, e^{-th}\psi)_{L^2(\RR,\rd \mu)}
=\int_{\RR} \EE_{\bar\rP^x}
\left[\overline{\phi(\hat X_0)} \psi(\hat X_t)\right] \rd \mu(x)
=\int_{\RR} \EE_{\bar\rP^x}
\left[\overline{\phi(\hat X_{-s})} \psi(\hat X_{t-s})\right] \rd \mu(x)
\end{align}
for any $0\leq s\leq t$. 

\subsubsection{Spin process}
In order to show the spin part by a path measure we introduce 
a Poisson process. 
Let $\pro N$ be a Poisson process on a probability space 
\[(\cY, \cB_\cY, \Pi)\] with the unit intensity, i.e., 
\[\EE_\Pi\left[\one_{\{N_t=n\}}\right] =\frac{t^n}{n!} e^{-t},\quad n\geq 0.\]
Note that $N_t$ is a nonnegative integer-valued random process, 
$N_0=0$ and 
$t\mapsto N_t$ is not decreasing. 
Furthermore $t\mapsto N_t$ is 
right continuous and its left limit exists (c\'adl\'ag). 

Let us consider $(N_{\la t})_{t\geq0}$ with $\la>0$. 
Let $0=t_0<t_1<t_2<\ldots$ 
be the jump points of $(N_{\la t})_{t\geq0}$ and 
we set $\delta_k=t_k-t_{k-1}$ for $k\in\NN$. 
Then 
$(\delta_k)_{k=1}^\infty$ are independent and identically distributed (i.i.d.) random variables such that 
the density function of $\delta_k$ is  
$\la e^{-\la t}\one_{[0,\infty)}(t)$. 
Hence the density function of $t_k=\sum_{i=1}^k\delta_i$ is
the $k$ fold convolutions of that of $\delta_i$ and 
it is actually given by 
 $\Gamma(\la,k)=\frac{\la^k}{(k-1)!}t^{k-1}e^{-\la t}\one_{[0,\infty)}(t)$. 
We shall use this in Section \ref{shirai}. 

Let 
\[{\ZZ}_2=\{-1,+1\}.\] 
Then 
for $u\in L^2(\ZZ_2)$, 
\[\|u\|^2_{L^2(\ZZ_2)}=\sum_{\alpha\in\ZZ_2}|u(\alpha)|^2.\] 
Introducing the norm on $\CC^2$ 
by 
$(u,v)_{\CC^2}=\sum_{i=1}^2 {\bar u_i v_i}$, 
we identify $\CC^2\cong L^2(\ZZ_2)$ 
by $\CC^2\ni u=\vvv{u_1\\u_2}\cong u(\alpha)$ with 
$u(+1)=u_1$ and $u(-1)=u_2$. 
Note that 
\[(u,v)_{\CC^2}= (u,v)_{L^2(\ZZ_2)}.\]
Under this identification 
$\s_x,\s_y$ and $\s_z$ are represented as 
the operators $U_x,U_y$ and $U_z$, respectively on $L^2(\ZZ_2)$ by 
\begin{align}
\label{SS}
U_xu(\alpha)=u(-\alpha),\quad U_yu(\alpha)=-i\alpha u(-\alpha),\quad U_zu(\alpha)=
\alpha u(\alpha),\quad u\in L^2(\ZZ_2).
\end{align}
We define 
\[S_t=(-1)^{N_t}\alpha,\quad \alpha\in{\ZZ}_2.\]
Here $\pro S$ is a dichotomous process which is called spin process in this paper. 
Let $\s_F=\half(\s_z+i\s_y)(\s_z-i\s_y)=-\s_x+\one$ be the fermionic 
harmonic oscillator. 
Then it is known that for $u,v\in\CC^2$, 
$
(u, e^{-t\s_F}v)_{\CC^2}=
 \sum_{\alpha\in\ZZ_2}\EE_\Pi[\overline{u(S_0)}v(S_t)]$. 
 Hence 
\begin{align}\label{F2}
(u, e^{t\s_x}v)_{\CC^2}=
e^t \sum_{\alpha\in\ZZ_2}\EE_\Pi[\overline{u(S_0)}v(S_t)].
\end{align}
We also extend Poisson process 
$\pro N$ to Poisson process $\proo {\hat N}$ on the whole real line on a probability space $(\bar \cY, \cB_{\bar \cY}, \bar \Pi)$.
Let 
$\bar \cY=\cY\times \cY$, 
$\cB_{\bar \cY}=
\cB_{ \cY}\times \cB_{\cY}$ and 
$\bar \Pi=\Pi\otimes\Pi$.
Let $\pro {\bar N}$ be a Poisson process on 
$(\cY, \cB_{\cY}, \Pi)$ such that 
$t\mapsto \bar N_t$ is 
left continuous and its right limit exists 
(c\`agl\`ad). 
Define 
for $w=(w_1,w_2)\in \cY\times\cY$, 
\[\hat N_t(w)=\left\{\begin{array}{ll}N_t(w_1),&t\geq0,\\
\bar N_{-t}(w_2),&t<0.
\end{array}
\right.\]
Then $\RR\ni t\mapsto \hat N_t$ is a c\`adl\`ag path. 
Note that $\hat N_t$ is independent of $\hat N_{-s}$ for any $s,t>0$. 
We define 
\[\hat S_t=(-1)^{\hat N_t}\alpha,\quad \alpha\in{\ZZ}_2.\]
By the shift invariance of $\hat S_s$ \cite[Proposition 3.44]{LHB20} we can see that for $u,v\in\CC^2$, 
\begin{align*}
(u, e^{t\s_x}v)_{\CC^2}=
e^t \sum_{\alpha\in\ZZ_2}\EE_{\bar \Pi}
[\overline{u(\hat S_0)}v(\hat S_t)]=
e^t \sum_{\alpha\in\ZZ_2}\EE_{\bar \Pi}
[\overline{u(\hat S_{-s})}v(\hat S_{t-s})]
\end{align*}
for any $0\leq s\leq t$. 

\subsubsection{Feynman-Kac formula}
Combining \eqref{F1} and \eqref{F2} we can represent 
$(\phi,e^{-tH}\psi)$ by a path measure. 
Let 
\[q_s=(S_s,X_s)\quad s\geq0\] be the $(\ZZ_2\times \RR)$-valued random process on the 
probability space 
$(\cX\otimes \cY,\cB_\cX\otimes\cB_\cY,\rP^x\otimes \Pi)$. 
We introduce the identification: 
\begin{align}
\label{HH}
\cH\cong L^2(\ZZ_2\times \RR)
\end{align}
by 
\begin{align}
\label{identification}
\vvv{\phi_+(x)\\ \phi_-(x)}\cong \delta_{+1\alpha}\phi_+(x)+
\delta_{-1 \alpha}\phi_-(x)=\phi(\alpha,x),\quad (\alpha,x)\in \ZZ_2\times\RR.
\end{align}
Here 
$\delta_{\alpha\beta}=\left\{\begin{array}{ll}1&\alpha=\beta\\
0&\alpha\neq \beta
\end{array}
\right.$ . 
We use identification \eqref{HH} without notices 
unless no confusion arises. 
Let $V:\ZZ_2\times \RR\to \RR$ be defined by 
\[V(\alpha,x)=2\sqrt 2\alpha x.\] Thus 
$V(q_{s-})=2\sqrt 2 S_{s-}X_s$. 
The Poisson integral 
$\int_0^{t+} V(q_{s-}) \rd N_s$ is a random process on 
the probability space 
$(\cX\otimes \cY,\cB_\cX\otimes\cB_\cY,\rP^x\otimes \Pi)$, which is defined by 
\[\left(\int_0^{t+} V(q_{s-}) \rd N_s\right) (w_1,w_2)=
\sum_{j=1}^n V(q_{s_j}(w_1,w_2))=
2\sqrt2\sum_{j=1}^n S_{s_j-}(w_1)X_{s_j}(w_2).\]
Here $\{s_j\}$ is the set of jump points 
such that 
$N_{s_j-}(w_1)\neq N_{s_j+}(w_1)$ for $0\leq s_j\leq t$. 
Also we set 
\[W(\alpha,x)=\sqrt 2\alpha x.\] 
Let 
\[{\bf E}\left[\ldots\right]=
\half\sum_{\alpha\in\ZZ_2}\int_{\RR}\EE_{\rP}^x\EE_\Pi\left[\ldots\right] \rd \mu(x) .\]

Let $a:\RR^3\to\RR^3$ be a vector potential and 
$b:\RR^3\to\RR^3$ be a magnetic field. 
Let $(B_t)_{t\geq0}$ be $3$D-Brownian motion starting from $x$ at $t=0$ 
on a probability space 
$(\cX, \cB_\cX, \cW^x)$. Let $\s\cdot b=\s_xb_1+\s_yb_2+\s_zb_3$. 
It is shown in \cite{HIL12} that for 
the self-adjoint operator 
\[h(a,b)=\half(-i\nabla-a)^2-\half 
\s\cdot b\] acting in $\CC^2\otimes L^2(\RR^3)$,
\begin{align}
(\phi, e^{-th(a,b)}\psi)
\label{ss}
=
e^t \sum_{\alpha\in\ZZ_2}
\int_{\RR^3} \EE_\Pi \EE_{\cW}^x\left[
\overline{\phi(S_0, B_0)}\psi(S_t,B_t) e^Z\right] 
\rd x,
\end{align}
where 
\[Z=-i\int_0^t a(B_s)\circ \rd B_s+\int_0^{t} S_s b_3(B_s)\rd s
+\int_0^{t+}
\log\left\{
\half(b_1(B_s)-iS_{s-} b_2(B_s))\right\}\rd N_s.\]
Here we used the identification $\CC^2\otimes L^2(\RR^3)\cong L^2(\ZZ_2\times\RR^3)$. 
From \eqref{ss} we can show the lemma below. 
\bl{FKF}
Let $\phi,\psi\in\cH$. 
Then under the identification \eqref{identification}, it follows that 
\begin{align}
\label{fkf}
&(\phi, e^{-tH}\psi)=
2e^te^{tg^2}
{\bf E}
\left[
\overline{\phi(q_0)}\psi(q_t) 
\triangle^{N_t}
e^{i g\int _0^{t+} V(q_{s-})\rd N_s}
\right],\\
&\label{fkfl}
(\phi, e^{-tL}\psi)=
2e^t
{\bf E}
\left[
\overline{\phi(q_0)}\psi(q_t) 
\triangle^{N_t}
e^{-g\int _0^{t} W(q_{s})\rd s}
\right].
\end{align}
\el
\proof 
It can be seen that 
\[\triangle \begin{pmatrix}0&e^{i2\sqrt 2gx}\\
e^{-i2\sqrt 2gx}&0
\end{pmatrix}=\half \s\cdot B,
\]
where 
$B=2(\triangle \cos(2\sqrt 2gx), -\triangle \sin(2\sqrt 2gx),0)$. 
Here $B$ can be regarded as a magnetic field. 
Replacing the dimension $d=3$, 
the Lebesgue measure $\rd x$, 
vector potential $a$, 
Brownian motion $\pro B$, 
and magnetic field $b$ in \eqref{ss} 
with 
$d=1$, 
probability measure 
$\rd \mu$, 
$a=0$, 
Orunstein-Uhrenbeck process $X_t$
and $B$, 
respectively, 
we can obtain \eqref{fkf}. 
Here we used 
\begin{align*}
\log\left\{
\triangle \left(
\cos(2\sqrt 2gX_s)+iS_{s-} \sin(2\sqrt 2gX_s)\right)\right\}
=\log\triangle+i 2\sqrt 2gS_{s-}X_s.
\end{align*}
Next 
we consider \eqref{fkfl} but the procedure is similar to that of \eqref{fkf}. 
Since 
\[\begin{pmatrix}-\sqrt 2 gx & \triangle\\ \triangle &\sqrt 2 gx
\end{pmatrix}= \half \s\cdot\tilde B,\]
where $\tilde B=2(\triangle,0,-\sqrt 2 gx)$, 
we can obtain \eqref{fkfl}. 
\qed
Since 
$\spec(H_0)=\{n\}_{n=0}^\infty$ and 
each $n$ is two fold degenerate, 
the Hurwitz type spectral zeta function of $H_0$ is 
$2\zeta(s;\tau)$. 
Let $\one_{\{N_t=0\}}$ be the indicator function on $\cY$ such that 
\[\one_{\{N_t=0\}}(\omega)=\left\{\begin{array}{ll}1,&N_t(\omega)=0,\\
0,& N_t(\omega)\geq1.
\end{array}
\right.\]
Similarly 
\[\one_{\{N_t\geq1\}}(\omega)=
\left\{\begin{array}{ll}
0,& N_t(\omega)=0,\\
1,&N_t(\omega)\geq 1.
\end{array}
\right.\]
The first main theorem in this paper is as follows. 
\bt{main1}
Let $\phi,\psi\in\cH$. Then 
\begin{align}
\label{dec}
(\phi, e^{-t(H+g^2)}\psi)=(\phi,e^{-tH_0}\psi)+
2e^t{\bf E}
\left[
\one_{\{N_t\geq1\}}
\overline{\phi(q_0)}\psi(q_t) 
\triangle^{N_t}
e^{i g\int _0^{t+} V(q_{s-})\rd N_s}
\right].
\end{align}
\et
\proof
By Lemma \ref{FKF} 
we have 
\begin{align*}
&(\phi, e^{-t(H+g^2)}\psi)=
2e^t
{\bf E}
\left[
\one_{\{N_t=0\}}
\overline{\phi(q_0)}\psi(q_t) 
\right]+
2e^t{\bf E}
\left[
\one_{\{N_t\geq1\}}
\overline{\phi(q_0)}\psi(q_t) 
\triangle^{N_t}
e^{i g\int _0^{t+} V(q_{s-})\rd N_s}
\right].
\end{align*}
The Feynman-Kac formula of $e^{-tH_0}$ is given by
\[
(\phi, e^{-tH_0}\psi)=
e^t \sum_{\alpha\in\ZZ_2}
\int_{\RR} \EE_{\rP}^x
\left[
\overline{\phi(\alpha,X_0)}\psi(\alpha,X_t) 
\right]\rd \mu(x)
=2e^t{\bf E}
\left[
\one_{\{N_t=0\}} \overline{\phi(q_0)}\psi(q_t) 
\right].
\]
Then the theorem follows. 
\qed
From Theorem \ref{main1} we immediately obtain inequalities. 
\bc{dia}
Let $\phi,\psi\in\cH$. 
Then 
\begin{align}
\label{d1}
&|(\phi, e^{-t(H+g^2)}\psi)|\leq 
|(\phi,e^{-tH_0}\psi)|+
2e^t{\bf E}
\left[
\one_{\{N_t\geq1\}}
|{\phi(q_0)}\psi(q_t)| 
\triangle^{N_t}
\right].\\
\label{d2}
&|(\phi, e^{-t(H+g^2)}\psi)|\leq (|\phi|, e^{-t(\triangle U_x+H_0)}|\psi|).
\end{align}
Here $U_x$ is given in \eqref{SS}. 
\ec
\proof
\eqref{d1} follows from Theorem \ref{main1}. 
From \eqref{fkf} it follows that 
\begin{align*}
|(\phi, e^{-t(H+g^2)}\psi)|
\leq
2e^t
{\bf E}
\left[
|{\phi(q_0)}\psi(q_t)| 
\triangle^{N_t}
\right]=
(|\phi|, e^{-t(\triangle U_x+H_0)}|\psi|).
\end{align*}
Then \eqref{d2} is obtained. 
\qed

\subsection{Euclidean Green functions}
Lemma \ref{FKF} can be extended to 
the path integral representations of Euclidean Green functions. 
Let $h=-\Delta/2$ and $\pro B$ be $1$D Brownian motion on $(\cX,\cB_{\cX},\cW^x)$. 
Suppose that $0< t_0< t_1<\ldots<t_n$. 
Let 
$
C^{\{t_0,t_1,\ldots,t_n\}}(A_0\times\cdots\times A_n)=
\{\omega\in \cX\mid \omega(t_j)\in A_j,j=0,1,\ldots,n\}$ 
be a cylinder set. Then it is known that 
\begin{align*}
\cW^x(C^{\{t_0,t_1,\ldots,t_n\}}(A_0\times\cdots\times A_n))
=\EE^x\left[
\left(
\prod_{j=0}^n\one_{A_j}(B_{t_j})\right)
\right].
\end{align*}
We know furthermore that for $f,g\in\LR$, 
\begin{align*}
\int_{\RR} 
\EE^x\left[
\left(
\prod_{j=0}^n\one_{A_j}(B_{t_j})\right)
\bar f(B_0) g(B_t)\right]
\rd x=
(f,e^{-t_0h}\one_{A_0}e^{-(t_1-t_0)h}\cdots e^{-(t_n-t_{n-1})h}\one_{A_n}e^{-(t-t_n)h}g).
\end{align*}
In Section \ref{s5}  
in order to construct a path measure ${\Pi_\infty}$ associated with the ground state of $L$, 
we also need the path integral representation 
of 
Euclidean Green functions of $L$ to compute finite dimensional distributions of $\Pi_\infty$. 
\bl{EG}
Let $f_j=f_j(\alpha,x)$ be bounded function on $\ZZ_2\times \RR$ for $j=0,1,\ldots,n$. 
Suppose that $0< t_0< t_1<\ldots<t_n$. 
Then 
\begin{align*}
&(\phi, e^{-t_0L}f_0e^{-(t_1-t_0)L} f_1 e^{-(t_2-t_1)L}\cdots
e^{-(t_n-t_{n-1})L}f_n
e^{-(t-t_{n})L}\psi)\\
&
=2e^{t} {\bf E}
\left[
\bar\phi(q_0)\psi(q_t)
\left(
\prod_{j=0}^n f_j(q_ {t_j})\right)
e^{-g \int_0^tW(q_s)\rd s}
\right].
\end{align*}
\el
\proof
Denote the natural filtrations of $\pro N$ and 
$\pro X$
by 
${\sN}_s=\boldsymbol{\s}(N_r,0\leq r\leq s)$ and 
${\sM}_s=\boldsymbol{\s}(X_r,0\leq r\leq s)$, respectively.
The Markov properties of $\pro N $ and $\pro X$ yield that
\begin{align*}
&
\left(e^{-sL}f e^{-tL} \phi\right) (\alpha,x)\\
&=
e^{s+t} 
\EE_\Pi
\EE_{\rP}^x
\left[
e^{-g\int_0^s W(q_r) \rd r} f(q_s)
\EE_\Pi^{S_s}\EE_{\rP}^{X_s} 
\left[
 e^{-g \int_0^t W(q_{r}) \rd r} 
\phi(q_{t})\right]\right]\\
&=
e^{s+t} 
\EE_\Pi
\EE_{\rP}^x
\left[e^{-g\int_0^s W(q_r) \rd r} f(q_s)
\EE_\Pi\EE_{\rP}^x \left[\left.
 e^{-g \int_0^t W(q_{r+s}) \rd r} 
\phi(q_{t+s})\right|{\sN}_s \times \sM_s\right] \right]\\
&=
e^{s+t} 
\EE_\Pi
\EE_{\rP}^x
\left[e^{-g\int_0^s W(q_r) \rd r} f(q_s)
 e^{-g \int_0^t W(q_{r+s}) \rd r} 
\phi(q_{t+s})\right]\\
&=
e^{s+t} 
\EE_\Pi
\EE_{\rP}^x
\left[e^{-g\int_0^{s+t} W(q_r) \rd r} f(q_s)
\phi(q_{t+s})\right].
\end{align*}
Repeating these procedures we have the lemma. 
\qed

\section{Asymptotic behaviors}
\label{s3}
\subsection{Asymptotic limits}
Let us consider  
$\lim_{|g|\to\infty}
\zeta_g(s;g^2+\tau)$ form now on. 
The  spectral zeta function can be represented as 
\[\zeta_g(s;g^2+\tau)=
\frac{1}{\Gamma(s)}\int_0^\infty t^{s-1}
{\rm Tr} \left(e^{-t(K+g^2+\tau)}\right) \rd t,\]
where $\Gamma$ denotes the gamma function.
Since
\[{\rm Tr} \left(e^{-t(K+g^2+\tau)}\right)=
{\rm Tr} \left(\cU_g e^{-t(H+g^2+\tau)}\cU_g\right)=
{\rm Tr} \left(e^{-t(H+g^2+\tau)}\right),\] 
the  spectral zeta function 
is identical to 
\begin{align}\label{zeta1}
\zeta_g(s;g^2+\tau)
=\frac{1}{\Gamma(s)}\int_0^\infty t^{s-1}
\sum_{\alpha\in\ZZ_2}
\sum_{n=0}^\infty (f_{\alpha n}, e^{-t(H+g^2+\tau)}f_{\alpha n}) \rd t,
\end{align}
where $\{f_{\alpha n}\}$ is a complete orthonormal system of $\cH$. 
Thus let us consider the asymptotic behaviour of 
$(\phi, e^{-t(H+g^2)}\psi)$ as $|g|\to\infty$. 
The idea is simple. We show that 
\begin{align*}
(\phi, e^{-t(H+g^2)}\psi)=(\phi,e^{-tH_0}\psi)+
2e^t{\bf E}
\left[
\one_{\{N_t\geq1\}}
\overline{\phi(q_0)}\psi(q_t) 
\triangle^{N_t}
e^{i g\int _0^{t+} V(q_{s-})\rd N_s}
\right]
\end{align*}
in Theorem \ref{main1}. 
Formally the most right-hand side above goes to zero as $|g|\to\infty$ by a version of Riemann-Lebesgue Lemma, 
and
by \eqref{zeta1} we can see that $\lim_{|g|\to\infty}
\zeta_g(s;g^2+\tau)
=2\zeta(s;\tau)$, since 
\[2\zeta(s;\tau)=
\frac{1}{\Gamma(s)}\int_0^\infty t^{s-1}
{\rm Tr} \left(e^{-t (H_0+\tau)}\right) \rd t.\] 
\bl{vacuum}
We have 
\begin{align}\label{V}
\lim_{|g|\to\infty}
(\one, e^{-t(H+g^2)}\one)
=
(\one, e^{-tH_0}\one).
\end{align}
\el
\proof
Let $\pro B$ be $1$D-Brownian motion. 
By \eqref{X} we have
\begin{align*}
e^{i g\int_0^{t+} V(q_{s-})\rd N_s}
\ \stackrel{d}{=}\ 
e^{i g2\sqrt 2\int_0^{t+} S_{s-}e^{-s}\left(x+\frac{1}{\sqrt{2}}B_{e^{2s}-1}\right) \rd N_s}
=e^{igA x }e^{i gB}.
\end{align*}
Here $A=2\sqrt 2 \alpha \int_0^{t+} (-1)^{N_{s-}} e^{-s}\rd N_s$ and 
$B=2 \alpha \int_0^{t+} (-1)^{N_{s-}}e^{-s} B_{e^{2s}-1} \rd N_s$. 
Both $A$ and $B$ are independent of $x$. 
We see that 
\begin{align*}
&\int_{\RR} e^{igA x }\rd \mu(x)=e^{-2g^2|\int_0^{t+} (-1)^{N_{s-}}e^{-s} \rd N_s|^2}\\
&\EE_{\cW}^0[e^{i gB}]=
e^{-2g^2\|\int_0^{t+} (-1)^{N_{s-}}e^{-s} \one_{[0,e^{2s}-1]}(\cdot) \rd N_s\|_{\LR}^2}. 
\end{align*}
Let
\[\xi=\left|\int_0^{t+} (-1)^{N_{s-}}e^{-s} \rd N_s\right|^2+
\left\|\int_0^{t+} (-1)^{N_{s-}}e^{-s} \one_{[0,e^{2s}-1]} (\cdot)\rd N_s\right\|_{\LR}^2.\]
We have 
\begin{align*}
(\one, e^{-t(H+g^2)}\one)
=\sum_{\alpha\in\ZZ_2}\EE_\Pi
\left[
\triangle^{N_t}
\EE_{\cW}^0[e^{i gB}]\int_{\RR} e^{igA x }\rd\mu(x) 
\right]
=
e^t
\sum_{\alpha\in\ZZ_2}
\EE_\Pi
\left[
\triangle^{N_t}e^{-2g^2\xi}
\right].
\end{align*}
Since $\xi=0$ on $\{w\in\cY\mid N_t(w)=0\}$, 
we have 
$e^t\sum_{\alpha\in\ZZ_2}\EE_\Pi[\one_{\{N_t=0\}}]=2=(\one,e^{-tH_0}\one)$. 
Then 
\begin{align*}
(\one, e^{-t(H+g^2)}\one)
=
(\one, e^{-tH_0}\one)
+
e^t
\sum_{\alpha\in\ZZ_2}
\EE_\Pi
\left[\one_{\{N_t\geq1\}}
\triangle^{N_t}e^{-g^2\xi }
\right].
\end{align*}
Since $\xi>0$ on $\{w\in\cY\mid N_t(w)\geq1\}$ 
by Lemma \ref{below} below, we obtain \eqref{V}.
Then the lemma follows. 
\qed
It remains to show $\xi\neq 0$ on $\{w\in\cY\mid N_t(w)\geq1\}$. 
\bl{below}
We see that $\xi>0$ on $\{w\in\cY\mid N_t(w)\geq1\}$. 
\el 
\proof
Let $\rho(k)=
\int_0^{t+} (-1)^{N_{s-}}e^{-s} \one_{[0,e^{2s}-1]} (k)\rd N_s$.
For each $w\in\{w\in\cY\mid N_t(w)\geq1\}$ 
there exist finite number of jump points $\{s_j\}_{j=1}^n$ such that 
$N_{s_j-}(w)\neq N_{s_j+}(w)$. Then 
\[\rho(k)=\sum_{j=1}^n (-1)^{N_{s_j-}}e^{-s_j} 
\one_{[0,e^{2s_j}-1]} (k).\]
Let $e^{2s_{N-1}}-1<k\leq e^{2s_{N}}-1$. Then $\one_{[0,e^{2s_j}-1]} (k)=0$ 
for $1\leq j\leq n-1$ and 
\[\rho(k)=(-1)^{N_{s_n-}}e^{-s_n}\neq 0.\]
Thus $\rho$ is nonzero function for each $w\in\cY$. 
Hence 
$\|\int_0^{t+} (-1)^{N_{s-}}e^{-s} \one_{[0,e^{2s}-1]} (\cdot)\rd N_s\|_{\LR}\neq0$ and $\xi>0$ is obtained. 
\qed
We extend Lemma \ref{vacuum} for general vectors $\phi,\psi\in\cH$. 
\bl{general}
Let $\phi,\psi\in\cH$. 
Then 
\begin{align*}
\lim_{|g|\to\infty}
(\phi, e^{-t(H+g^2)}\psi)
=
(\phi, e^{-tH_0}\psi).
\end{align*}
In particular 
\[s\text{-}\lim_{|g|\to\infty} e^{-t(H+g^2)}=e^{-tH_0}.\]
\el
\proof
By Theorem \ref{main1} it is enough to show that 
\begin{align}
\label{conv2}
\lim_{|g|\to\infty}
{\bf E}
\left[
\one_{\{N_t\geq1\}}
\overline{\phi(q_0)}\psi(q_t) 
\triangle^{N_t}
e^{i g\int _0^{t+} V(q_{s-})\rd N_s}
\right]=0.
\end{align}
It is seen that 
\begin{align*}
&{\bf E}
\left[
\one_{\{N_t\geq1\}}
\overline{\phi(q_0)}\psi(q_t) 
\triangle^{N_t}
e^{i g\int _0^{t+} V(q_{s-})\rd N_s}
\right]\\
&=
\half \sum_{\alpha\in\ZZ_2}
\EE_\Pi \left[
\one_{\{N_t\geq1\}}
\triangle^{N_t}
\int_{\RR} \overline{\phi(q_0)}
e^{-igAx}
\EE_{\cW}^0
\left[
\psi\left(S_t,e^{-t}x+\frac{e^{-t}}{\sqrt2}B_{e^{2t}-1}\right) e^{igB}
\right] \rd \mu(x)
\right]
\end{align*}
We shall show that 
for each $(w,x)\in\{w\in\cY\mid N_t(w)\geq1\}\times\RR$, 
\begin{align}
\label{conv1}
\lim_{|g|\to\infty}
\EE_{\cW}^0
\left[
\psi\left(
S_t,e^{-t}x+\frac{e^{-t}}{\sqrt2}B_{e^{2t}-1}\right)
e^{igB}
\right] =0.
\end{align}
Set 
$
G(X)=\psi(S_t,e^{-t}x+X)$, and 
\begin{align}
\label{P}&P(\cdot)={e^{-t}} \one_{[0,e^{2t}-1]}(\cdot),\\
\label{Q}
&Q(\cdot)=2 \alpha \int_0^{t+} (-1)^{N_{s-}}e^{-s} \one_{[0,e^{2s}-1]}(\cdot) \rd N_s
\end{align}
Let $\phi(f)$ be the Gaussian random variable indexed by 
a real-valued 
$f\in \LR$ 
on a probability space $(\cQ, \lambda)$ such that 
$\int_\cQ \phi(f)\phi(g)\rd \lambda(\phi)=\half (f,g)$ and $\int_\cQ\phi(f)\rd \lambda(\phi)=0$. 
In particular 
$\int_\cQ e^{z\phi(f)}\rd \lambda(\phi)=e^{\frac{z^2}{4}\|f\|^2}$ for $z\in\CC$ and 
$f\in\LR$. 
Note that $f\mapsto\phi(f)$ is linear. 
Then $B_t$ can be identified with $\sqrt 2 \phi(\one_{[0,t]})$. 
Thus we have
\begin{align*}
\EE_{\cW}^0
\left[
\psi\left(S_t,e^{-t}x+\frac{e^{-t}}{\sqrt2}B_{e^{2t}-1}\right) e^{igB}
\right] =
\EE_{\cW}^0
\left[
G\left(\frac{e^{-t}}{\sqrt2}B_{e^{2t}-1}\right) e^{igB}
\right]
=\int_\cQ G(\phi(P))e^{ig\phi(Q)}\rd \lambda(\phi). 
\end{align*}
Then by a version of Riemann-Lebesgue Lemma (see Lemma \ref{a1} below) 
it follows that 
\begin{align}
\label{RL}
\lim_{|g|\to\infty}
\int_\cQ G(\phi(P))e^{ig\phi(Q)}\rd \lambda(\phi) 
=0.
\end{align}
Then \eqref{conv1} holds true, and hence the Lebesgue dominated convergence theorem yields \eqref{conv2}. 
\qed

\begin{lemma}[Riemann-Lebesgue Lemma]
\label{a1}
Let $P$ and $Q$ be \eqref{P} and \eqref{Q}, 
and 
suppose that 
$(w,x)\in\{w\in\cY\mid N_t(w)\geq1\}\times\RR$. 
Then 
\eqref{RL} holds true. 
\el 
\proof
Since 
$(w,x)\in\{w\in\cY\mid N_t(w)\geq1\}\times\RR$, 
note that 
$\|Q\|\neq 0$ by Lemma \ref{below}. 
Let $G\in \sS(\RR)$ and $\check G$ be the inverse Fourier transform of $G$. 
We see that 
\begin{align*}
&\lim_{|g|\to\infty}
\int_\cQ G(\phi(P))e^{ig\phi(Q)}\rd \lambda(\phi),\\
 &=
\frac{1}{\sqrt{2\pi}}
\int_{\RR} 
 \check G(z)\left(\int_\cQ e^{iz\phi(P)} e^{ig\phi(Q)}\rd \lambda(\phi)\right)\rd z\\
&=\frac{1}{\sqrt{2\pi}}
\int_{\RR} 
 \check G(z) e^{-\frac{1}{4}(z^2\|P\|^2+2\Re zg(P,Q)+g^2\|Q\|^2)}
 \rd z.
\end{align*}
Since $\|Q\|\neq 0$, by the Lebesgue dominated convergence theorem 
it holds that 
\[\lim_{|g|\to\infty}
\int_\cQ G(\phi(P))e^{ig\phi(Q)}\rd \lambda(\phi)=0.\]
Let us consider a general $G$. 
It is known that 
$G(\phi(P))$ can be approximated in $L^2(\cQ)$ by 
functions of the form 
$G_\epsilon(\phi(P_1),\ldots,\phi(P_m))$ with $G_\epsilon\in \sS(\RR^m)$ 
so that 
\[\|G(\phi(P))-
G_\epsilon(\phi(P_1),\ldots,\phi(P_m))\|_{L^2(\cQ)}<\epsilon.\] 
See \cite[Lemma 1.5]{sim74}.
Hence 
\begin{align*}
&
\left|
\int_\cQ G(\phi(P))e^{ig\phi(Q)}\rd \lambda(\phi)\right|
\\
&\leq
\int_\cQ\left|
 G(\phi(P))-G_\epsilon(\phi(P_1),\ldots,\phi(P_m)) 
 \right|
 \rd \lambda(\phi)
+ 
 \left|
 \int_\cQ G_\epsilon(\phi(P_1),\ldots,\phi(P_m)) e^{ig\phi(Q)}\rd \lambda(\phi)\right|
 \\
&
\leq
\epsilon
+ \left|
\int_\cQ G_\epsilon(\phi(P_1),\ldots,\phi(P_m)) e^{ig\phi(Q)}\rd \lambda(\phi)\right|, 
\end{align*}
and 
$
\lim_{|g|\to\infty}
\left|
\int _\cQ G(\phi(P))e^{ig\phi(Q)}\rd \lambda(\phi)\right|\leq
\epsilon$ follows.
Then the proof is complete. 
\qed


We prepare two simple lemmas to estimate $\zeta_g^{\rm ran}(s;\tau)$. 
\bl{p1}
Suppose that $0<\triangle <\tau$. 
Let $0\leq s\leq1$ and $r=n+a$ with $n\in\NN$ and $0<a<1$.
Then 
\begin{align*}
&\|(H+g^2+\tau)^{-s}\phi\|\leq \left(1+\frac{\triangle}{\tau-\triangle}\right)^s\|(H_0+\tau)^{-s}\phi\|,\\
&\|(H+g^2+\tau)^{-r}\phi\|
\leq
\left(\frac{1}{\tau-\triangle}\right)^n
\left(1+\frac{\triangle}{\tau-\triangle}\right)^a
\|(H_0+\tau)^{-a}\phi\|.
\end{align*}
\el 
\proof
Notice that $\inf\spec(H+g^2+\tau)\geq\tau-\triangle$ 
and 
$\|(H_0+\tau)\phi\|\leq \|(H+g^2+\tau)\phi\|+\triangle\|\phi\|$. 
Then 
\[\|(H_0+\tau)(H+g^2+\tau)^{-1}\phi\|\leq 
\|\phi\|+\triangle\|(H+g^2+\tau)^{-1}\phi\|
\leq
\left(1+\frac{\triangle}{\tau-\triangle}\right)\|\phi\|.\] 
It concludes that 
$\|(H+g^2+\tau)^{-1}(H_0+\tau)\phi\|\leq 
\left(1+\frac{\triangle}{\tau-\triangle}\right)\|\phi\|$ and hence
\begin{align}
\|(H+g^2+\tau)^{-1}\phi\|\leq 
\left(1+\frac{\triangle}{\tau-\triangle}\right)\|(H_0+\tau)^{-1}\phi\|.
\end{align} 
Then L\"owen-Hainz inequality \cite{kat52} we see that 
\[\|(H+g^2+\tau)^{-s}\phi\|\leq \left(1+\frac{\triangle}{\tau-\triangle}\right)^s\|(H_0+\tau)^{-s}\phi\|\] for any $0\leq s\leq 1$. 
We show the second inequality. 
Since $\inf\spec(H+g^2+\tau)\geq\tau-\triangle$, 
we see that 
\begin{align*}
\|(H+g^2+\tau)^{-r}\phi\|
\leq 
\left(\frac{1}{\tau-\triangle}\right)^n
\|(H+g^2+\tau)^{-a}\phi\|.
\end{align*}
By the first inequality, 
the second one follows. 
\qed
We can also obtain an inequality between 
the semigroup $e^{-t(H+g^2+\tau)}$ and 
the resolvent $(H_0+\tau)^{-1}$. 
\bl{p2}
Let $0<s\leq 2$ and $r=2n+a$ with $n\in\NN$ and $0<a<2$. 
Then 
\begin{align*}
&(\phi, e^{-t(H+g^2+\tau)}\phi)\leq 
\frac{1}{t^{s}}
\left(\frac{s}{e}\right)^s
\left(1+\frac{\triangle}{\tau-\triangle}\right)^s\|(H_0+\tau)^{-s/2}\phi\|^2,\\
&(\phi, e^{-t(H+g^2+\tau)}\phi)\leq 
\frac{1}{t^{r}}
\left(\frac{r}{e}\right)^r
\left(\frac{\triangle}{\tau-\triangle}\right)^{2n}\left(1+\frac{\triangle}{\tau-\triangle}\right)^a\|(H_0+\tau)^{-a/2}\phi\|^2.
\end{align*}
\el 
\proof
Since 
$\sup_{\la\geq0}e^{-t\la}\la^p=\left(\frac{p}{t}\right)^pe^{-p}$ for any $p>0$, 
we have by Lemma \ref{p1}
\[(\phi, e^{-t(H+g^2+\tau)}\phi)\leq 
\left(\frac{p}{t}\right)^pe^{-p}(\phi, (H+g^2+\tau)^{-p}\phi).\]
From Lemma \ref{p1} we can show the inequalities. 
\qed

Let $\Psi_{\alpha\,n}\in \cH$, $n\geq0$, $\alpha\in\ZZ_2$, be given by 
\[\Psi_{\alpha\,n} (\beta,x)=\delta_{\alpha \beta}h_n(x)\cong 
\left\{\begin{array}{ll}
\vvv{0\\ 1}\otimes h_n(x)& \alpha=-1,\\
\ \\
\vvv{1\\ 0}\otimes h_n(x)&\alpha=+1,\end{array}
\right.\] 
where $h_n$ denotes the $n$th Hermite polynomial. 
$\{\Psi_{\alpha\,n}\}_{n, \alpha}$ is a complete orthonormal system of $\cH$ 
and 
\[H_0\Psi_{\alpha\,n}=n\Psi_{\alpha\,n},\quad \alpha\in\ZZ_2.\]
The second main theorem in this paper is as follows. 
\bt{main2}
Suppose that $\Re(s)>1$ and $0<\triangle<\tau$. 
Then it follows that 
\[\lim_{|g|\to\infty} \zeta_g(s;g^2+\tau)=2\zeta(s;\tau).\]
\et
\proof
We have
\begin{align*}
\zeta_g(s;g^2+\tau)
=\frac{1}{\Gamma(s)}
\int_0^\infty t^{s-1}
\sum_{\alpha\in\ZZ_2}
\sum_{n=0}^\infty (\Psi_{\alpha\,n}, e^{-t(H+g^2+\tau)}\Psi_{\alpha\,n}) \rd t. 
\end{align*}
We shall show that 
one can exchange $\lim_{|g|\to\infty}$ and 
$\int_0^\infty 
\sum_{\alpha\in\ZZ_2}\sum_{n=0}^\infty\ldots\rd t$. 
To show this we construct a function $\rho(t,n)$ independent of $g$ 
such that 
\begin{align*}
&(1)\ (\Psi_{\alpha\,n}, e^{-t(H+g^2+\tau)}\Psi_{\alpha\,n}) \leq \rho(t,n),\\
&(2)\ \int_0^\infty\sum_{\alpha\in\ZZ_2}\sum_{n=0}^\infty t^{|s|-1}\rho(t,n) \rd t<\infty.
\end{align*} 
Set $c_s=\left(\frac{|s|}{e}\right)^{|s|}\left(1+\frac{\triangle}
{\tau-\triangle}\right)^{|s|}$ and 
$a_k=\|(H_0+\tau)^{-k}\Psi_{\alpha\,n}\|^2=\frac{1}{(n+\tau)^{2k}}$ for simplicity. 
Let $1<|s|\leq2$ and $1<r<|s|$. 
By Lemma \ref{p2} we obtain that 
\begin{align*}
(\Psi_{\alpha\,n}, e^{-t(H+g^2+\tau)}\Psi_{\alpha\,n})
\leq
\frac{c_{r}a_{r/2}}{t^r}
\one_{[0,1)}(t) 
+
\frac{c_{2}a_1}{t^2}\one_{[1,\infty)}(t).
\end{align*}
Set the right-hand side above as $\rho(t,n)$. 
Then 
\begin{align*}
&\int_0^\infty 
\sum_{\alpha\in\ZZ_2}
\sum_{n=0}^\infty t^{|s|-1}
\rho(t,n)\rd t\leq
{c_{r}
\zeta(r)}
\int_0^1
t^{|s|-r-1}
\rd t
+
{c_{2} 
\zeta(2)}
\int_1^\infty
t^{|s|-3}
 \rd t
<\infty.
\end{align*}
Next let $|s|>2$ and $|s|<r=2n+a$, where $n\in\NN$ and $0\leq a<2$. 
By Lemma \ref{p2} again we see that 
\begin{align*}
&(\Psi_{\alpha\,n}, e^{-t(H+g^2+\tau)}\Psi_{\alpha\,n}) \leq
\frac{c_{2}a_1}{t^2}\one_{[0,1)}(t) 
+\frac{a_{a/2}}{t^r}
\left(\frac{r}{e}\right)^r
\left(\frac{1}{\tau-\triangle}\right)^{2n}
\left(1+\frac{\triangle}{\tau-\triangle}\right)^a
\one_{[1,\infty)}(t).
\end{align*}
Set the right-hand side above as $\rho(t,n)$. 
Then 
\begin{align*}
&\int_0^\infty 
\sum_{\alpha\in\ZZ_2}
\sum_{n=0}^\infty t^{|s|-1}
\rho(t,n)\rd t\\
&\leq
{c_2}
\zeta(2)
\int_0^1
t^{|s|-3} 
\rd t+
\left(\frac{r}{e}\right)^r 
\left(\frac{1}{\tau-\triangle}\right)^n
\left(1+\frac{\triangle}{\tau-\triangle}\right)^a
\zeta(a)\int_1^\infty
t^{|s|-r-1} 
\rd t<\infty.
\end{align*}
Hence by the Lebesgue dominated convergence theorem 
one can exchange 
$\lim_{|g|\to\infty}$ and $\int_0^\infty \sum_{\alpha\in\ZZ_2}\sum_{n=0}^\infty\ldots \rd t$, and 
we have
\begin{align*}
\lim_{|g|\to\infty}
\zeta_g(s;g^2+\tau)
&=\frac{1}{\Gamma(s)}
\int_0^\infty t^{s-1}
\sum_{\alpha\in\ZZ_2}
\sum_{n=0}^\infty (\Psi_{\alpha\,n}, e^{-t(H_0+\tau)}\Psi_{\alpha\,n}) \rd t\\
&=\frac{2}{\Gamma(s)}
\int_0^\infty t^{s-1}
\sum_{n=0}^\infty e^{-t(n+\tau)} \rd t=2\zeta(s;\tau). 
\end{align*}
Then the theorem follows. 
\qed
Let 
\begin{align}\label{hana}
P=\cU_g(\s_z\otimes (-\one)^{\add a}) \cU_g^{-1}=\s_x\otimes (-\one)^{\bdd b}
\end{align} be the parity operator. 
Let 
$\cH=\cH_+\oplus \cH_-$, where 
$\cH_\pm$ denotes the party $\pm1$ subspace. 
Since $H$ has the parity symmetry, 
$H$ is reduced by $\cH_\pm$. 
Define $H_\pm=H\lceil_{\cH_\pm}$. 
Let $\{E_{\pm, n}(g)\}$ be the eigenvalues of $H_\pm$ and 
$\zeta_{\pm,g}(s;g^2+\tau)$ be the  
spectral zeta function of 
$H_\pm$: 
\[\zeta_{\pm,g}(s;g^2+\tau)=\sum_{n=0}^\infty \frac{1}{(E_{\pm, n}(g)+g^2+\tau)^s}.\] 

\bc{parity}
Let $\Re (s)>1$ and $0<\triangle<\tau$. 
Then 
\[\lim_{|g|\to\infty}\zeta_{\pm,g}(s;g^2+\tau)=\zeta(s;\tau).\]
\ec
\proof
Let 
$\{\Psi_{n\,\alpha}\}_{(\alpha,n)=(-1,2m+1), (1,2m),m\geq0}$ be 
a complete orthonormal system of $\cH_+$ and 
$\{\Psi_{n\,\alpha}\}_{(\alpha,n)=(1,2m+1), (-1,2m),m\geq0}$ 
that of $\cH_-$. 
We have
\begin{align*}
&\zeta_{+,g}(s;g^2+\tau)
=\frac{1}{\Gamma(s)}
\int_0^\infty t^{s-1}
\sum_{m=0}^\infty\sum_{(\alpha,n)=(-1,2m+1),(1,2m)}
(\Psi_{\alpha\,n}, e^{-t(H_{+}+g^2+\tau)}\Psi_{\alpha\,n}) \rd t,\\
&\zeta_{-,g}(s;g^2+\tau)
=\frac{1}{\Gamma(s)}
\int_0^\infty t^{s-1}
\sum_{m=0}^\infty\sum_{(\alpha,n)=(-1,2m),(-1,2m)}
(\Psi_{\alpha\,n}, e^{-t(H_{-}+g^2+\tau)}\Psi_{\alpha\,n}) \rd t.
\end{align*}
Then 
$\lim_{|g|\to\infty}\zeta_{\pm,g}(s;g^2+\tau)=\zeta(s;\tau)$ can be derived in the same way as Theorem \ref{main2}. 
\qed
Let us consider  other asymptotic behaviors of $\zeta_g(s;g^2+\tau)$. 
\bc{zeta2}Suppose that $\Re(s)>1$. 
\bi\item[(1)] Suppose that $\tau- \triangle>0$. 
Then 
\[
\lim_{g\to0}\zeta_g(s;g^2+\tau)=
\zeta(s;\tau+\triangle)
+
\zeta(s;\tau-\triangle).
\]
\item[(2)]
 We have 
\[
\lim_{\triangle\to0} \zeta_g(s;g^2+\tau)=2\zeta(s;\tau).\]
\ei
\ec
\proof
We shall show 
that 
\begin{align}
\label{CC1}&\lim_{g\to0}
 (\phi, e^{-t(H+g^2)}\psi)=
 (\phi,e^{-t(\triangle U_x+\add a)}\psi),\\
\label{CC2}&\lim_{\triangle\to0}
 (\phi, e^{-t(H+g^2)}\psi)=(\phi,e^{-tH_0}\psi).
\end{align}
The corollary follows from \kak{CC1} and \kak{CC2}. 
\eqref{dec} implies that 
\begin{align}\label{i1}
 (\phi, e^{-t(H+g^2)}\psi)=
 (\phi,e^{-tH_0}\psi)+ 
 2e^t{\bf E}
\left[
\one_{\{N_t\geq1\}}
\overline{\phi(q_0)}\psi(q_t) 
\triangle^{N_t}
e^{i g\int _0^{t+} V(q_{s-})\rd N_s}
\right].
\end{align}
Since
\[
\lim_{g\to0}
 (\phi,e^{-t(H+g^2)}\psi)
=
 (\phi,e^{-tH_0}\psi)+
2e^t{\bf E}
\left[
\one_{\{N_t\geq1\}}
\overline{\phi(q_0)}\psi(q_t) 
\triangle^{N_t}
\right],
\]
\kak{CC1} follows. 
Since $N_t\neq0$ in the integrand of \eqref{i1}, 
\kak{CC2}  follows from $\lim_{\triangle\to0}\triangle^{N_t}=0$. 
\qed

\subsection{Convergence of eigenvalues}
In Corollary \ref{parity} we shown that 
$\lim_{|g|\to\infty}\zeta_{\pm,g}(s;g^2+\tau)=\zeta(s;\tau)$. 
Next let us consider the convergence of $E_n(g)+g^2$ as $|g|\to\infty$ for each $n$. 
\bc{conv}
For any $n\geq0$, 
we have 
$\lim_{|g|\to\infty} E_{\pm,n}(g)+g^2=n$.
\ec
\proof
We consider the case of parity $=+1$. 
The proof for parity $=-1$ is the same. 
Let 
$\xi(g)=(1/(E_{+,n}(g)+g^2+\tau))_n$ and 
$\xi(\infty)=(1/(n+\tau))_n$.
We have $\xi(g)\in \bigcap_{p>1}\ell_p$ for $0\leq g\leq \infty$. 
We have shown that $\|\xi(g)\|_{\ell_p}\to\|\xi(\infty)\|_{\ell_p}$ as $|g|\to\infty$ for $p>1$ in 
Corollary~\ref{parity}. Then for sufficiently large $R>0$, 
\[\{\xi(g)\mid 0\leq g\leq\infty\}\subset B_R,\]
where $B_R$ denotes the closed ball in $\ell_2$ centered at $0$ with radius $R$. 
$B_R$ is weak-$\ast$ compact. 
Then for any sequence $\{\xi(g_j)\}_j$ such that $g_j\to\infty$ as $j\to\infty$, 
there exists a subsequence $j'$ 
such that $\xi(g_{j'})$ converges to an $\ell_2$-sequence 
$a=(a_n)_n$ in the weak $\ast$ topology as $j'\to\infty$. 
Let $e_k=(\delta_{kn})_n$. 
Hence
\[\lim_{j'\to\infty} \frac{1}{E_{+,k}(g_{j'})+g_{j'}^2+\tau}=
\lim_{j'\to\infty} (e_k, \xi(g_{j'}))_{\ell^2}
=(e_k,a)_{\ell^2}=a_k.\] 
This implies that \[\lim_{|g|\to\infty} \frac{1}{E_{+,k}(g)+g^2+\tau}=a_k\] for each $k$. 
We shall show that $a_k=1/(k+\tau)$. 
We show that 
$e^{-t(H_{+}+g^2+\tau)}\to
 e^{-t(H_0+\tau)}$ as $|g|\to\infty$ in the weak sense in Lemma \ref{general}. 
It implies that 
$e^{-t(H_{+}+g^2+\tau)}\to e^{-t(H_0+\tau)}$ in the strong sense, 
 and furthermore it is equivalent to that 
$H_{+}+g^2+\tau \to H_0+\tau$ in the strong resolvent sense. 

Let $A$ be self-adjoint. 
Let $a<b$ and $\la=\frac{a+b}{2}+i(\frac{b-a}{2})$. Then 
it is a fundamental fact that 
$\|(A-\la)^{-1}\|\leq \frac{\sqrt 2}{b-a}$ if and only if 
$(a,b)\cap \spec(A)=\emptyset$. 

Suppose that 
$(a,b)\cap \spec(H_{+}+g^2+\tau)=\emptyset$ for any $g>N$ with some $N$. 
Then Banach-Steinhaus theorem yields that 
\[\|(H_0+\tau-\la)^{-1}\|\leq \liminf_{|g|\to\infty}
\|(H_{+}+g^2+\tau-\la)^{-1}\|
\leq\frac{\sqrt 2}{b-a}.\] 
Then
$(a,b)\cap \spec(H_0+\tau)=\emptyset$. 
By the contraposition 
if $x\in \spec(H_0+\tau)$, there exists $x_g\in 
\spec(H_{+}+g^2+\tau)$ such that $\lim_{|g|\to\infty} x_g=x$. 
Since $\spec(H_0+\tau)=\{n+\tau\}_{n=0}^\infty$, 
there exist eigenvalues 
$e(n,g)+\tau\in\spec(H_{+}+g^2+\tau)$ such that 
$\lim_{|g|\to\infty}e(n,g)=n$. 
Let $\cE=\{e(n,\cdot)\mid n\in {\NN}\cup\{0\}\}$. 
Note that for any fixed $n$, $e(0,g),\ldots, e(n,g)$ are different each others for all sufficiently large $g$. 
Suppose that 
there exists $m\geq0$ 
such that 
$\lim_{|g|\to\infty} E_{+,m}(g)+g^2=p\not\in {\NN}\cup\{0\}$ and 
set $e(p,g)=E_{+,m}(g)+g^2$. 
Note that $e(p,\cdot)\not\in \cE$. Then we have
\begin{align*}
\frac{1}{(p+\tau)^s}+\sum_{n=0} ^N\frac{1}{(n+\tau)^s}
&=
\lim_{|g|\to\infty}
\left(
\frac{1}{(e(p,g)+\tau)^s}+
\sum_{n=0}^N \frac{1}{(e(n,g)+\tau)^s}\right)\\
&\leq
\lim_{|g|\to\infty}
\sum_{n=0} ^\infty\frac{1}{(E_{+,n}(g)+g^2+\tau)^s}
=
 \sum_{n=0}^\infty \frac{1}{(n+\tau)^s}.
 \end{align*}
Hence
\begin{align*}
\frac{1}{(p+\tau)^s}<
 \sum_{n=N+1}^\infty \frac{1}{(n+\tau)^s}
 \end{align*}
for any $N$. 
Then it contradicts. 
Thus $\lim_{|g|\to\infty} {E_{+,m}(g)}+g^2\in{\NN}\cup\{0\}$ for any $m$. 
Let $\cE_n=\{m\mid \lim_{|g|\to\infty} {E_{+,m}(g)}+g^2=n\}$. 
Note that $\# \cE_n\geq1$ for $n\geq0$. 
Then \[\sum_{n=0} ^N\frac{\#\cE_n}{(n+\tau)^s}=
\lim_{|g|\to\infty}\!\!\!\!\sum_{m\in \cup_{n=0}^N\cE_n} \frac{1}{(e(m,g)+\tau)^s}
\leq
\lim_{|g|\to\infty}\sum_{m=0}^\infty \frac{1}{(E_{+,m}(g)+g^2+\tau)^s}
=\sum_{n=0}^\infty \frac{1}{(n+\tau)^s}.\]
Let $\epsilon>0$. Taking sufficiently large $N$, we have 
\[0<\sum_{n=0} ^N \frac{\#\cE_n}{(n+\tau)^s}- \sum_{n=0}^N \frac{1}{(n+\tau)^s}<\sum_{n=N+1}^\infty \frac{1}{(n+\tau)^s}<\epsilon.\] 
Thus we conclude that $\#\cE_n=1$ for any $n$, and
$\lim_{|g|\to\infty} {E_{+,k}(g)+g^2}=k$. 
Then 
$a_k=1/(k+\tau)$ for any $k\in{\NN}\cup\{0\}$. 
\qed

{\it Alternative proof of Theorem \ref{conv}}: \\
\proof
Let $a_n(g)=E_{+,n}(g)+g^2+\tau$. 
Spectral zeta function 
$\zeta_{+,g}(s)$ can be regarded as Laplace transform of measures:
\begin{align*}
&\zeta_{+,g}(s+u)=\int_0^\infty e^{-st}\mu_g^{(u)}(\rd t),\\
&\zeta(s+u)=\int_0^\infty e^{-st}\mu_\infty^{(u)}(\rd t),
 \end{align*}
where
\[
\mu_g^{(u)}(\rd t)=\sum_{n=0}^\infty \frac{1}{a_n(g)^u} \delta_{\log a_n(g)},\quad
\mu_\infty^{(u)}(\rd t)=\sum_{n=0}^\infty \frac{1}{n^u}\delta_{n}.
\]
Since 
$\lim_{|g|\to\infty} \zeta_{+,g}(s+u)=
\zeta(s+u)$ by Corollary \ref{parity}, 
the Laplace transform of 
$\mu_g^{(u)}(\rd t)$ converges to 
$\mu_\infty^{(u)}(\rd t)$.
Thus 
\[\lim_{|g|\to\infty}\mu_g^{(u)}(\rd t)=\mu_\infty^{(u)}(\rd t)\] 
in the weak sense. 
Take the $\epsilon$-neighborhood of $n$, 
which is denoted by 
$U_\epsilon(n)$. 
Suppose that $U_\epsilon(n)\cap ({\NN}\cup\{0\})=n$. 
By the weak convergence we have
\[
\sum_{m=0}^\infty \frac{1}{a_m(g)^u}\one_{U_\epsilon}(\log a_m(g))=
\mu_g^{(u)}(U_\epsilon(n))
\to
\mu_\infty^{(u)}(U_\epsilon(n))=\frac{1}{n^u}.
\]
In other words, 
$n^u\mu_g^{(u)}(U_\epsilon(n))
\to 1$ and hence 
\[
\sum_{m=0}^\infty \left(\frac{n}{a_m(g)}\right)^u 
\one_{[n^ue^{-\epsilon u}, n^ue^{u\epsilon }]}
\left(\left(\frac{n}{a_m(g)}\right)^u\right)\to 1.
\]
Suppose that $1/2<n^ue^{-\epsilon u}$. 
Only the term such that $\left(\frac{n}{a_m(g)}\right)^u>1/2$ is left, but the limit is $1$. 
Then there exists a unique $m(g)$ such that 
$\frac{n}{a_{m(g)}(g)}\in [n^u e^{-u\epsilon},n^ue^{u\epsilon}]$. 
Then we have 
$\lim_{|g|\to\infty} \frac{n}{a_{m(g)}(g)}=1$. 
Thus in the same way as the proof of Theorem \ref{conv} we can show that 
 $\lim_{|g|\to\infty} {E_{+,k}(g)+g^2}=k$. 
 \qed

\subsection{Asymmetric quantum Rabi model}
For $\epsilon>0$, we can define the asymmetric quantum Rabi Hamiltonian by 
\[K_\epsilon=\triangle\s_z\otimes \one+\one\otimes\add a+g\s_x\otimes(\add+a)+\epsilon\s_x.\]
The term $\epsilon \s_x$ breaks $\ZZ_2$-symmetry of $K_\epsilon$ but numerically it is shown that \cite[Figure 2]{LB15} 
\[\lim_{|g|\to\infty}E_{2m}(g)+g^2=m-\epsilon,\quad 
\lim_{|g|\to\infty}E_{2m+1}(g)+g^2=m+\epsilon.\] 
Let $\spec(K_\epsilon)=\{E_{n,\epsilon}(g)\}_{n=0}^\infty$. 
We define
\[\zeta_{\epsilon,g}(s;g^2+\tau)
=\sum_{n=0}^\infty \frac{1}{(E_{n,\epsilon}(g)+g^2+\tau)^s}.\]
In $\cH$,  
by $\cU_g$, $K_\epsilon$ is transformed to 
\begin{align}
\label{Heps}
H_\epsilon={\cal U}_g K_\epsilon {\cal U}_g^{-1}=
\begin{pmatrix}-\half \frac{\rd^2}{\rd x^2}+x\frac{\rd}{\rd x} +\epsilon&0\\
0& -\half \frac{\rd^2}{\rd x^2}+x\frac{\rd}{\rd x}-\epsilon
\end{pmatrix}
-g^2-\triangle 
\begin{pmatrix}0&e^{i2\sqrt 2gx}\\
e^{-i2\sqrt 2gx}&0
\end{pmatrix}.
\end{align}
\bl{FKFeps}
Let $\phi,\psi\in\cH$. 
Then under the identification \eqref{identification}, it follows that 
\begin{align*}
(\phi, e^{-t(H_\epsilon+g^2)}\psi)=
2e^t
{\bf E}
\left[
\overline{\phi(q_0)}\psi(q_t) 
\triangle^{N_t}
e^{-\epsilon\int_0^t S_s\rd s}
e^{i g\int _0^{t+} V(q_{s-})\rd N_s}
\right].
\end{align*}
Furthermore
\begin{align*}
(\phi, e^{-t(H_\epsilon+g^2)}\psi)=
(\phi, e^{-t(H_0+\epsilon U_z)}\psi)
+
2e^t
{\bf E}
\left[
\one_{\{N_t\geq1\}}
\overline{\phi(q_0)}\psi(q_t) 
\triangle^{N_t}
e^{-\epsilon\int_0^t S_s\rd s}
e^{i g\int _0^{t+} V(q_{s-})\rd N_s}
\right].
\end{align*}
Here $U_z$ is given by \eqref{SS}. 
\el
\proof
The proof is similar to that of Lemma \ref{FKF}. 
Then we omit it. 
\qed
Since $\epsilon U_z$ is bounded with $\|\epsilon U_z\|=|\epsilon|$. notice that $\inf\spec(H_\epsilon+g^2+\tau)\geq\tau-\triangle-|\epsilon|$ 
and 
$\|(H_0+\tau)\phi\|\leq \|(H_\epsilon+g^2+\tau)\phi\|+(\triangle+|\epsilon|)\|\phi\|$. 
In a similar manner to Lemmas \ref{p1} and \ref{p2}, 
we obtain lemmas below. 
\bl{p1eps}
Suppose that $0<\triangle+|\epsilon| <\tau$. 
Let $0\leq s\leq1$ and $r=n+a$ with $n\in\NN$ and $0<a<1$.
Then 
\begin{align*}
&\|(H_\epsilon+g^2+\tau)^{-s}\phi\|\leq \left(1+\frac{\triangle+|\epsilon|}{\tau-\triangle-|\epsilon|}\right)^s\|(H_0+\tau)^{-s}\phi\|,\\
&\|(H_\epsilon+g^2+\tau)^{-r}\phi\|
\leq
\left(\frac{1}{\tau-\triangle-|\epsilon|}\right)^n
\left(1+\frac{\triangle+|\epsilon|}{\tau-\triangle-|\epsilon|}\right)^a
\|(H_0+\tau)^{-a}\phi\|.
\end{align*}
\el 
\bl{p2eps}
Suppose that $0<\triangle+|\epsilon| <\tau$. 
Let $0<s\leq 2$ and $r=2n+a$ with $n\in\NN$ and $0<a<2$. 
Then 
\begin{align*}
&(\phi, e^{-t(H_\epsilon+g^2+\tau)}\phi)\leq 
\frac{1}{t^{s}}
\left(\frac{s}{e}\right)^s
\left(1+\frac{\triangle+|\epsilon|}{\tau-\triangle-|\epsilon|}\right)^s\|(H_0+\tau)^
{-s/2}\phi\|^2,\\
&(\phi, e^{-t(H_\epsilon+g^2+\tau)}\phi)\leq 
\frac{1}{t^r}
\left(\frac{r}{e}\right)^r
\left(\frac{\triangle}{\tau-\triangle-|\epsilon|}\right)^{2n}\left(1+\frac{\triangle+|\epsilon|}{\tau-\triangle-|\epsilon|}\right)^a\|(H_0+\tau)^{-a/2}\phi\|^2.
\end{align*}
\el 
It is parallel with that of $\zeta_g(s;g^2+\tau)$ 
to discuss the asymptotic behavior of the  spectral zeta function $\zeta_{\epsilon,g}(s;g^2+\tau)$ of $H_\epsilon$. 
\bt{SSS}
Suppose that $\Re(s)>1$ and $0<\triangle +|\eps|<\tau$. 
Then 
\begin{align*}
\lim_{|g|\to\infty}
\zeta_{\epsilon,g}(s;g^2+\tau)=
\zeta(s;\tau+\epsilon)+\zeta(s;\tau-\epsilon).
\end{align*}
\et
\proof
The proof is similar to that of Theorem \ref{main2}. 
Firstly we can show that
by Lemma \ref{FKF} 
\begin{align*}
s\text{-}\lim_{|g|\to\infty} e^{-t(H_\epsilon+g^2)}=
e^{-t(H_0+\epsilon U_z)}.
\end{align*}
Secondly, 
since
$
\zeta_{\epsilon,g}(s;g^2+\tau)
=\frac{1}{\Gamma(s)}
\int_0^\infty t^{s-1}
\sum_{\alpha\in\ZZ_2}
\sum_{n=0}^\infty (\Psi_{\alpha\,n}, e^{-t(H_\epsilon+g^2+\tau)}\Psi_{\alpha\,n}) \rd t$
and by Lemmas \ref{p1eps} and \ref{p2eps}, 
one can exchange 
$\lim_{|g|\to\infty}$ and 
$\int_0^\infty 
\sum_{\alpha\in\ZZ_2}
\sum_{n=0}^\infty \rd t$, 
and 
\begin{align*}
\lim_{|g|\to\infty}
\zeta_{\epsilon,g}(s;g^2+\tau)
=\frac{1}{\Gamma(s)}
\int_0^\infty t^{s-1}
\sum_{\alpha\in\ZZ_2}
\sum_{n=0}^\infty (\Psi_{\alpha\,n}, e^{-t(H_0+\epsilon U_z)}\Psi_{\alpha\,n}) \rd t
=\zeta(s;\tau+\epsilon)+\zeta(s;\tau-\epsilon).
\end{align*}
Then the theorem follows. 
\qed
In a similar way to Corollary \ref{zeta2} we can also consider 
other asymptotic behaviors of $\zeta_{\eps,g}(s;g^2+\tau)$. 
\bc{zeta3}Suppose that $\Re(s)>1$. 
\bi\item[(1)] 
Suppose that $\sqrt{\triangle^2+\eps^2}<\tau$. 
Then 
\[
\lim_{g\to0}\zeta_{\eps, g}(s;g^2+\tau)=
\zeta(s;\tau+\sqrt{\triangle^2+\eps^2})
+
\zeta(s;\tau-\sqrt{\triangle^2+\eps^2}).
\]
\item[(2)]
Suppose that $|\eps|<\tau$. 
Then 
\[
\lim_{\triangle\to0} \zeta_{\eps, g}(s;g^2+\tau)=\zeta(s;\tau+\eps)+\zeta(s;\tau-\eps).\]
\ei
\ec
\proof
In a similar manner to the proof of Theorem \ref{SSS}
we can see that 
\begin{align*}
&\lim_{g\to0}
 (\phi, e^{-t(H+g^2)}\psi)=
 (\phi,e^{-t(\triangle U_x+\add a+\eps U_z)}\psi),\\
&\lim_{\triangle\to0}
 (\phi, e^{-t(H+g^2)}\psi)=(\phi,e^{-t(\add a+\eps U_z)}\psi).
\end{align*}
Then the corollary  follows. 
\qed

\subsection{Meromorphic continuations}
The Hurwitz zeta function $\zeta (s;\tau)$ is defined for $s$ such that $\Re(s)>1$ 
but can be extended to the whole complex plane $\CC$ except for $s=1$. 
It has a simple pole at $s=1$ with residue~$1$. 

The meromorphic continuation of $\zeta$ is denoted by $\tilde \zeta$ in this paper. 
In \cite[Theorem 1.1]{sug18} and \cite[Theorem 4.1]{RW21}
the meromorphic continuation of 
$\zeta_g$ is also shown  for $0<\triangle<\tau$. 
Moreover the meromorphic continuation of 
$\zeta_{\pm,g}$ for $0<\triangle<\tau$, and 
that of $\zeta_{\eps,g}$ for $0<\triangle+|\eps|<\tau$ are proven in 
\cite[Corollary 4.2]{RW21} and 
\cite[Theorem 3.1]{rey23}, respectively. 
The meromorphic continuation of 
$\zeta_g$, $\zeta_{\pm,g}$ and $\zeta_{\eps,g}$ 
are denoted by 
$\tilde \zeta_g$, $\tilde \zeta_{\pm,g}$ and $\tilde \zeta_{\eps,g}$, 
respectively. 
It is also proven that all of them have a unique pole at $s=1$, and its residue is $2$ for 
$\tilde \zeta_g$ and $\tilde \zeta_{\eps,g}$, and~$1$ for 
$\tilde \zeta_{\pm,g}$. 
Theorems \ref{main2},\ref{SSS} and Corollary \ref{parity},\ref{zeta2}, \ref{zeta3} can be extended to $s\in \CC\setminus\{1\}$.  
\bc{cor}
Suppose $s\in\CC\setminus\{1\}$. 
Let 
$0<\triangle<\tau$. 
Then 
\bi
\item[(1)]
$\lim_{|g|\to\infty} \tilde \zeta_g(s;g^2+\tau)=2\tilde \zeta(s;\tau)$, 
\item[(2)]
$\lim_{|g|\to\infty}\tilde \zeta_{\pm,g}(s;g^2+\tau)=\tilde \zeta(s;\tau)$,
\item[(3)] 
$
\lim_{g\to0}\tilde \zeta_g(s;g^2+\tau)=
\tilde \zeta(s;\tau+\triangle)
+
\tilde \zeta(s;\tau-\triangle)$, 
\item[(4)]
$
\lim_{\triangle\to0} \tilde \zeta_g(s;g^2+\tau)=2\tilde \zeta(s;\tau)$.
\ei
For the asymmetric quantum Rabi model, similar results hold true. 
Suppose  that
$\triangle +|\eps|<\tau$. 
Then \bi
\item[(5)]
$\lim_{|g|\to\infty}
\tilde \zeta_{\epsilon,g}(s;g^2+\tau)=
\tilde \zeta(s;\tau+\epsilon)+\tilde \zeta(s;\tau-\epsilon)$, 
\item[(6)] 
$\lim_{g\to0}\tilde\zeta_g(s;g^2+\tau)=
\tilde\zeta(s;\tau+\sqrt{\triangle^2+\eps^2})
+
\tilde\zeta(s;\tau-\sqrt{\triangle^2+\eps^2})$, 
\item[(7)]
$\lim_{\triangle\to0} \tilde\zeta_g(s;g^2+\tau)=\tilde\zeta(s;\tau+\eps)+\tilde\zeta(s;\tau-\eps)$. 
\ei
\ec
\proof We shall prove (1).  
The proofs of (2)-(7) are similar. 
Note that the unique pole of $\tilde \zeta_g(s;g^2+\tau)$ is $s=1$. 
Let $\cD_0=\{s\in\CC\mid \Re(s)>1\}$ and $B_r(s)=\{w\in\CC\mid |s-w|<r\}$ be the open disk of radius $r$ with center $s$.   
Let $s_0\in \cD_0$. 
Then the Taylor expansion of $\tilde \zeta_g(s;g^2+\tau)$ at $s=s_0$ 
is given by 
$\tilde \zeta_g(s;g^2+\tau)=\sum_{n=0}^\infty a_n(g,s_0)(s-s_0)^n$ for any 
$s\in B_{|s_0-1|}(s_0)$, where  
\begin{align}\label{CR}
a_n(g,s_0)=\frac{1}{2\pi i }
\int_{|s-s_0|=\eps}
\frac{\tilde \zeta_g(s;g^2+\tau)}
{(s-s_0)^{n+1}}\rd s.\end{align}  
Suppose that $\eps$ is sufficiently small so that 
$\{s_0+\eps e^{i\theta}\mid \theta\in[0,2\pi]\}\subset\cD_0$. 
Then the integral path in \kak{CR} is contained in region $\cD_0$. 
Similarly 
$\tilde \zeta(s;\tau)=\sum_{n=0}^\infty a_n(s_0)(s-s_0)^n$ for any 
$s\in B_{|s_0-1|}(s_0)$ with some constant $a_n(s_0)$. 
Since by Theorem \ref{main2} we show that $\lim_{|g|\to\infty}\tilde \zeta_g(s;g^2+\tau)=
2 \tilde \zeta(s;\tau)$ for any $s\in \cD_0$, we can show that $\lim_{|g|\to\infty }a_n(g,s_0)= 
2a_n(s_0)$ by \kak{CR}. 
Hence 
$\lim_{|g|\to\infty}\tilde \zeta_g(s;g^2+\tau)
=2 \tilde \zeta(s; \tau)$ for any not only $s\in \cD_0$ but also 
$s\in B_{|s_0-1|}(s_0)$. 
Let $\cD_1=\bigcup_{s\in \cD_0}B_{|s-1|}(s)$. 
Then we proved that 
$\lim_{|g|\to\infty}\tilde \zeta_g(s;g^2+\tau)
=2 \tilde \zeta(s; \tau)$ for any 
$s\in \cD_1$. 
Replacing $\cD_0$ with $\cD_1$ in the above argument, we can show that 
$\lim_{|g|\to\infty}\tilde \zeta_g(s;g^2+\tau)
=2 \tilde \zeta(s; \tau)$ for any 
$s\in \cD_2$, where 
$\cD_2=\bigcup_{s\in \cD_1}B_{|s-1|}(s)$. 
Repeating these procedures, we can see that 
$\cD_0\subset \cD_1\subset \cD_2\cdots=\CC\setminus\{1\}$ and show (1) for any $s\in \CC\setminus\{1\}$.  
\qed

\subsection{Heat kernels}
Since 
$\EE_{\rP}^x[\ldots]=\int_{\RR} \EE_{\rP}^x[\ldots|X_t=y]\kappa_t(y,x)\rd y$ 
and $\rd \mu(x)=\gr^2(x)\rd x$, 
we see that 
\begin{align*}
(f, e^{-t\add a }g)_{\LR}&=(\cU_{\gr}^{-1}f, e^{-th}\cU_{\gr}^{-1}g)_{L^2(\RR,\rd \mu)}=
\int_{\RR} \frac{\bar f(x)}{\gr(x)}
\EE_{\rP}^x\left[\frac{g(X_t)}{\gr(X_t)}
\right]\rd \mu(x)\\
&=
\int _{\RR\times\RR} \bar f(x) g(y) M_t(x,y) \rd x\rd y.
\end{align*}
Here $M_t$ is the Mehler kernel. 
Let 
\begin{align}
\label{tildeH}
\tilde H=\cS_gU K U^{-1} \cS_g^{-1}=
\begin{pmatrix}-\half \frac{\rd^2}{\rd x^2}+\half x^2 +\half&0\\
0& -\half \frac{\rd^2}{\rd x^2}+\half x^2+\half
\end{pmatrix}
-g^2-\triangle 
\begin{pmatrix}0&e^{i2\sqrt 2gx}\\
e^{-i2\sqrt 2gx}&0
\end{pmatrix}
\end{align}
In this section we construct 
the heat kernel $H_t(\alpha,m, x,y)$ of $e^{-t\tilde H}$: 
\begin{align}
\label{kernel}
(\phi,e^{-t\tilde H}\psi)_{\CC^2\otimes\LR}=
\sum_{\alpha\in\ZZ_2}
\sum_{m=0}^\infty
\int_{\RR\times\RR}
 \overline{\phi(\alpha,x)}H_t(\alpha,m, x,y)
\psi((-1)^m\alpha,y)\rd x \rd y
\end{align}
and we see the asymptotic behavior of $H_t(\alpha,m, x,y)$ as $|g|\to\infty$. 

We give a remark on the conditional expectation: 
$\EE_\Pi\EE_{\rP}^x
\left[e^{ig \int_0^{t+} V(q_{-s})\rd N_s}
\left| X_t=y\atop N_t=m\right.\right] $. 
Let $Y_t=(N_t,X_t)$.  Thus 
the conditional expectation 
$\EE_\Pi\EE_{\rP}^x
\left[e^{ig \int_0^{t+} V(q_{-s})\rd N_s}
\left| \boldsymbol{\s}(Y_t)\right.\right] $
is measurable with respect to $Y_t$. It is known that any measurable function with respect to $\boldsymbol{\s}(Y_t)$ can  be represented as $h(Y_t)$ with some function $h=h(m,y)$ on $({\NN}\cup\{0\})\times \RR$. We denote $h(m,y)$ by 
$h(m,y)=\EE_\Pi\EE_{\rP}^x
\left[e^{ig \int_0^{t+} V(q_{-s})\rd N_s}
\left| X_t=y\atop N_t=m\right.\right]$. 
Similarly we can also define
the conditional expectation:  
$\EE_\Pi\EE_{\rP}^x
\left[e^{ig \int_0^{t+} V(q_{-s})\rd N_s}
\left| X_t=y\right.\right]$. 

\bt{kernel2}
The heat kernel of $e^{-t(\tilde H+g^2)}$ is given by 
\begin{align*}
H_t(\alpha,m, x,y)=\left\{
\begin{array}{ll}
M_t(x,y),&m=0,\\
\frac{\triangle^m t^m}{m!} 
\EE_\Pi\EE_{\rP}^x
\left[
e^{ig \int_0^{t+} V(q_{-s})\rd N_s}
\left| X_t=y\atop N_t=m\right.\right] 
M_t(x,y),&m\geq1,
\end{array}\right.
\end{align*}
and 
\begin{align}
\label{g}
\lim_{|g|\to\infty}\sum_{m=0}^\infty H_t(\alpha,m, x,y)=
M_t(x,y),\quad {\rm a.e.}\ x,y.
\end{align}
\et
\proof
We have
\begin{align*}
(\phi, e^{-t(\tilde H+g^2)}\psi)
=e^t \sum_{\alpha\in\ZZ_2} 
\int_{\RR} \frac{\overline{\phi(\alpha,x)}}{\gr(x)}
\EE_\Pi\EE_{\rP}^x
\left[\frac{\psi(S_t, X_t)}{\gr(X_t)} \triangle^{N_t}
e^{ig \int _0^{t+} V(q_{-s})\rd s}
\right]\rd \mu(x).
\end{align*}
We divid the right-hand side above as 
\begin{align*}
(\phi, e^{-t(\tilde H+g^2)}\psi)
&=e^t \sum_{\alpha\in\ZZ_2} 
\int_{\RR} \frac{\overline{\phi(\alpha,x)}}{\gr(x)}
\EE_\Pi\EE_{\rP}^x
\left[\frac{\psi(\alpha, X_t)}{\gr(X_t)} 
\one_{\{N_t=0\}}
\right]\rd \mu(x)\\
&+e^t \sum_{\alpha\in\ZZ_2} 
\int_{\RR} \frac{\overline{\phi(\alpha,x)}}{\gr(x)}
\EE_\Pi\EE_{\rP}^x
\left[\frac{\psi(S_t, X_t)}{\gr(X_t)} \triangle^{N_t}
e^{ig \int _0^{t+} V(q_{-s})\rd s}
\one_{\{N_t\geq1\}}
\right]\rd \mu(x).
\end{align*}
Since $X_t$ and $N_t$ are independent, we have
\begin{align*}
&
\int_{\RR} \frac{\overline{\phi(\alpha,x)}}{\gr(x)}
\EE_\Pi\EE_{\rP}^x
\left[\frac{\psi(\alpha, X_t)}{\gr(X_t)} 
\one_{\{N_t=0\}}
\right]\rd \mu(x)\\
&=
\int_{\RR} {\overline{\phi(\alpha,x)}}
{\psi(\alpha, y)} \EE_\Pi\EE_{\rP}^x
\left[
\one_{\{N_t=0\}}|X_t=y
\right] M_t(x,y)\rd x\rd y\\
&=
e^{-t}\int_{\RR} {\overline{\phi(\alpha,x)}}
{\psi(\alpha, y)} 
 M_t(x,y)\rd x\rd y.
\end{align*}
On the other hand
we have 
\begin{align*}
&
e^t\int_{\RR} \frac{\overline{\phi(\alpha,x)}}{\gr(x)}
\EE_\Pi\EE_{\rP}^x
\left[\frac{\psi(S_t, X_t)}{\gr(X_t)} \triangle^{N_t}
e^{ig \int _0^{t+} V(q_{-s})\rd s}
\one_{\{N_t\geq1\}}
\right]\rd \mu(x)\\
&=
\sum_{m=1}^\infty 
\frac{\triangle^mt^m}{m!}
\int_{\RR} {\overline{\phi(\alpha,x)}}
{\psi((-1)^m \alpha, y)}
\EE_\Pi\EE_{\rP}^x\left[
e^{ig \int _0^{t+} V(q_{-s})\rd s}
\left| X_t=y\atop N_t=m\right.\right]
M_t(x,y)
\rd x \rd y\\
&=
\sum_{m=1}^\infty 
\int_{\RR} {\overline{\phi(\alpha,x)}}
{\psi((-1)^m \alpha, y)}
H_t(\alpha,m,x,y) 
M_t(x,y)
\rd x \rd y.
\end{align*}
Together with them we have 
\begin{align*}
(\phi, e^{-t(\tilde H+g^2)}\psi)&=
\sum_{\alpha\in\ZZ_2} 
\int_{\RR} {\overline{\phi(\alpha,x)}}
{\psi(\alpha, y)} 
 M_t(x,y)\rd x\rd y\\
 &+
\sum_{\alpha\in\ZZ_2} 
\sum_{m=1}^\infty 
\int_{\RR} {\overline{\phi(\alpha,x)}}
{\psi((-1)^m \alpha, y)}
H_t(\alpha,m,x,y) 
M_t(x,y)
\rd x \rd y. 
\end{align*}
Then the heat kernel is $H_t(\alpha,m,x,y)$. 
We can also see that 
\begin{align*}
(\phi, e^{-t(\tilde H+g^2)}\psi)\to 
(\phi, e^{-t H_0}\psi)=
\sum_{\alpha\in\ZZ_2} 
\int_{\RR} {\overline{\phi(\alpha,x)}}
{\psi(\alpha, y)} 
 M_t(x,y)\rd x\rd y
 \end{align*}
as $|g|\to\infty$. 
It implies that 
$\sum_{m=0}^\infty H_t(\alpha,2m+1,x,y)\to0$ and 
$\sum_{m=1}^\infty H_t(\alpha,2m,x,y)\to0$ a.e.~$x,y$ as $|g|\to\infty$. 
Hence \eqref{g} follows. 
\qed

\section{Path measure associated with the ground state}
\label{s5}
Spin-boson model in quantum field theory describes a linear interaction between a two-level atom and a scalar quantum field. 
The spin boson Hamiltonian is defined by
\[\triangle \s_z\otimes\one+\one\otimes\rd \Gamma(\omega)+g\s_x\otimes\phi(h)\]
on $\CC^2\otimes \cF(L^2(\RR^d))$. 
Here 
$\cF(L^2(\RR^d))=\bigoplus_{n=0}^\infty L^2_{\rm sym} (\RR^{nd})$ denotes the boson Fock space over $L^2(\RR^d)$, 
where we set $L^2_{\rm sym} (\RR^{0})=\CC$. 
Vector $\Phi\in \cF(L^2(\RR^d))$ is described as $\Phi=\{\Phi^{(0)}, \Phi^{(1)}, \Phi^{(2)},\ldots,\}$, 
where $\Phi^{(n)}\in L^2_{\rm sym}(\RR^{nd})$, and $\sum_{n=0}^\infty \|\Phi^{(n)}\|^2_{L^2_{\rm sym}(\RR^{nd})}<\infty$. The field operator is defined by    
\[\phi(h)=\frac{1}{\sqrt2}(\add(h)+a(h)),\] where 
$a(h)$ and $\add(h)$ are the annihilation operator and the creation operator smeared by $h\in L^2(\RR^d)$, respectively. Note that  
$f\mapsto \add(f)$ and $h\mapsto a(h)$ are linear, and 
$a(h):L^2_{\rm sym} (\RR^{nd})\to L^2_{\rm sym} (\RR^{(n-1)d})$ 
and $\add(h):L^2_{\rm sym} (\RR^{nd})\to L^2_{\rm sym} (\RR^{(n+1)d})$. 
They satisfy $a(h)^\ast =\add(\bar h)$ and 
canonical commutation relation: \[ [a(f),\add(g)]=(\bar f, g)_{L^2(\RR^d)}\one.\] 
Finally 
$\rd \Gamma(\om)$ denotes the second quantization of the multiplication by  
$\om(k)=\sqrt{|k|^2+m^2}$, which leaves $L^2_{\rm sym} (\RR^{nd})$ invariant and acts as 
\[(\rd \Gamma(\om)\Psi)^{(n)}(k_1,\ldots,k_n)=\lk \sum_{j=1}^n \omega(k_j)\rk \Psi^{(n)}(k_1,\ldots,k_n).\] 
I.e., $ \rd \Gamma(\om)$ restricted on $L_{\rm sym}^2(\RR^{dn})$ is the multiplication by  
$\sum_{j=1}^n \omega(k_j)$. 

One mode version of spin-boson model is defined on $\CC^2\otimes \cF(\CC)$ instead of 
$\CC^2\otimes \cF(L^2(\RR^d))$. 
Regarding $\CC$ as the one-dimensional Hilbert space with the scalar product  $(z,w)_\CC=\bar z w$,  
it can be seen that 
\begin{align}
\label{SB}
\cF(\CC)\cong\LR.
\end{align}
Let $\add=\add(1)$ and $a=a(1)$ be the creation operator and the annihilation operator on $\cF(\CC)$, which satisfy 
$[a,\add]=\one$. 
The field operator on $\cF(\CC)$ is given by 
\[\phi=\frac{1}{\sqrt2}(\add +a).\] 
Let $\omega_0\geq0$ be a constant, which can be regarded as the multiplication operator on the Hilbert space $\CC$.  
The the second quantization $\rd \Gamma(\omega_0)$ of $\omega_0$ can be represented as 
\[ \rd \Gamma(\omega_0)=\omega_0\add a.\] 
Hence the one mode version of spin-boson model on $\CC^2\otimes \cF(\CC)$ is given by   
\[\triangle \s_z\otimes\one+\one\otimes \omega_0\add a +\frac{g}{\sqrt 2}\s_x\otimes(a+\add).\]
Under the identification \kak{SB}, it is just the quantum Rabi model with coupling constant $g/\sqrt2$ and $\omega_0=1$.  

In \cite{HHL14} the path measure associated with the ground state of the spin-boson model is constructed. 
In this section we construct the path measure associated with the ground state of the quantum Rabi model. 
In order to do that, instead of $H$ we investigate $L$ given in \eqref{L}. 
We recall that 
$L=-\triangle \s_x\otimes\one+
\one\otimes \bdd b+g\s_z\otimes (b+\bdd)$. 
Let $\grr$ be the ground state of $L$ such that 
\[L\grr=E\grr\] 
with $E=\inf \spec(L)$. It is shown that $\grr>0$ in \cite{HH14} under the identification \eqref{HH}. 
Hence $(\one,\grr)_\cH\neq 0$. 
Then 
\[\grr=\limt \frac{e^{-tL}\one}{\|e^{-tL}\one\|_\cH}.\]
Let us set 
\[\lr{\cO}=(\grr, \cO \grr)_\cH\] 
for a bounded operator $\cO$. 
Then 
we have 
\[\lr{\cO}=
\limt \frac{(e^{-tL}\one,\cO e^{-tL}\one)_\cH}{\|e^{-tL}\one\|_\cH^2}.\]
The right-hand side can be represented in terms of Feynman-Kac formula, 
and under some condition we can also see that 
\[\lr{\cO}=\EE_{{\Pi_\infty}}[f_{\cO}]\]
with some probability measure ${\Pi_\infty}$ and a function $f_{\cO}$. 
The probability measure ${\Pi_\infty}$ is called the path measure associated with the ground state $\grr$. 
The similar results are investigated in models in quantum field theory 
\cite{spo89, BHLMS01, HHL14,hir14,HM21}, but as far as we know there is no example in quantum mechanics. 

\subsection{Probability measure ${\Pi_\infty}$ associated with the ground state}
We set 
$T_s=S_{\triangle s}$ and 
$q^\triangle_s=(T_s, X_s)$.
In Section \ref{s5} we assume that 
\[\triangle>0.\]
\bl{G0}
Let $\phi,\psi\in\cH$. Then
\begin{align}
\label{g0}
(\phi, e^{-tL}\psi)=
e^{\triangle t}
\sum_{\alpha\in\ZZ_2}\int_{\RR} 
\EE_\Pi
\EE_{\rP}^x
\left[
\overline{\phi(q^\triangle_0)}
\psi(q^\triangle_t)e^{-g\int_0^t W(q^\triangle_s)\rd s}\right]
\rd \mu(x).
\end{align}
\el
\proof
Since 
\[\frac{1}{\triangle}L=-\s_x\otimes\one+\one\otimes \frac{1}{\triangle}\bdd b +\frac{g}{\triangle}\s_z\otimes (\bdd+b),\]
the Feynman-Kac formula \eqref{fkfl} yields that 
\begin{align*}
(\phi, e^{-tL}\phi)&=
(\phi, e^{-\triangle t\frac{1}{\triangle}L}\phi)\\
&=
e^{\triangle t}
\sum_{\alpha\in\ZZ_2}\int_{\RR} 
\EE_\Pi
\EE_{\rP}^x
\left[
\overline{\phi(S_0,X_0)}
\psi(S_{\triangle t},X_{t})
e^{-\frac{g}{\triangle}\int_0^{\triangle t} \sqrt 2 S_{s} X_{s/\triangle}
\rd s}\right]\rd \mu(x).
\end{align*}
By the change of variable $s$ to $\triangle s$ in 
$\frac{g}{\triangle}\int_0^{\triangle t} \sqrt 2 S_{s} X_{s/\triangle}
\rd s$, we see \eqref{g0}. 
\qed
For the later use we have a technical lemma below. 
\bl{c}
We have
\begin{align*}
\EE_{\rP }^x\left[ 
e^{-g\int_0^tW(\hat q^\triangle_s)\rd s}
\right]=
e^{-g\left(\int_0^te^{-s}(-1)^{N_{\triangle s}}\rd s\right) x}
e^{\frac{g^2}{4}\int_0^{(1-e^{-2t})/2}
\left|\int_y^t(-1)^{N_{\triangle s}}\rd s\right|^2\rd y}.
\end{align*}
In particular
\begin{align*}
\EE_{\rP }^x\left[ 
e^{-g\int_0^tW(\hat q^\triangle_s)\rd s}
\right]\leq
e^{|g|(1-e^{-t})x}
e^{\frac{g^2}{4}\int_0^{(1-e^{-2t})/2}|t-y|^2\rd y}.
\end{align*}
\el
\proof
The proof is similar to that of Lemma \ref{vacuum}. 
We have
\begin{align*}
\EE_{\rP }^x\left[ 
e^{-g\int_0^tW(\hat q^\triangle_s)\rd s}
\right]
&=
\EE_\cW^0\left[ 
e^{-g\int_0^te^{-s}(x+\frac{1}{\sqrt2} B_{e^{2s}-1})(-1)^{N_{\triangle s}}\rd s}
\right]\\
&=
e^{-g\left(\int_0^te^{-s}(-1)^{N_{\triangle s}}\rd s\right) x}
\EE_\cW^0\left[ 
e^{-g\int_0^t B_{(1-e^{-2s})/2}(-1)^{N_{\triangle s}}\rd s}\right]\\
&=
e^{-g\left(\int_0^te^{-s}(-1)^{N_{\triangle s}}\rd s\right) x}
e^{\frac{g^2}{4}\left \|
\int_0^t \one_{(1-e^{-2s})/2}(\cdot)
(-1)^{N_{\triangle s}}\rd s\right\|_{\LR}^2}.
\end{align*}
Then the lemma is proven. 
\qed
Now we extend $\pro T$ to the process on the whole real line. 
Let \[\hat T_t=(-1)^{\hat N_{\triangle t}}\alpha\quad t\in\RR.\] 
We can realize $\proo {\hat T}$ as a coordinate process as usual. 
Let $\cD=D(\RR)$ be 
the space of c\`adl\`ag paths on $\RR$. 
There exists a topology $d^\circ$ on $\cD$ such that 
$(\cD,d^\circ)$ is a separable and complete metric space (e.g. \cite[Section 3.5]{EK86} and \cite[Section 16]{bil68}). 
Let $\cB_\cD$ be the Borel sigma-field of $\cD$. 
Thus 
\[\hat T_ \bullet :
(\bar \cY, \cB_{\bar \cY}, \bar \Pi)\to (\cD, \cB_\cD)\] 
is an 
$\cD $-valued random variable. We denote its image measure on $(\cD, \cB_\cD)$ by $\rQ^\alpha$, i.e., 
$\rQ^\alpha(A)=\bar \Pi(\hat T_\bullet^{-1}(A))$ for 
$A\in\cB_\cD$, and the coordinate process on $(\cD,\cB_\cD)$ 
by the same symbol $\pro {\hat T}$, i.e., $\hat T_t(\omega)=\omega(t)$ for $\omega \in \cD$.
Let $\pi_\Lambda: \cD \to \RR^\Lambda$ be the projection defined by $\pi_\Lambda(\omega)=
(\omega(t_0),\ldots,\omega(t_n))$ for $\omega\in \cD$ and 
$\Lambda=\{t_0,\ldots,t_n\}$. Then 
\[{\cA}= \{\pi^{-1}_\Lambda(E)\,|\,\Lambda\subset \RR,\#\Lambda<\infty,E\in \cB(\RR^\Lambda)\}\]
is the  family of cylinder sets. 
It is known that the sigma-field generated by cylinder sets coincides with $\cB_\cD$. 
Moreover let 
$\cD_T=D([-T,T])$ be the space of c\`adl\`ag paths on $[-T,T]$ and $\pi_T:\cD\to \cD_T$ be the projection defined by $\pi_T \omega=\omega\lceil_{[-T,T]}$. Let 
$\cB_T$ be the Borel sigma-field of $\cD_T$. 
Let
$\pi_\Lambda: \cD_T \to \RR^\Lambda$ be the projection defined by $\pi_\Lambda(\omega)=
(\omega(t_0),\ldots,\omega(t_n))$ for $\omega\in \cD_T$ and 
$\Lambda=\{t_0,\ldots,t_n\}$. 
Note that we use the same notation $\pi$ as the projection 
from $\cD$ to $\RR^\Lambda$. 
Then 
\[{\cA}_T= \{\pi^{-1}_\Lambda(E)\,|\,\Lambda\subset 
[-T,T],\#\Lambda<\infty,E\in \cB(\RR^\Lambda)\}\]
is the family of cylinder sets. 
We set 
\[
\cBc=\bigcup_{s\geq0}\pi_s^{-1}(\cB_s),\quad
\cBc_T=\bigcup_{0\leq s\leq T}\pi_s^{-1}(\cB_s).\] 
It is also seen that the sigma-field generated by $\stackrel{\circ}{\cB}$ (resp. $\cBc_T$) 
coincides with $\cB_\cD$ (resp. $\cB_T$). 
Together with them we have 
\begin{align}
\label{good}
\cB_\cD=\boldsymbol{\s}(\cA)=\boldsymbol{\s}(\cBc),\quad 
\cB_T=\boldsymbol{\s}(\cA_T)=\boldsymbol{\s}(\cBc_T).
\end{align}
Hence 
\eqref{fkfl} can be reformulated in terms of 
the coordinate process $\pro {\hat T}$ on $(\cD, \cB_\cD, \rQ^\alpha)$ 
instead of $(\bar \cY,\cB_{\bar \cY},\bar \Pi)$ as 
\begin{align}\label{rep}
(\phi, e^{-tL}\psi) = e^{\triangle t} 
\sum_{\alpha\in\ZZ_2}\int_{\RR} 
\EE_\rQ^\alpha\EE_{\bar\rP}^x \left[\overline{\phi(\hat q^\triangle_0)}
e^{-g\int_0^t W(\hat q^\triangle_s) \rd s} \psi(\hat q^\triangle_t) \right]\rd \mu(x).
\end{align}
Here \[\hat q^\triangle_s=(\hat T_s, \hat X_s)\quad s\in\RR,\]
where $\hat X_t$ is the Ornstein-Uhlenbeck process on the whole real line. 
The advantage of \eqref{g0} is that $\triangle^{N_t}$ disappears. 
$\triangle^{N_t}$ is not shift invariant but 
$\hat T_s$ in \eqref{g0} is shift invariant. 
Then 
\begin{align*}
&\sum_{\alpha\in\ZZ_2}\int_{\RR} \EE_\rQ^\alpha 
\EE_{\bar\rP}^x \left[\overline{\phi(\hat q^\triangle_0)}
e^{-g\int_0^t W(\hat q^\triangle_s) \rd s} \psi(\hat q^\triangle_t) \right]\rd \mu(x)\\
&= 
\sum_{\alpha\in\ZZ_2}\int_{\RR} \EE_\rQ^\alpha 
\EE_{\bar\rP}^x \left[\overline{\phi(\hat q^\triangle_{-r})}
e^{-g\int_0^t W(\hat q^\triangle_{s-r}) \rd s} \psi(\hat q^\triangle_{t-r}) \right]\rd \mu(x)
\end{align*}
for any $0\leq r\leq t$. 
Let 
\begin{align}\label{W}
W_{\triangle}(t,s)=\hat T_t\hat T_se^{-|t-s|}.
\end{align} 

\bl{G1}
We have 
\begin{align*}
(\one, e^{-tL}\one )
=
2e^{\triangle t}
\EE_{\rQ}^\alpha
\left[
\exp\left(
\frac{g^2}{2}
\int_0^{t} \rd s \int_0^{t} \rd r 
W_{\triangle}(s,r)\right)
\right].
\end{align*}
\el 
\proof
By the Feynman-Kac formula given by \eqref{g0} and inserting \eqref{X}, 
we can see that 
\begin{align*}
(\one, e^{-tL}\one )
=
e^{\triangle t}
\sum_{\alpha\in\ZZ_2} 
 \EE_{\rQ}^\alpha
 \left[
\EE_{\bar\rP}^x
\left[
 e^{-g\int _0^t \hat T_s {e^{-s} B_{e^{2s}-1}}\rd s}
\right] 
\int_{\RR}
\frac{1}{\sqrt{\pi}}
e^{- x^2-\left(\sqrt{2}g\int_0^t\hat T_s e^{-s} \rd s\right) {x}}\rd {x}
\right]\end{align*}
Since
\begin{align*}
&\EE_{\bar \rP}^x
\left[
 e^{-g\int _0^t \hat T_s e^{-s} B_{e^{2s}-1}\rd s}
\right]
=\exp\left(
\frac{g^2}{2}
\int_0^t \rd s \int_0^t \rd r 
\hat T_s \hat T_r 
e^{-(s+r)}
(e^{2(s\wedge r)}-1)\right)\\
&
\int_{\RR}
e^{-\left(\sqrt{2}g\int_0^t\hat T_s e^{-s} \rd s \right){x}}\rd \mu(x)
=\exp\left(\frac{g^2}{2}\left(\int_0^t \hat T_s e^{-s}\rd s\right)^2\right), 
\end{align*}
we obtain that 
\begin{align*}
(\one, e^{-tL}\one )
=
e^{\triangle t}
\sum_{\alpha\in\ZZ_2} 
\EE_{\rQ}^\alpha
\left[
\exp\left(
\frac{g^2}{2}
\int_0^t \rd s \int_0^t \rd r 
\hat T_s \hat T_r 
e^{-|s-r|}\right)
\right]. 
\end{align*}
Hence the lemma follows. 
\qed
\begin{remark}
(1) Since $W_{\triangle}(s,r)$ is independent of $\alpha$, 
$
\EE_{\rQ}^\alpha
\left[
\exp\left(
\frac{g^2}{2}
\int_0^{t} \rd s \int_0^{t} \rd r 
W_{\triangle}(s,r)\right)
\right]$ is also independent of $\s$. I.e., 
\[
\EE_{\rQ}^{\s}
\left[
\exp\left(
\frac{g^2}{2}
\int_0^{t} \rd s \int_0^{t} \rd r 
W_{\triangle}(s,r)\right)
\right]=\EE_{\rQ}^{+1}
\left[
\exp\left(
\frac{g^2}{2}
\int_0^{t} \rd s \int_0^{t} \rd r 
W_{\triangle}(s,r)\right)
\right]\]
for $\alpha\in\ZZ_2$. \\
(2) By the shift invariance of $\hat T_s$ 
we can also see that 
\begin{align*}
\EE_{\rQ}^\alpha
\left[
\exp\left(
\frac{g^2}{2}
\int_0^{t} \rd s \int_0^{t} \rd r 
W_{\triangle}(s,r)\right)
\right]
=
\EE_{\rQ}^\alpha
\left[
\exp\left(
\frac{g^2}{2}
\int_{-u}^{t-u} \rd s \int_{-u}^{t-u} \rd r 
W_{\triangle}(s,r)\right)
\right]
\end{align*}
for any $0\leq u\leq t$. 
Thus
 we see that
\begin{align}
(e^{-tL}\one, e^{-tL}\one )
&=
2e^{2\triangle t}
\EE_{\rQ}^\alpha
\left[
\exp\left(
\frac{g^2}{2}
\int_0^{2t} \rd s \int_0^{2t} \rd r 
W_{\triangle}(s,r)\right)
\right]\nonumber\\
\label{shift}&=
2e^{2\triangle t}
\EE_{\rQ}^\alpha
\left[
\exp\left(
\frac{g^2}{2}
\int_{-t}^{t} \rd s \int_{-t}^{t} \rd r 
W_{\triangle}(s,r)\right)
\right].
\end{align}
\end{remark}
We can also compute 
$(e^{-tL}\one, e^{-\beta\bdd b}e^{-tL}\one )$ for $\beta>0$. 
\bl{G2}
Let $\beta>0$. Then 
\begin{align*}
&(e^{-tL}\one, e^{-\beta \bdd b}e^{-tL}\one )\\
&=2e^{2\triangle t}
\EE_{\rQ}^\alpha
\left[
\exp\left(
\frac{g^2}{2}
\int_{-t}^t 
\int_{-t}^t
W_{\triangle}(s,r)
\rd s\rd r
-g^2(1-e^{-\beta})\int_{-t}^0\int_0^tW_{\triangle}(s,r)\rd s\rd r\right)\right].
\end{align*}
\el
\proof
Since 
\begin{align*}
(\phi, e^{-\beta \bdd b}\psi)=
\sum_{\alpha\in\ZZ_2} 
\int _{\RR}
\bar \phi(\alpha,X_0) \EE_{\bar\rP}^x
[\psi(\alpha,X_\beta)]\rd \mu(x),
\end{align*}
we see that 
\begin{align*}
(e^{-tL}\one, e^{-\beta \bdd b}e^{-tL}\one )
=
\sum_{\alpha\in\ZZ_2} 
\int _{\RR} 
(e^{-tL}\one)(\alpha,X_0) \EE_{\bar\rP}^x
[(e^{-tL}\one)(\alpha,X_\beta)]\rd \mu(x).
\end{align*}
It is straightforward to compute 
$(e^{-tL}\one)(\alpha,X_0)$ and $(e^{-tL}\one)(\alpha,X_\beta)$. 
We have
\begin{align*}
(e^{-tL}\one)(\alpha,X_0)&=
e^{\triangle t}\EE_{\rQ}^\alpha
\EE_{\bar\rP}^x
\left[e^{-\sqrt{2}g\int_0^t\hat T_s X_s^{ x} \rd s}
\right]\\
&=e^{\triangle t}
\EE_{\rQ}^\alpha 
\left[
e^{-\sqrt{2}g\int_0^t\hat T_s e^{-s} \rd s {x}}
\EE_{\cW}^0
\left[
e^{-g\int_0^t \hat T_s e^{-s}B_{e^{2s}-1}\rd s} 
\right]
\right]\\
&=e^{\triangle t}
\EE_{\rQ}^\alpha 
\left[
e^{-\sqrt{2}g\int_0^t\hat T_s e^{-s} \rd s {x}}
e^{\frac{g^2}{2}
\int_0^t \rd s \int_0^t \rd r 
\hat T_s \hat T_r 
e^{-(s+r)}
(e^{2(s\wedge r)}-1)
}
\right].
\end{align*}
The computation of
$\EE_{\bar\rP}^x\left[
(e^{-tL}\one)(\alpha,X_\beta)\right]$ is more complicated than 
that of 
$(e^{-tL}\one)(\alpha,X_0)$. 
We have 
\begin{align*}
\EE_{\bar\rP}^x\left[
(e^{-tL}\one)(\alpha,X_\beta)\right]
=e^{\triangle t}
\EE_{\bar\rP}^x\left[
\EE_{\rQ}^\alpha 
\left[
e^{-\sqrt{2}g\int_0^t\hat T_s e^{-s} \rd s {X_\beta}}
e^{\frac{g^2}{2}
\int_0^t \rd s \int_0^t \rd r 
\hat T_s \hat T_r 
e^{-(s+r)}
(e^{2(s\wedge r)}-1)
}
\right]\right].
\end{align*}
Inserting \eqref{X} to $X_\beta$ above again, we obtain that
 \begin{align*}
&=e^{\triangle t}
\EE_{\cW}^0\left[
\EE_{\rQ}^\alpha 
\left[
e^{-\sqrt{2}g\int_0^t\hat T_s e^{-s} \rd s 
e^{-\beta}
\left(
x+\frac{1}{\sqrt 2}B_{e^{2\beta}-1}\right)
}
e^{\frac{g^2}{2}
\int_0^t \rd s \int_0^t \rd r 
\hat T_s \hat T_r 
e^{-(s+r)}
(e^{2(s\wedge r)}-1)
}
\right]\right]\\
&=e^{\triangle t}
\EE_{\rQ}^\alpha 
\left[
e^{-\left(\sqrt{2}g\int_0^t\hat T_s e^{-s} \rd s 
\right) e^{-\beta}
x}
e^{\frac{g^2}{2}
\int_0^t \rd s \int_0^t \rd r 
\hat T_s \hat T_r 
e^{-(s+r)}
(1-e^{-2\beta })
}
e^{\frac{g^2}{2}
\int_0^t \rd s \int_0^t \rd r 
\hat T_s \hat T_r 
e^{-(s+r)}
(e^{2(s\wedge r)}-1)
}
\right]\\
&=e^{\triangle t}
\EE_{\rQ}^\alpha 
\left[
e^{-\left(\sqrt{2}g\int_0^t\hat T_{s-t} e^{-s} \rd s 
\right) e^{-\beta}
x}
e^{\frac{g^2}{2}
\int_0^t \rd s \int_0^t \rd r 
\hat T_{s-t} \hat T_{r-t} 
e^{-(s+r)}
(1-e^{-2\beta })
}
e^{\frac{g^2}{2}
\int_0^t \rd s \int_0^t \rd r 
\hat T_{s-t} \hat T_r 
e^{-(s+r)}
(e^{2(s\wedge r)}-1)
}
\right].
\end{align*}
In the last line above we shift 
$\hat T_s$ by $t$. 
Since $\hat T_u$ for $0\leq u\leq t$ 
and $\hat T_{s-t}$ for $0\leq s\leq t$ are independent, 
combining above computations, we have 
\begin{align}
&(e^{-tL}\one, e^{-\beta \bdd b}e^{-tL}\one )\nonumber \\
&=
\sum_{\alpha\in\ZZ_2} 
e^{2\triangle t}
\int _{\RR} \frac{e^{-x^2}}{\sqrt\pi}
\EE_{\rQ}^\alpha
\left[
e^{-\left(\sqrt{2}g\int_0^t\hat T_s e^{-s} \rd s\right) {x}}
e^{-\left(\sqrt{2}g\int_0^t\hat T_{s-t} e^{-s} \rd s \right) 
e^{-\beta}
x}e^{\frac{g^2}{2}
\int_0^t \rd s \int_0^t \rd r 
\hat T_s \hat T_r 
e^{-(s+r)}
(e^{2(s\wedge r)}-1)
}
\right.\nonumber \\
&\label{K3}\left.
\hspace{2cm}\times 
e^{\frac{g^2}{2}
\int_0^t \rd s \int_0^t \rd r 
\hat T_{s-t} \hat T_{r-t} 
e^{-(s+r)}
(1-e^{-2\beta })
}
e^{\frac{g^2}{2}
\int_0^t \rd s \int_0^t \rd r 
\hat T_{s-t} \hat T_{r-t} 
e^{-(s+r)}
(e^{2(s\wedge r)}-1)
}
\right]\rd x.
\end{align}
Terms dependent on $x$ on the exponent above 
can be computed as 
\begin{align*}
&- x^2
-\sqrt{2}g\left(
e^{-\beta}\int_0^t\hat T_{s-t} e^{-s} \rd s 
+\int_0^t \hat T_s e^{-s} \rd s\right) 
x\\
&=
- \left(
x
+\frac{g}{\sqrt{2}} 
\int_0^t \hat T_s e^{-s} \rd s 
+\frac{g}{\sqrt{2}} 
e^{-\beta}\int_0^t\hat T_{s-t} e^{-s} \rd s 
\right)^2 +
\frac{g^2}{2}
\left(
\int_0^t \hat T_s e^{-s} \rd s 
+
e^{-\beta}\int_0^t\hat T_{s-t} e^{-s} \rd s 
\right)^2.
\end{align*}
The first term on the right-hand side 
can be integrated with respect to $\rd x$ as 
\[
\frac{1}{\sqrt\pi}
\int _{\RR} e^{-\left(
x
+\frac{g}{\sqrt{2}} 
\int_0^t \hat T_s e^{-s} \rd s 
+\frac{g}{\sqrt{2}} 
e^{-\beta}\int_0^t\hat T_{s-t} e^{-s} \rd s 
\right)^2 }
\rd x=1.\]
The second term on the right-hand side can be computed as 
\begin{align}
&\left(
\int_0^t \hat T_s e^{-s} \rd s 
+
e^{-\beta}\int_0^t\hat T_{s-t} e^{-s} \rd s 
\right)^2\nonumber \\
\label{K1}
&=
\int_0^t\int_0^t \hat T_s \hat T_r
e^{-(s+r)} \rd s \rd r
+
2e^{-\beta}
\int_0^t\int_0^t \hat T_{s-t} \hat T_r
e^{-(s+r)} \rd s \rd r
+
e^{-2\beta}
\int_0^t\int_0^t \hat T_{s-t} \hat T_{r-t}
e^{-(s+r)} \rd s \rd r.
\end{align}
Terms independent of $x$ on \eqref{K3} are 
\begin{align}
\int_0^t \rd s \int_0^t 
\hat T_s \hat T_r 
e^{-(s+r)}
(e^{2(s\wedge r)}-1)\rd r&+
\int_0^t \rd s \int_0^t 
\hat T_{s-t} \hat T_{r-t} 
e^{-(s+r)}
(1-e^{-2\beta })\rd r\nonumber\\
&\label{K2}
+
\int_0^t \rd s \int_0^t 
\hat T_{s-t} \hat T_{r-t} 
e^{-(s+r)}
(e^{2(s\wedge r)}-1)\rd r. 
\end{align}
Then the sum of \eqref{K1} and \eqref{K2} is 
\begin{align*}
&\eqref{K1}+\eqref{K2}\\
&=
\int_0^t \rd s \int_0^t 
\hat T_{s-t} \hat T_{r-t} 
e^{-(s+r)}
e^{2(s\wedge r)}\rd r+
\int_0^t \rd s \int_0^t 
\hat T_s \hat T_r 
e^{-(s+r)}
e^{2(s\wedge r)}\rd r\\
&\hspace{6.3cm}+
2e^{-\beta}
\int_0^t\int_0^t \hat T_{s-t} \hat T_r
e^{-(s+r)} \rd s \rd r\\
&=
\int_0^t \rd s \int_0^t 
\hat T_{s-t} \hat T_{r-t} 
e^{-|s-r|}\rd r
+
\int_0^t \rd s \int_0^t 
\hat T_s \hat T_r 
e^{-|s-r|}\rd r
+
2e^{-\beta}
\int_0^t\int_0^t \hat T_{s-t} \hat T_r
e^{-(s+r)} \rd s \rd r\\
&=
\int_{-t}^0 \rd s \int_{-t}^0 
\hat T_s \hat T_r 
e^{-|s-r|}\rd r
+
\int_0^t \rd s \int_0^t 
\hat T_s \hat T_r 
e^{-|s-r|}\rd r
+
2e^{-\beta}
\int_{-t}^0\int_0^t \hat T_s \hat T_r
e^{-|s-r|} \rd s \rd r.
\end{align*}
By the trick 
$\int_{-t}^t \int_{-t}^t =
\int_{-t}^0 \int_{-t}^0 
+\int_0^t \int_0^t 
+2\int_{-t}^0 \int_0^t$, 
we see that 
\[
\eqref{K1}+\eqref{K2}
=
\int_{-t}^t \rd s \int_{-t}^t 
\hat T_s \hat T_r 
e^{-|s-r|}\rd r
-2(1-e^{-\beta})
\int_{-t}^0\int_0^t \hat T_s \hat T_r
e^{-|s-r|} \rd s \rd r.\]
Then the lemma follows. 
\qed

\bc{PARITY}
Let $P$ be the parity operator given by \eqref{hana}. 
Then 
$P\grr=-\grr$, i.e., $\grr\in \cH_-$ and 
$g\mapsto E(g)$ is concave and differentiable in a.e. $g$. 
\ec
\proof
From $E(g)=\limt -\frac{1}{t}\log (\one,e^{-tL}\one )$ it follows that 
$g\mapsto E(g)$ is concave and differentiable a.e. in $g$. 
From $P\one=-\one$ and 
$\grr=\limt e^{-tL}\one /\|e^{-tL}\one \|$, the second statement follows. 
\qed
Define the probability measure $\Pi_T$ on $(\cD, \cB_{\cD})$ by
\begin{align}\label{measure0}
\Pi_T(A)=
\frac{
\EE_\rQ^\alpha
\left[\one_A
e^{\frac{g^2}{2}\int_{-T}^T \rd t\int_{-T}^T \rd s 
W_\triangle(t,s) }\right]}
{\EE_\rQ^\alpha
\left[
e^{\frac{g^2}{2}\int_{-T}^T \rd t\int_{-T}^T \rd s 
W_\triangle(t,s) }\right]}
, \quad A\in \cB_{\cD}.
\end{align}
The following proposition is shown for spin boson model in \cite[Theorem 3.8]{HHL14} and 
for relativistic Pauli-Fierz model in \cite[Lemma 7.6]{hir14}, and the proof for the quantum Rabi Hamiltonian is a  
minor modification of \cite{HHL14,hir14}. 

\begin{proposition}\label{hir}
There exists a probability measure 
${\Pi_\infty}$ on $(\cD, \cB_{\cD})$ 
such that 
\[\lim_{T\to\infty}\Pi_T(A)={\Pi_\infty}(A)\quad 
A\in\cBc.\] 
\end{proposition}
\proof
We leave the proof in Appendix \ref{A}. 
\qed

\subsection{Expectations by ${\Pi_\infty}$}
In this section we give some examples of application of ${\Pi_\infty}$. 
These examples are one mode versions of the spin boson model \cite{spo89,HHL14}. 
Then we show only outlines of proofs. 

The sequence of probability measures 
$(\Pi_T)_{T>0}$ is said to locally converge to the probability measure 
${\Pi_\infty}$ whenever
$\lim_{T\to\infty} \Pi_{T}(A)= {\Pi_\infty}(A)$ for all $A\in \pi_t^{-1}(\cB_t)$ and for all $t\geq 0$.
\bc{bounded}
Let $f$ be a $\cB_t$-measurable and bounded function. 
Then 
\[
\lim_{T\to\infty}
\EE_{\Pi_T}[f]=
\EE_{{\Pi_\infty}}[f].\] 
\ec
\proof
It is enough to show the corollary for a nonnegative function $f$.
Since $f$ is bounded and $\cB_t$-measurable, 
there exists a sequence $\{f_n\}$ 
 such that 
$\lim_{n\to\infty} \sup_{x\in \cD} |f_n(x)-f(x)|=~0$. 
Here $f_n$ is of the form 
$f_n=\sum_{j=1}^{m_n} a_j\one_{A_j}$ with 
$A_j\in \cB_t$ and $a_j>0$. 
Let $\epsilon>0$ be arbitrary. We assume that 
$\sup_{x\in \cD} |f_n(x)-f(x)|\leq \epsilon$. 
Then 
we see that 
\begin{align*}
|\EE_{\Pi_T}[f]-\EE_{{\Pi_\infty}}[f]|
&\leq
\EE_{\Pi_T}[|f-f_n|]
+|\EE_{\Pi_T}[f_n]-\EE_{{\Pi_\infty}}[f_n]|
+\EE_{{\Pi_\infty}}[|f_n-f|]\\
&\leq
2\epsilon +
|\EE_{\Pi_T}[f_n]-\EE_{{\Pi_\infty}}[f_n]|
\end{align*}
and from Proposition \ref{hir} 
it follows that 
$\lim_{T\to\infty} |\EE_{\Pi_T}[f]-\EE_{{\Pi_\infty}}[f]|\leq 2\epsilon$. 
Then the corollary follows. 
\qed

\subsubsection{Number operator $\bdd b$}
\bt{main3}
Let $\beta\in\CC$. 
Then 
\begin{align}
\label{GG1}
\lr{e^{\beta \bdd b}}
&=\EE_{{\Pi_\infty}}
\left[
e^{-g^2(1-e^{\beta}) 
\int_{-\infty}^0 \int_{0}^\infty
W_{\triangle}(s,r)
\rd s\rd r
}\right],\\
\label{GG2}
\lr{(\bdd b)^m}&=\sum_{l=1}^m a_l(m)
g^{2l}
\EE_{{\Pi_\infty}}\left[
\left(
\int_{-\infty}^0 \int_{0}^\infty
W_{\triangle}(s,r)
\rd s\rd r\right)^l\right].
\end{align}
Here $a_l(m)=\frac{(-1)^l}{l!}\sum_{s=1}^l(-1)^s\binom{l}{s}$ are the Stirling numbers. 
In particular 
$\lr{(\bdd b)^m}\leq e^{2g^2}-1$
for any $m\geq0$. 
\et
\proof
Since $(\one,\grr)>0$, 
we see that 
$
\lr{e^{-\beta \bdd b}}=
\limt 
\frac{(e^{-tL}\one, e^{-\beta \bdd b} e^{-tL}\one )}{(e^{-tL}\one, e^{-tL}\one )}$. 
By Lemmas \ref{G1} and \ref{G2} we have
\begin{align*}
\lr{e^{-\beta \bdd b}}
&=\limt 
\frac{
\EE_{\rQ}^\alpha
\left[
e^{
\frac{g^2}{2}
\int_{-t}^t 
\int_{-t}^t
W_{\triangle}(s,r)
\rd s\rd r
{-g^2(1-e^{-\beta}) 
\int_{-t}^0 \int_{0}^t
W_{\triangle}(s,r)
\rd s\rd r}
}
\right]}
{
\EE_{\rQ}^\alpha
\left[
e^{
\frac{g^2}{2}
\int_{-t}^t 
\int_{-t}^t
W_{\triangle}(s,r)
\rd s\rd r
}
\right]}\\
&=
\limt 
\EE_{\Pi_t}
{\left[
e^{
-g^2(1-e^{-\beta}) 
\int_{-t}^0 \int_{0}^t
W_{\triangle}(s,r)
\rd s\rd r
}\right]}.
\end{align*}
Note that 
$\left|
\int_{-\infty}^0 \int_{0}^\infty
W_{\triangle}(s,r)
\rd s\rd r\right|\leq 1$ and hence 
\[\left|
e^{-g^2(1-e^{-\beta})\int_{-\infty}^0 \int_{0}^\infty
W_{\triangle}(s,r)
\rd s\rd r}
-
e^{-g^2(1-e^{-\beta})\int_{-S}^0 \int_{0}^SW_{\triangle}(s,r)
\rd s\rd r}\right|\leq \epsilon\] 
uniformly in paths for sufficiently large $S>0$. 
Then by Corollary \ref{bounded} 
we see that 
\begin{align*}
\limt \left|\EE_{\Pi_t}\left[e^{-g^2(1-e^{-\beta}) \int_{-t}^0 \int_{0}^tW_{\triangle}(s,r)\rd s\rd r}\right]
-
\EE_{{\Pi_\infty}}
\left[e^{-g^2(1-e^{-\beta}) \int_{-\infty}^0 \int_{0}^\infty W_{\triangle}(s,r)\rd s\rd r}\right]\right|=0.
\end{align*}
We conclude that 
\begin{align}\label{gibbs}
\lr{e^{-\beta \bdd b}}
=\EE_{{\Pi_\infty}}
\left[
e^{-g^2(1-e^{-\beta}) 
\int_{-\infty}^0 \int_{0}^\infty
W_{\triangle}(s,r)
\rd s\rd r
}\right]
\quad \beta>0.
\end{align}
\eqref{GG1} follows from analytic continuation of \eqref{gibbs} in $\beta$. 
\eqref{GG2} can be derived by taking the derivative 
$d/d\beta$ of \eqref{GG1} at $\beta=0$. 
\qed
Simple but non trivial application is as follows. 
We know that 
$\lr{\s_x\otimes (-\one)^{\bdd b}}<0$ since the parity of $\grr$ is $-1$. 
As a corollary of Theorem \ref{main3} we can show that 
$\lr{ (-\one)^{\bdd b}}>0$. 
\bc{p}
We have
\begin{align*}
\lr{ (-\one)^{\bdd b}}
=\EE_{{\Pi_\infty}}
\left[
e^{-2g^2
\int_{-\infty}^0 \int_{0}^\infty
W_{\triangle}(s,r)
\rd s\rd r
}\right]>0.
\end{align*}
\ec
\proof
Put $\beta=i\pi$ in Theorem \ref{main3}. 
Then the corollary follows. 
\qed

\subsubsection{Gaussian functions}
We construct a path integral representation of  $\lr{e^{i\beta x}}$. 
\bt{x}
We have
\[\lr{e^{i\beta x}}=e^{-\beta^2/4}\EE_{{\Pi_\infty}}\left[
e^{i\beta K}\right],\] 
where 
\[K=-\frac{g}{\sqrt 2}\int_{-\infty}^\infty \hat T_s e^{-|s|}\rd s.\]
\et
\proof
The proof is similar to that of Lemma \ref{G1}. 
Since 
\begin{align*}
(\phi, e^{i\beta x}\psi)=
\sum_{\alpha\in\ZZ_2} 
\int _{\RR}
\bar \phi(\alpha,x) 
e^{i\beta x}
\psi(\alpha,x)\rd \mu(x),
\end{align*}
we have 
\begin{align*}
(e^{-tL}\one, e^{i\beta x}e^{-tL}\one )
&=
\sum_{\alpha\in\ZZ_2} 
e^{2\triangle t}
\int _{\RR} 
\EE_{\Pi} 
\left[
e^{-\sqrt{2}g\int_0^t\hat T_s e^{-s} \rd s {x}}
e^{-\sqrt{2}g\int_0^t\hat T_{s-t} e^{-s} \rd s 
x}e^{i\beta x}
\right. \\
&
\left.
\times e^{\frac{g^2}{2}
\int_0^t \rd s \int_0^t \rd r 
\hat T_s \hat T_r 
e^{-(s+r)}
(e^{2(s\wedge r)}-1)
}
e^{\frac{g^2}{2}
\int_0^t \rd s \int_0^t \rd r 
\hat T_{s-t} \hat T_{r-t} 
e^{-(s+r)}
(e^{2(s\wedge r)}-1)
}
\right]\frac{e^{-x^2}}{\sqrt\pi}\rd x
\end{align*}
Terms dependent on $x$ on the exponent above can be computed as 
\begin{align*}
i\beta x- \left(
x
+\frac{g}{\sqrt{2}} 
\int_0^t \hat T_s e^{-s} \rd s 
+\frac{g}{\sqrt{2}} 
\int_0^t\hat T_{s-t} e^{-s} \rd s 
\right)^2 +
\frac{g^2}{2}
\left(
\int_0^t \hat T_s e^{-s} \rd s 
+
\int_0^t\hat T_{s-t} e^{-s} \rd s 
\right)^2
\end{align*}
The first term on the right-hand side 
can be integrated with respect to $\rd x$ as 
\begin{align*}
&\frac{1}{\sqrt\pi}
\int _{\RR} e^{i\beta x
-\left(
x
+\frac{g}{\sqrt{2}} 
\int_0^t \hat T_s e^{-s} \rd s 
+\frac{g}{\sqrt{2}} 
\int_0^t\hat T_{s-t} e^{-s} \rd s 
\right)^2 }
\rd x=e^{-\beta^2/4}
e^{-i\beta(\frac{g}{\sqrt{2}} 
\int_0^t \hat T_s e^{-s} \rd s 
+\frac{g}{\sqrt{2}} 
\int_0^t\hat T_{s-t} e^{-s} \rd s)}.
\end{align*}
The remaining computations are the same as that of the proof of 
Lemma \ref{p}. 
Hence we obtain that 
\[\lr{e^{i\beta x}}=\limt 
e^{-\beta^2/4}\EE_{\Pi_t}\left[
e^{i\beta K_t}\right],\] 
where 
\[K_t=-\frac{g}{\sqrt 2}\int_{-t}^t \hat T_s e^{-|s|}\rd s.\]
Then the theorem follows 
from Corollary \ref{bounded}.
\qed
\bc{xx}
Let $\beta\in \CC$ such that $|\beta|<1$. 
Then 
\begin{align}\label{i2}
\lr{e^{\beta x^2}}=
\frac{1}{\sqrt{1-\beta}}
\EE_{{\Pi_\infty}}\left[
e^{\frac{\beta K^2}{1-\beta}}
\right].
\end{align}
In particular 
$\lim_{\beta\uparrow 1}
\|e^{\beta x^2/2}\grr\|^2=\infty$.
\ec
\proof
By Theorem \ref{p} we see that 
\begin{align*}
\lr{e^{-\beta^2x^2/2}}&=
\frac{1}{\sqrt{2\pi}}\int_{\RR} (\grr, e^{ik\beta x}\grr)e^{-k^2/2}\rd k
=
\frac{1}{\sqrt{2\pi}}\int_{\RR} 
e^{-k^2\beta^2/4}
\EE_{{\Pi_\infty}}\left[
e^{ik\beta K}\right]
e^{-k^2/2}\rd k\\
&=
\EE_{{\Pi_\infty}}\left[
\frac{1}{\sqrt{2\pi}}\int_{\RR} 
e^{-k^2\beta^2/4}
e^{ik\beta K}e^{-k^2/2}\rd k\right]
=
\frac{1}{\sqrt{1+\beta^2/2}}
\EE_{{\Pi_\infty}}\left[
e^{-\frac{\beta^2 K^2}{\beta^2+2}}
\right].
\end{align*}
By an analytic continuation we obtain \eqref{i2} 
for $\beta\in\CC$ such that $|\beta|<1$. 
Then the corollary follows. 
\qed

\subsubsection{Spin $\s_z$}
Let $\bar L=L-E$.  
Path integral representations of 
Euclidean Green functions 
by Lemma \ref{EG} can be rewritten as follows. 
\bc{PI}
(1) Suppose that  $\phi,\psi\in\cH$ and $f_j=f_j(\alpha,x)\in L^\infty(\ZZ_2\times \RR)$ for $j=0,1,\ldots,n$, 
and 
 $0< t_0< t_1<\ldots<t_n<t$. 
Then 
\begin{align}
&(\phi, e^{-t_0\bar L}f_0e^{-(t_1-t_0)\bar L} f_1 e^{-(t_2-t_1)\bar L}\cdots
e^{-(t_n-t_{n-1})\bar L}f_n
e^{-(t-t_{n})\bar L}\psi)\nonumber\\
&\label{EGQ}
=e^{\triangle t}e^{Et}
\sum_{\alpha\in\ZZ_2}\int_{\RR}
\EE_{\rQ}^\alpha\EE_{\bar\rP}^x
\left[
\bar\phi(\hat q^\triangle_0)\psi(\hat q^\triangle_t)
\left(
\prod_{j=0}^n f_j(\hat q^\triangle_ {t_j})\right)
e^{-g \int_0^tW(\hat q^\triangle_s)\rd s}
\right].
\end{align}
(2)
Suppose that $g_j=g_j(\alpha)\in L^\infty(\ZZ_2)$ for $j=0,1,\ldots,n$ and  $0< t_0< t_1<\ldots<t_n<t$. 
Then 
\begin{align}
&(\one, e^{-t_0\bar L}g_0(\s_z)e^{-(t_1-t_0)\bar L} g_1 (\s_z)e^{-(t_2-t_1)\bar L}\cdots
e^{-(t_n-t_{n-1})\bar L}g_n(\s_z)
e^{-(t-t_{n})\bar L}\one)\nonumber\\
&\label{EGQS}
=e^{\triangle t}e^{Et}
\sum_{\alpha\in\ZZ_2}
\EE_{\rQ}^\alpha
\left[
\left(
\prod_{j=0}^n g_j(\hat T_ {t_j})
\right)
\int_{\RR}
\EE_{\bar\rP}^x
\left[
e^{-g \int_0^tW(\hat q^\triangle_s)\rd s}
\right]\rd\mu(x)\right].
\end{align}
\ec
\proof
(1) is a simple reworking of Lemma \ref{EG} and (2) is a special case of (1).  
\qed
One can see that the integrand in \kak{EGQS} 
is 
\begin{align*}
\EE_{\rP }^x\left[ 
e^{-g\int_0^tW(\hat q^\triangle_s)\rd s}
\right]=
e^{-g\left(\int_0^te^{-s}(-1)^{N_{\triangle s}}\rd s\right) x}
e^{\frac{g^2}{4}\int_0^{(1-e^{-2t})/2}
\left|\int_y^t(-1)^{N_{\triangle s}}\rd s\right|^2\rd y}.
\end{align*}
by Lemma \ref{c}. 
\bt{pi}
We have
$\lr{\s_ze^{-|t-s|\bar L}\s_z}=
\EE_{{\Pi_\infty}}[\hat T_t \hat T_s]$ for any $t,s\in\RR$.
\et
\proof
By Lemma \ref{PI} and a limiting argument, 
we see that 
 \begin{align*}
(\s_z\grr, e^{-t\bar L}\s_z \grr)&
=
\lim_{T\to\infty}
\frac{1}{\|\one_{T-t/2}\|^2}
(\s_z\one_{T-t/2}, e^{-t\bar L}\s_z\one_{T-t/2})\\
&=
\lim_{T\to\infty}
\frac{e^{2ET}e^{2T\triangle}}{\|\one_{T-t/2}\|^2}
\sum_{\alpha\in\ZZ_2}
\EE_{\rQ}^\alpha\left[
\hat T_{-t/2}\hat T_{t/2}
e^{\frac{g^2}{2}\int_{-T}^T \rd t\int_{-T}^T \rd s W_{\triangle}(t,s) }\right].
\end{align*}
Then we have
\begin{align*}
(\s_z\grr, e^{-t\bar L}\s_z \grr)
&=
\lim_{T\to\infty}
\frac{\|\one_T\|^2}{\|\one_{T-t/2}\|^2}
\frac{e^{2ET}e^{2T\triangle}}{\|\one_T\|^2}
\sum_{\alpha\in\ZZ_2}
\EE_{\rQ}^\alpha\left[
\hat T_{-t/2}\hat T_{t/2}
e^{\frac{g^2}{2}\int_{-T}^T \rd t\int_{-T}^T \rd s W_{\triangle}(t,s) }\right]\\
&=
\lim_{T\to\infty}
\frac{\|\one_T\|^2}{\|\one_{T-t/2}\|^2}
\frac{
\EE_{\rQ}^\alpha\left[
\hat T_{-t/2}\hat T_{t/2}
e^{\frac{g^2}{2}\int_{-T}^T \rd t\int_{-T}^T \rd s W_{\triangle}(t,s) }\right]}
{
\EE_{\rQ}^\alpha\left[
e^{\frac{g^2}{2}\int_{-T}^T \rd t\int_{-T}^T \rd s W_{\triangle}(t,s) }\right]}
=\EE_{{\Pi_\infty}}[\hat T_{-t/2}\hat T_{t/2}
].
\end{align*}
Hence for $t>s$, 
\begin{align*}
(\s_z\grr, e^{-(t-s)\bar L}\s_z \grr)
=\EE_{{\Pi_\infty}}[\hat T_{-(t-s)/2}\hat T_{(t-s)/2}]
=\EE_{{\Pi_\infty}}[\hat T_{t} \hat T_{s}]
\end{align*}
by the shift invariance. 
\qed
We have 
$\|(L-E+\one)^{-1} \grr\|^2=\|\grr\|^2=1$. 
We can also 
estimate $\|(L-E+\one)^{-1}\s_z\grr\|$ by ${\Pi_\infty}$. 
\bc{res}
We have
\[
\|(L-E+\one)^{-1}\s_z\grr\|_\cH^2=
\EE_{{\Pi_\infty}}
\left[
\int_{-\infty}^0 \int_{0}^\infty
W_{\triangle}(s,r)
\rd s\rd r
\right].\]
\ec
\proof
By Theorem \ref{pi}  we have 
$\EE_{{\Pi_\infty}}[\hat T_t\hat T_s]=
(\s_z\grr, e^{-|t-s|(L-E)}\s_z\grr)$, and hence
\begin{align*}
\EE_{{\Pi_\infty}}\left[
\int_{-\infty}^0 \int_{0}^\infty
W_{\triangle}(t,s)
\rd s\rd t\right]
&=
\int_{-\infty}^0 \int_{0}^\infty
(\s_z\grr, e^{-|t-s|(L-E)}\s_z\grr)e^{-|t-s|}
\rd s\rd t\\
&=
\|(L-E+1)^{-1}\s_z\grr\|^2.
\end{align*}
Then the corollary follows. 
\qed
We can also give an alternative proof of Corollary \ref{res}. 
From commutation relations 
$Lb \grr=bE \grr-b\grr-\s_zg\grr$, 
it follows that 
$(L-E+\one)b\grr=-g\s_z \grr$. 
Then 
\begin{align}
\label{pt}
\|b\grr\|^2=g^2\|(L-E+\one)^{-1}\s_z \grr\|^2.
\end{align}
On the other hand 
$\|b\grr\|^2=\lr{b^\dagger b}$. 
\eqref{pt} is called the pull-through formula \cite{GJ68} in quantum field theory. 

\section{Pair interactions}
\label{shirai}
In Section \ref{s5} the path measure $\Pi_\infty$ associated 
with the ground state is discussed. 
The random process 
$
\int_{-\infty}^0 \int_{0}^\infty
W_{\triangle}(s,r)
\rd s\rd r
$
plays an important role, which gives the expectations of the number of bosons in the ground state: 
\[
\lr{\bdd b}=\EE_{\Pi_\infty}\left[\int_{-\infty}^0 \int_{0}^\infty
W_{\triangle}(s,r)
\rd s\rd r\right].\] 
We see that 
\[
\EE_{\Pi_\infty}
\left[ 
\int_{-\infty}^0 \int_{0}^\infty
\!\!W_{\triangle}(s,r)
\rd s\rd r\right]
=
\int_{-\infty}^0 \int_{0}^\infty
\!\!\EE_{\Pi_\infty} [
\hat T_s
\hat T_t
e^{-|t-s|}]
\rd s\rd r
=
\int_{-\infty}^0 \int_{0}^\infty
\!\!\EE_{\Pi_\infty} [
\hat T_{|s-t|}
e^{-|t-s|}]
\rd s\rd r
\]
by the shift invariance. Then it can be represented as 
\[
\EE_{\Pi_\infty}\left[
\int_{-\infty}^0 \int_{0}^\infty W_{\triangle}(s,r)\rd s\rd r\right]=
\EE_{\Pi_\infty}\left[
\int_0^\infty t(-1)^{N_{\triangle t}}e^{-t} \rd t\right].\]
In this section we investigate a random process
$\int_0^\infty t(-1)^{N_{\triangle t}}e^{-t} \rd t$ under 
the probability measure $\Pi$ but not $\Pi_\infty$. 
To do that we introduce a dummy random variable
$\int_0^\infty (-1)^{N_{\triangle t}}e^{-t} \rd t$, and 
we set 
\begin{align*}
X_1&=
\int_0^\infty (-1)^{N_{\triangle t}}e^{-t} \rd t,\\
X_2&=
\int_0^\infty t(-1)^{N_{\triangle t}}e^{-t} \rd t.
\end{align*}
Let $0=t_0<t_1<t_2<\ldots$ be jump points of 
$(N_{\triangle t})_{t\geq0}$ and 
$\delta_k=t_k-t_{k-1}$ the time difference between adjacent jumps.
Then $(\delta_{k})_k$ is i.i.d.and the density function of $\delta_k$ 
is given by 
$\triangle e^{-t\triangle}\one_{[0,\infty)}(t)$. 
Since $t_k=\delta_1+\cdots+\delta_k$, the density function  of $t_k$ is 
$\frac{t^{k-1}}{(k-1)!}e^{-\triangle t}$ for $k\geq1$. 

\bt{shirai1}
Means and covariances of $X_j$ are given by 
\begin{align*}
&\EE_{\Pi}[X_1]=\EE_{\Pi}[X_1^2]=\frac{1}{1+2\triangle},\\ 
&
\EE_{\Pi}[X_2]=\frac{1}{(1+2\triangle)^2},\\
&
\EE_{\Pi}[X_2^2]=\EE_{\Pi}[X_1X_2]=\frac{1+\triangle}{(1+2\triangle)^2}.
\end{align*}
\et
\proof
Since $X_1$ and $X_2$ are 
\begin{align*}
X_1&=\sum_{k=0}^\infty (-1)^k (e^{-t_k}-e^{-t_{k+1}})=1+2\sum_{k=1}^\infty (-1)^k e^{-t_k},\\
X_2&=1+2\sum_{k=1}^\infty (-1)^k(1+ t_k)e^{-t_k}, 
\end{align*}
we define two random variables:
\begin{align*}
Y_1=\sum_{k=1}^\infty (-1)^k e^{-t_k},\quad
Y_2=\sum_{k=1}^\infty (-1)^k t_ke^{-t_k}.
\end{align*}
From this it follows that 
\begin{align*}
&X_1=1+2Y_1,\quad 
X_2=1+2Y_1+2Y_2,\\
&X_1^2=1+4Y_1+4Y_1^2,\quad 
X_2^2=
1+4Y_1+4Y_1^2+4Y_2+8Y_1Y_2+4Y_2^2,\\
&X_1X_2=1+4Y_1+2Y_2+4Y_1^2+4Y_1Y_2.
\end{align*}
Then it is enough to investigate 
$\EE_\Pi[Y_iY_j]$ instead of $\EE_\Pi[X_iX_j]$ to show the theorem. 
Let $\delta=\delta_1$ for simplicity and 
we define two additional random variables: 
\begin{align*}
Z_1&= \sum_{k=2}^\infty (-1)^{k-1} e^{-(t_k-\delta) },\\
Z_2&= \sum_{k=2}^\infty (-1)^{k-1} (t_k-\delta )e^{-(t_k-\delta) }.
\end{align*}
Since $(\delta_k)_k$ 
is i.i.d.and $t_k=\sum_{i=1}^k \delta_i$, 
 $Z_i$ and $\delta $ are independent and 
the joint laws of $(Y_1,Y_2)$ and $(Z_1,Z_2)$ are identical. 
In particular 
$\EE_{\Pi}[P(Z_1)Q(Z_2)]=\EE_{\Pi}[P(Y_1)Q(Y_2)]$ for any polynomials $P$ and $Q$. 
Moreover $Z_i$ and $Y_i$ satisfy the following identities: 
\begin{align*}
Y_1&=-e^{-\delta }(1+Z_1),\\
Y_2&=-\delta e^{-\delta }(1+Z_1)-e^{-\delta }Z_2.
\end{align*}
Since 
\[\EE_{\Pi}[e^{-m\delta }\delta ^n]=
\frac{n!\triangle}{(m+\triangle)^{n+1}}.\] 
we can compute $\EE_{\Pi}[Y_1^m]$ from the identity:
\begin{align*}
\EE_{\Pi}[Y_1^m]=
\EE_{\Pi}[(-1)^m e^{-m\delta }]\EE_{\Pi}[(1+Z_1)^m]
=\frac{(-1)^m\triangle}{m+\triangle}\EE_{\Pi}[(1+Y_1)^m].
\end{align*}
Expanding $(1+Y_1)^m$ on the right-hand side above, 
we have 
\begin{align}\label{YY}
\EE_{\Pi}[Y_1^m]
=\frac{(-1)^mm!\triangle}{m+(1-(-1)^m)\triangle}
\sum_{k=0}^{m-1}\frac{1}{(m-k)!k!}\EE_{\Pi}[Y_1^k].
\end{align}
For example one can see that 
\begin{align}
\label{y1}&\EE_{\Pi}[Y_1]=-\frac{\triangle}{1+2\triangle},\\
\label{y12}&\EE_{\Pi}[Y_1^2]=\frac{\triangle}{2(1+2\triangle)}.
\end{align}
Then 
$\EE_{\Pi}[X_1]=\frac{1}{1+2\triangle}$ and 
$\EE_{\Pi}[X_1^2]=\frac{1}{1+2\triangle}$ follow. 
Similarly we have 
\begin{align*}
\EE_{\Pi}[Y_2^m]&=
(-1)^m 
\EE_{\Pi}[(\delta e^{-\delta }(1+Z_1)+e^{-\delta }Z_2)^m]\\
&=(-1)^m 
\sum_{k=0}^m \binom{m}{k}
\EE_{\Pi}[\delta ^{m-k}e^{-m\delta }]
\EE_{\Pi}[(1+Z_1)^{m-k}Z_2^k]\\
&=(-1)^m 
\sum_{k=0}^m 
\frac{m!\triangle}{k!(m+\triangle)^{m-k+1}}
\EE_{\Pi}[(1+Z_1)^{m-k}Z_2^k].
\end{align*}
In a similar manner to \eqref{YY} we also have 
\begin{align}\label{jfa}
\EE_{\Pi}[Y_2^m]
=\frac{(-1)^mm!\triangle }{m+(1-(-1)^m)\triangle}
\sum_{k=0}^{m-1} 
\frac{1}{k!(m+\triangle)^{m-k}}
\EE_{\Pi}[(1+Y_1)^{m-k}Y_2^k].
\end{align}
Putting $m=1$ above, we see that 
\begin{align}
\label{y2}
\EE_{\Pi}[Y_2]=-\frac{\triangle}{(1+2\triangle)^2}.
\end{align}
Then 
$\EE_{\Pi}[X_2]=\frac{1}{(1+2\triangle)^2}$ follows. 
Next let $m=2$ in \eqref{jfa}. 
Then 
\begin{align*}
\EE_{\Pi}[Y_2^2]=&
\frac{\triangle }{2+\triangle }\EE_{\Pi}[(1+Y_1)Y_2]
+
\frac{\triangle }{(2+\triangle )^2}\EE_{\Pi}[(1+Y_1)^2]\\
=&\frac{-\triangle ^2}{(2+\triangle )(1+2\triangle )^2}+\frac{\triangle }{2+\triangle }\EE_{\Pi}[Y_1Y_2]+
\frac{\triangle }{(2+\triangle )^2}\\
&+
\frac{-2\triangle ^2}{(2+\triangle )^2(1+2\triangle )}+\frac{\triangle^2 }{2(2+\triangle )^2(1+2\triangle )}
\end{align*}
and one can derive that 
\begin{align}
\label{ma1}
&\EE_{\Pi}[Y_2^2]-
\frac{\triangle }{2+\triangle }\EE_{\Pi}[Y_1Y_2]
=\frac{\triangle }{2(\triangle +2)(1+2\triangle )^2}.
\end{align}
To see $\EE_\Pi[Y_1Y_2]$ we consider 
\begin{align}
\EE_{\Pi}[(Y_2-Y_1)^m]&
=(-1)^m 
\EE_{\Pi}[
(e^{-\delta }(\delta -1)(1+Z_1)+e^{-\delta }Z_2)^m]\nonumber \\
&=(-1)^m 
\sum_{k=0}^m 
\binom{m}{k}
\EE_{\Pi}[e^{-m\delta }(\delta -1)^{m-k}]
\EE_{\Pi}[(1+Z_1)^{m-k}(Z_2)^k]\nonumber \\
&\label{jfa2}= 
\sum_{k=0}^m 
\sum_{l=0}^{m-k}
\frac{(-1)^{k+l}m!}{k!(m-k-l)!}
\frac{\triangle}{(m+\triangle)^{l+1}}
\EE_{\Pi}[(1+Z_1)^{m-k}(Z_2)^k].
\end{align}
Let $m=2$ in \eqref{jfa2}. Then 
\begin{align*}
\EE_{\Pi}[(Y_2-Y_1)^2]
=&
\left(\frac{\triangle }{2+\triangle }+\frac{-2\triangle }{(2+\triangle )^2}+\frac{2\triangle }{(2+\triangle )^3}\right)\EE_{\Pi}[(1+Y_1)^2]\\
&+
\left(\frac{-2\triangle }{2+\triangle }+\frac{2\triangle }{(2+\triangle )^2}\right)\EE_{\Pi}[(1+Y_1)Y_2]
+\frac{\triangle }{2+\triangle }\EE_{\Pi}[Y_2^2].
\end{align*}
Inserting $\EE_\Pi[Y_1]$ and $\EE_\Pi[Y_2]$ above, we have 
\begin{align*}
&\frac{\triangle }{2(1+2\triangle )}-2\EE_{\Pi}[Y_1Y_2]+\EE_{\Pi}[Y_2^2]\\
&=
\left(\frac{\triangle }{2+\triangle }+\frac{-2\triangle }{(2+\triangle )^2}+\frac{2\triangle }{(2+\triangle )^3}\right)
\left(1+\frac{-2\triangle }{1+2\triangle }+\frac{\triangle }{2(1+2\triangle )}\right)
\\
&\ \ \ \ +
\left(\frac{-2\triangle }{2+\triangle }+\frac{2\triangle }{(2+\triangle )^2}\right)\left(\frac{-\triangle }{(1+2\triangle )^2}+\EE_{\Pi}[Y_1Y_2]\right) +\frac{\triangle }{2+\triangle }\EE_{\Pi}[Y_2^2],
\end{align*}
which implies that 
\begin{align}
\label{ma2}
\EE_{\Pi}[Y_2^2]-\frac{(4+3\triangle)}{(2+\triangle )}
\EE_{\Pi}[Y_1Y_2]
=\frac{-\triangle (1+\triangle)}
{2(2+\triangle )(1+2\triangle )^2}.
\end{align}
Relations \eqref{ma1} and \eqref{ma2} imply that 
\begin{align}
\label{y1y2}
\EE_{\Pi}[Y_1Y_2]=
\frac{\triangle }{4(1+2\triangle )^2}=
\EE_{\Pi}[Y_2^2].
\end{align}
By \eqref{y1},\eqref{y12}, \eqref{y2} and \eqref{y1y2}, 
we have
\begin{align*}
&\EE_{\Pi}[X_2^2]=
1+4\EE_{\Pi}[Y_1+Y_2+Y_1^2+2Y_1Y_2+Y_2^2]
=\frac{1+\triangle }{(1+2\triangle )^2},\\
&\EE_{\Pi}[X_1X_2]=
1+\EE_{\Pi}[4Y_1+2Y_2+4Y_1^2+4Y_1Y_2]
=\frac{1+\triangle }{(1+2\triangle )^2}.
\end{align*}
Then the proof is complete. 
\qed
\begin{remark}
From Theorem  \ref{shirai1} one can see that 
the covariance of $X_1$ and $X_2$ is positively correlated and it is actually given by 
$$
{\rm cov}(X_1,X_2) = \EE_\Pi
[X_1X_2]- \EE_\Pi[X_1]\EE_\Pi[X_2]=\frac{\triangle (3+2\triangle)}{(1+2\triangle)^3}>0.$$
\end{remark}
In Theorem \ref{shirai1} we show the mean and the covariance of $X_1$. 
We can have a more strong statement. 
\bt{shiraiconj}
The density function $g_{\triangle}$ of $X_1$ exists and 
is given by 
\begin{equation}
g_{\triangle}(t) = \frac{1}{Z_{\triangle}} (1+t)(1-t^2)^{\triangle-1} \quad (-1<t<1), 
\label{eq:ga}
\end{equation}
where $Z_{\triangle}$ is the normalizing constant given by 
$Z_{\triangle} = B(\triangle,1/2)$.  
In particular 
\begin{align}
\label{moment}
 \EE_{\Pi}[X_1^{2m-1}] =  \EE_{\Pi}[X_1^{2m}] =  \prod_{j=1}^m \frac{2j-1}{2j-1+2\triangle}. 
\end{align}
\et
\proof
Set 
\[\tilde X_1=\sum_{k=1}^\infty(-1)^{k-1} (e^{-(t_k-\delta)}-e^{-(t_{k+1}-\delta)}).\]
Then we notice very crucial relations:
\begin{align}
&\label{es}X_1\stackrel{d}{=}\tilde X_1,\\
&\label{ES}
X_1\stackrel{d}{=}
1-e^{-\delta}-e^{-\delta}\tilde X_1.
\end{align}
Let $1-e^{-\delta}=\eta$. 
Then 
the density function of $\eta$ is given
by $\triangle(1-\xi)^{\triangle-1}\one_{[0,1]}(\xi)$. 
From \eqref{ES}, for any
bounded measurable function $f$, 
\[
 \EE_\Pi[f(X_1)] = \EE_\Pi[f(\eta - (1-\eta) \tilde X_1)]. 
\]
Suppose that the density function of $X_1$ exists, and it is denoted by $g$. 
Then by \eqref{es} we see that 
\begin{align*}
\EE_\Pi[f(\eta - (1-\eta) \tilde X_1)] 
=\int_{-1}^1 g(x)\rd x \int_{0}^1 \triangle(1-\xi)^{\triangle-1}  
f(\xi - (1-\xi)x)\rd\xi.
\end{align*}
By change of variables $s=x$ and $t=\xi-(1-\xi)x$, 
we see
that 
\begin{align*}
\EE_\Pi[f(\eta - (1-\eta) \tilde X_1)] 
 = \int_{-1}^1 f(t) \left(\triangle(1-t)^{\triangle-1}\int_{-t}^1 
 \frac{g(s)}{(1+s)^\triangle}\rd s\right) \rd t.  
\end{align*}
The right-hand side is also 
\[\EE_\Pi[f(X_1)]=\int_{-1}^1 f(t) g(t)\rd t.\] 
Thus we  have the
following equation:  
\[
 g(t) = \triangle (1-t)^{\triangle-1} \int_{-t}^1 
 \frac{g(s)}{(1+s)^\triangle}\rd s . 
\]
It is easy to see that $g_\triangle(t)$ given by \eqref{eq:ga}
satisfies the above equation. 
Since $(1-t^2)^{\triangle-1}$ is even, momenta can be directly computed as   
\[
 \EE_{\Pi}[X_1^{2m-1}] =  \EE_{\Pi}[X_1^{2m}] = 
\frac{1}{Z_{\triangle}} \int_{-1}^1 t^{2m} (1-t^2)^{\triangle-1} \rd t.
\]
Then, by change of variables $s=t^2$, we obtain 
\[
 \EE_{\Pi}[X_1^{2m-1}] =  \EE_{\Pi}[X_1^{2m}] = 
\frac{1}{Z_{\triangle}} \int_0^1 s^m (1-s)^{\triangle-1} s^{-1/2}\rd s 
= \frac{B(m+1/2, \triangle)}{B(1/2,\triangle)}, 
\]
which equals to \eqref{moment}. 
\qed

\appendix
\section{Proof of Proposition \ref{hir}}
\label{A}
Redefine the probability measure $\Pi_T$ on 
$(\cD, \cB_{\cD})$ by
\begin{align}\label{measure}
\Pi_T(A)=\frac{1}{Z_T} \half e^{2T\triangle} \sum_{\alpha\in\ZZ_2} \EE_\rQ^\alpha
\left[\one_A
e^{\frac{g^2}{2}\int_{-T}^T \rd t\int_{-T}^T \rd s 
W_{\triangle}(t,s) 
 }\right], \quad A\in \cB_{\cD},
\end{align}
where 
$Z_T=
\half e^{2T\triangle} \sum_{\alpha\in\ZZ_2} \EE_\rQ^\alpha
\left[e^{\frac{g^2}{2}\int_{-T}^T \rd t\int_{-T}^T \rd s W_{\triangle}(t,s) }\right]$ 
is the normalizing constant. 
Note that pair interaction $W_{\triangle}(t,s) $ is independent of $\s$ and hence one can replace 
$\sum_{\alpha\in\ZZ_2} \EE_\rQ^\alpha $ with 
$2\EE_\rQ^\alpha $ in \eqref{measure}. 
We also notice that 
$1=\|\grr\|_{\cH}^2=\sum_{\alpha\in\ZZ_2}\int_{\RR} |\grr(\alpha,x)|^2\rd\mu(x)$, 
$2=\|\one\|^2_{\cH}=\sum_{\alpha\in\ZZ_2}\int_{\RR} \rd\mu(x)$ 
and 
$2Z_T=\|e^{-TL}\one\|^2$. 

Let $A_j\in\cB(\RR)$ for $j=0,1,\ldots,n$ and 
$\Lambda=\{t_0,t_1,\ldots,t_n\}\subset[-T,T]$. 
The cylinder set is defined by 
\[
C_T^{\Lambda}(A_0\times\cdots\times A_n)=
\{\omega\in \cD_T\mid \omega(t_j)\in A_j,j=0,1,\ldots,n\}. \]
Recall that the family of cylinder sets is denoted by 
${\cA}_T$. 
We also note that $\boldsymbol{\s}({\cA}_T)=\cB_T$.

The idea of the proof of Proposition \ref{hir} is to apply the fact that finite dimensional distribution 
$\Pi_T\circ\pi_t^{-1}
(C_t^\Lambda(A_0\times\cdots\times A_n))$ with 
$\Lambda=\{t_0,...,t_n\}\subset[-t,t]\subset[-T,T]$
is represented as 
\begin{align*}
&\Pi_T\circ\pi_t^{-1}(C_t^\Lambda(A_0\times\cdots\times A_n))\\
&=\frac{e^{2Et}(e^{-(T-t)\bar L}\one,e^{-(t_0+t){L}}
\one_{A_0}e^{-(t_1-t_0){L}}\one_{A_1}\cdots \one_{A_n}e^{-(t-t_n){L}} e^{-(T-t)\bar L}\one)}{\|e^{-T\bar L}\one\|^2}
\end{align*}
by \eqref{EGQ}. 
Formally it converges to 
$
(\grr,
\one_{A_0}e^{-(t_1-t_0){\bar L}}\one_{A_1}\cdots 
e^{-(t_{n-1}-t_n){\bar L}}\one_{A_n}\grr)$
as $T\to\infty$. 
Since 
\begin{align*}
&(\grr,
\one_{A_0}e^{-(t_1-t_0){\bar L}}\one_{A_1}\cdots 
e^{-(t_{n-1}-t_n){\bar L}}\one_{A_n}\grr)\\
&=
e^{E(t_n-t_0)}e^{\triangle (t_n-t_0)}\sum_{\alpha\in\ZZ_2} 
\int_{\RR} 
\EE_\rQ^\alpha\EE_{\bar\rP}^x
\left[
\left(\prod_{j=0}^n\one_{A_j}(\hat T_j)\right)
\grr(\hat q^\triangle_ {t_0}) \grr(\hat q^\triangle_{t_n})
e^{-g\int_{t_0}^{t_n} W(\hat q^\triangle_s)\rd s}
\right] \rd \mu(x), 
\end{align*}
we expect that $\Pi_T(A)$ converges to 
\begin{align}
\label{m}
{\Pi_\infty}(A)=
e^{2Et}e^{2\triangle t}
\sum_{\alpha\in\ZZ_2} 
\int_{\RR} 
\EE_\rQ^\alpha\EE_{\bar\rP}^x
\left[
\one_{A}
\grr(\hat q^\triangle_ {-t}) \grr(\hat q^\triangle_{t})
e^{-g\int_{-t}^{t} W(\hat q^\triangle_s)\rd s}
\right] \rd \mu(x). 
\end{align}
We shall show this explicitly below.

We set the right-hand side of \eqref{m} by $m_t(A)$. 
Since $\cBc$ is a finitely additive family of sets, 
we define the finitely additive set function $\nu$ on $(\cD, {\cBc})$
by
$\nu(A)= m_t(A)$
for 
$A\in \pi_t^{-1}(\cB_t)$.

\bl{A1}
$\nu$ is well defined, i.e.,
$m_t(A)=m_s(A)$ 
for $A\in \pi_t^{-1}(\cB_t)\subset \pi_s^{-1}(\cB_s)$. 
\el
\proof
Notice that $m_t\circ \pi_t^{-1}$ and $m_s\circ \pi_t^{-1}$ are probability measures on 
$(\cD_t,\cB_t)$. 
We compute finite dimensional distributions of 
$m_t\circ \pi_t^{-1}$ and $m_s\circ \pi_t^{-1}$. 
Let $\Lambda=\{t_0,t_1,\ldots,t_n\}\subset[-t,t]\subset[-s,s]$. 
Since $e^{-r\bar L}
\grr=\grr$ for any $r\geq0$, 
we have by \eqref{EGQ}, 
\begin{align*}
&m_t\circ \pi_t^{-1}(C_t^{\Lambda}(A_0\times\cdots\times A_n))\\
&=
e^{2Et}e^{2\triangle t}\sum_{\alpha\in\ZZ_2} 
\int_{\RR} 
\EE_\rQ^\alpha\EE_{\bar\rP}^x
\left[
\left(\prod_{j=0}^n \one_{A_j}(\hat T_{t_j})\right) 
\grr(\hat q^\triangle_ {-t}) \grr(\hat q^\triangle_t)
e^{-g\int_{-t}^t W(\hat q^\triangle_s)\rd s}
\right] \rd \mu(x)\\
&=(
e^{-(t_0+t)\bar L}
\grr, 
\one_{A_0}
e^{-(t_1-t_0)\bar L}
\one_{A_1}
\cdots
e^{-(t_n-t_{n-1})\bar L}
\one_{A_n}
e^{-(t-t_{n})\bar L}
\grr)\\
&=(
\grr, 
\one_{A_0}
e^{-(t_1-t_0)\bar L}
\one_{A_1}
\cdots
e^{-(t_n-t_{n-1})\bar L}
\one_{A_n}
\grr)\\
&=(
e^{-(t_0+s)\bar L}
\grr, 
\one_{A_0}
e^{-(t_1-t_0)\bar L}
\one_{A_1}
\cdots
e^{-(t_n-t_{n-1})\bar L}
\one_{A_n}
e^{-(s-t_{n})\bar L}
\grr)\\
&=m_s\circ \pi_t^{-1}(C_t^{\Lambda}
(A_0\times\cdots\times A_n)).
\end{align*}
It is straightforward to see that 
the Kolmogorov consistency condition also holds true:
\[
m_t\circ \pi_t^{-1}\left(
C_t^{\{\Lambda,s_1,...,s_m\}}
\left(A_0\times\cdots\times A_n\times \prod^m \RR
\right)\right)
= 
m_t\circ \pi_t^{-1}(C_t^\Lambda
(A_0\times\cdots\times A_n)).\]
Let $\pi_\Lambda:[-t,t]^\RR\to \RR^{\Lambda}$ be the projection such that 
for $\omega\in [-t,t]^{\RR}$, 
$\pi_\Lambda\omega=(\omega(t_0),\ldots, \omega(t_n))$. 
Thus by the Kolmogorov extension theorem 
there exists a
unique probability measure $\bar m_t$ on 
$([-t,t]^{\RR}, \boldsymbol{\s}({\cA}_t))$ such that
\begin{align}
\bar m_t(\pi_\Lambda^{-1}(A_0\times\cdots\times A_n))=m_t\circ \pi_t^{-1}(C_t^\Lambda(A_0\times\cdots\times A_n))
\end{align}
for all $\Lambda \subset [-t,t]$ with $\#\Lambda <\infty$ 
and $A_j\in \cB(\RR) $.
Since the extension is unique, $m_t\circ \pi_t^{-1}=\bar m_t$. 
Similarly 
there exists a
unique probability measure $\bar m_s$ on 
$([-t,t]^{\RR}, \boldsymbol{\s}({\cA}_t))$ such that 
$m_s\circ \pi_t^{-1}=\bar m_s$. 
Then $m_s\circ \pi_t^{-1}=m_t\circ \pi_t^{-1}$ on $\cB_t$, 
which implies the lemma. 
\qed

The first task is to extend $\nu$ to a probability measure 
by the Hopf extension theorem. 

\bl{existence2}
$\nu$ can be uniquely extended 
to a probability measure ${\Pi_\infty}$ on $(\cD, \cB_{\cD})$. \el
\proof
Suppose that $E_n\in \cBc$ such that 
$E_n\supset E_{n+1}\supset\ldots$ and 
$\lim_{n\to\infty} \nu(E_n)=\alpha>0$. 
It is enough to show that 
$\bigcap_n E_n\neq \emptyset$ by the Hopf extension theorem. 
Let $E_n=\pi_{T_n}^{-1}(E_n')$ with $E_n'\in \cB_{T_n}$. 
We can assume that $T_n< T_{n+1}<\rightarrow \infty$. 
Let $\mu_T=\nu\circ \pi_T^{-1}$ be a probability measure on $\cD_T$. 
Since $\cD_T$ is a Polish space, 
$\mu_T$ is regular, i.e., 
for $A\in \cB_T$ and $\epsilon>0$ 
there exist a compact set $K$ and an open set $O$ in $\cD_T$ such that 
$K\subset A\subset O$ and $\mu_T(O\setminus K)<\epsilon$. There exists a compact set $K_n'\subset \cD_{T_n}$ such that 
$\mu_{T_n}(E_n'\setminus K_n')<\alpha/2^n$. 
Let $K_n=\pi_{T_n}^{-1}(K_n')$, 
$D_n=\bigcap_{j=1}^n K_j$ and 
$D=\bigcap_{n=1}^\infty D_n$. 
Since $D\subset \bigcap_n E_n$, it is enough to show 
that $D\neq \emptyset$. 
We see that 
\begin{align*}
\alpha-\nu(D_n)
&\leq \nu(E_n)-\nu(D_n)
\leq \nu(E_n\setminus D_n)\\
&=
\nu(\cup_{j=1}^n E_n\setminus K_j)
=
\nu(\pi_{T_n}^{-1}\cup_{j=1}^n E_n'\setminus K_j')
=
\mu_{T_n}(\cup_{j=1}^n E_n'\setminus K_j')\\
&=
\sum_{j=1}^n \mu_{T_n}(E_n'\setminus K_j')
\leq
\sum_{j=1}^n \mu_{T_n}(E_j'\setminus K_j')
\leq\sum_{j=1}^n \alpha/2^j.
\end{align*}
Then 
$0<\nu(D_n)$ and we see that $D_n\neq\emptyset$. 
Let $f_n\in D_n$, i.e., 
$f_n\in \bigcap_{j=1}^n K_j$. Thus 
\[f_n\in K_\ell \ 
\mbox{ for any }\ n\geq \ell.\] 
Let $\ell=1$. Then $\pi_{T_1}(f_n)\in K_1'$ for any $n\geq 1$. 
Taking a subsequence $n'$, we see that 
$\lim_{n'\to\infty} \pi_{T_1}(f_{n'})=h_1\in K_1'$ exists. 
Let $\ell=2$. Then $\pi_{T_2}(f_{n'})\in K_2'$ for any $n'\geq 2$.
Take a subsequence $n''$ of $n'$ again, then 
$\lim_{n''\to\infty} \pi_{T_2}(f_{n''})=h_2\in K_2'$ exists. 
Proceeding this procedure, we can obtain 
a subsequence $\{m\}$ that 
$ \lim_{m\to\infty} \pi_{T_\ell }(f_{m})=h_\ell \in K_\ell '$ exists for any $\ell$. 
Let $g_\ell =\pi_{T_\ell }^{-1}(h_\ell )\in L_\ell$. Define $g\in \cD$ 
by 
$g(x)=g_\ell (x)$ for $x\in[-T_\ell ,T_\ell ]$. 
By the construction this is well defined, i.e., 
$g_\ell (x)=g_{\ell+1}(x)$ for $x\in[-T_\ell ,T_\ell ]$.
We see that $g\in D$ and $D\neq\emptyset$.
\qed
For probability measures $\Pi_T$ and ${\Pi_\infty}$ on $(\cD,\cB_{\cD})$ in order to show that 
$\Pi_T(A)\to {\Pi_\infty}(A)$ for every 
 $A\in {\cBc}$, we define the finitely additive set 
 function $\rho_T$ on $(\cD_T, \cBc_T)$.
Let 
$\one_T=e^{-T\bar L}\one$ for $t\geq0$. 
Then 
$s\text{-}\lim_{T\to\infty} \one_T=\grr$ 
and
$\|\one_T\|^2=2e^{2TE}Z_T$. 
The finitely additive set function $\rho_T$ on $(\cD_T, \cBc_T)$ is defined by
\begin{align}\label{rho}
 \rho_T(A)= 
 e^{2Et}e^{2t\triangle}\frac{1}{\|\one_T\|^2}
 \sum_{\alpha\in\ZZ_2}\int_{\RR} \EE_\rQ^\alpha
 \EE_{\bar\rP}^x \left[
 \one_A
{\one_{T-t}(\hat q^\triangle_ {-t})} 
{\one_{T-t}(\hat q^\triangle_ {t})}
e^{-g\int_{-t}^tW(\hat q^\triangle_s)\rd s} \right]\rd \mu(x)
\end{align}
for 
$A\in \pi_t^{-1}(\cB_t)$ but $t\leq T$. 
The right-hand side of \eqref{rho} is denoted by $M_{T,t}(A)$. 

\bl{A2}
$\rho_T$ is well defined, i.e, 
$M_{T,t}(A)=M_{T,s}(A)$ 
for $A\in \pi_t^{-1}(\cB_t)
\subset \pi_s^{-1}(\cB_r)$. 
\el
\proof
This is shown in a similar manner to Lemma \ref{A1}. 
Let 
\[M_{T,t}(A)=
e^{2Et}e^{2t\triangle}
\sum_{\alpha\in\ZZ_2}\int_{\RR} \EE_\rQ^\alpha
\EE_{\bar\rP}^x
\left[
 \one_A
{\one_{T-t}(\hat q^\triangle_ {-t})} 
{\one_{T-t}(\hat q^\triangle_ {t})}
e^{-g\int_{-t}^tW(\hat q^\triangle_s)\rd s} \right]\rd\mu(x).\]
Then $M_{T,t}\circ \pi_t^{-1}$ and $M_{T,s}\circ \pi_t^{-1}$ are probability measures on 
$(\cD_t,\cB_t)$. 
Let $\Lambda=\{t_0,t_1,\ldots,t_n\}\subset[-t,t]\subset[-s,s]$. 
We have by \eqref{EGQ}, 
\begin{align*}
&M_{T,t}\circ \pi_t^{-1}(C_t^{\Lambda}(A_0\times\cdots\times A_n))\\
&=
e^{2Et}e^{2\triangle t}\sum_{\alpha\in\ZZ_2} 
\int_{\RR} 
\EE_\rQ^\alpha\EE_{\bar\rP}^x
\left[
\left(\prod_{j=0}^n \one_{A_j}(\hat T_{t_j})\right) 
\one_{T-t}(\hat q^\triangle_ {-t}) 
\one_{T-t}(\hat q^\triangle_t)
e^{-g\int_{-t}^t W(\hat q^\triangle_r)\rd r}
\right] \rd \mu(x)\\
&=(
e^{-(t_0+t)\bar L}
\one_{T-t}, 
\one_{A_0}
e^{-(t_1-t_0)\bar L}
\one_{A_1}
\cdots
e^{-(t_n-t_{n-1})\bar L}
\one_{A_n}
e^{-(t-t_{n})\bar L}
\one_{T-t})\\
&
=(
e^{-(t_0+s)\bar L}
\one_{T-s}, 
\one_{A_0}
e^{-(t_1-t_0)\bar L}
\one_{A_1}
\cdots
e^{-(t_n-t_{n-1})\bar L}
\one_{A_n}
e^{-(s-t_{n})\bar L}
\one_{T-s})\\
&=M_{T,s}\circ \pi_t^{-1}(C_t^{\Lambda}
(A_0\times\cdots\times A_n)).
\end{align*}
It is straightforward to see that 
the Kolmogorov consistency condition also holds true:
\[
M_{T,t}\circ \pi_t^{-1}\left(
C_t^{\{\Lambda,s_1,...,s_m\}}
\left(A_0\times\cdots\times A_n\times \prod^m \RR
\right)\right)
= 
M_{T,t}\circ \pi_t^{-1}(C_t^\Lambda
(A_0\times\cdots\times A_n)).\]
Thus by the Kolmogorov extension theorem 
there exists a
unique probability measure $\bar M_{T,t}$ on 
$([-t,t]^{\RR}, \boldsymbol{\s}({\cA}_t))$ such that
\begin{align}
\bar M_{T,t}(\pi_\Lambda^{-1}(A_0\times\cdots\times A_n))=M_{T,t}\circ \pi_t^{-1}(C_t^\Lambda(A_0\times\cdots\times A_n))
\end{align}
for all $\Lambda \subset [-T,T]$ with $\#\Lambda <\infty$ 
and $A_j\in \cB(\RR) $.
Since the extension is unique, $M_{T,t}\circ \pi_t^{-1}=\bar M_{T,t}$. 
Similarly 
there exists a
unique probability measure $\bar M_{T,s}$ on 
$([-t,t]^{\RR}, \boldsymbol{\s}({\cA}_t))$ such that 
$M_{T,s}\circ \pi_t^{-1}=\bar M_{T,s}$. 
Then $M_{T,s}\circ \pi_t^{-1}=M_{T,t}\circ \pi_t^{-1}$ on $\cB_t$, 
which implies the lemma. 
\qed
We shall show that $\rho_T=\Pi_T$ on $\cBc_T$ for any $T>0$.
\bl{identity}
We have
$\rho_T=\Pi_T$ on $\cBc_T$. 
\el
\proof
Let $t\leq T$. It is enough to show that
$\Pi_T(A)=\rho_T(A)$ for $A\in \pi_t^{-1}(\cB_t)$. 
Let $\Lambda =\{t_0,t_1,...,t_n\}\subset[-t,t]\subset[-T,T]$ and $A_0\times\cdots\times A_n\in \cB(\RR^\Lambda)$. 
We have
\begin{align}\label{finite}&\Pi_T\circ\pi_t^{-1}(C_t^\Lambda(A_0\times\cdots\times A_n))= 
\frac{1}{Z_T}e^{2T\triangle}\half \sum_{\alpha\in\ZZ_2} \EE_\rQ^\alpha\left[\left(\prod_{j=0}^n \one _{A_{j}}(\hat T_ {t_j}) \right) 
e^{\frac{g^2}{2}\int_{-T}^T \rd t\int_{-T}^T \rd s W_{\triangle}(t,s) }\right],\\
&\rho_T\circ\pi_t^{-1}(C_t^\Lambda(A_0\times\cdots\times A_n)) 
\nonumber\\
&\label{finiterho}= e^{2Et}e^{2\triangle t}\frac{1}{\|\one_T\|^2} \sum_{\alpha\in\ZZ_2}\int_{\RR} \EE_\rQ^\alpha \EE_{\bar\rP}^x \left[\left(\prod_{j=0}^n\one_{A_{j}}(\hat T_ {t_j})\right){\one_{T-t}(\hat q^\triangle_ {-t})} {\one_{T-t}(\hat q^\triangle_ {t})} e^{-g\int_{-t}^tW(\hat q^\triangle_s)\rd s} \right]\rd \mu(x).
\end{align}
By \eqref{EGQ} we see that
\begin{align*}
\eqref{finite}
&= \frac{1}{\|\one_T\|^2}
(\one,e^{-(t_0+T){L}}\one_{A_0}e^{-(t_1-t_0){L}}\one_{A_1}\cdots \one_{A_n}e^{-(T-t_n){L}}\one)\\
&= \frac{e^{2Et}}{\|\one_T\|^2} (\one_{T-t},e^{-(t_0+t){L}}
\one_{A_0}e^{-(t_1-t_0){L}}\one_{A_1}\cdots \one_{A_n}e^{-(t-t_n){L}} \one_{T-t})=\eqref{finiterho}.
\end{align*}
Then we have
\begin{align}\label{eq}
\Pi_T\circ\pi_t^{-1}
(C_t^\Lambda(A_0\times\cdots\times A_n))
=
\rho_T\circ\pi_t^{-1}(C_t^\Lambda(A_0\times\cdots\times A_n)).
\end{align}
Since both sides of \eqref{eq} satisfy the Kolmogorov consistency condition, there exists a unique probability measure $\mu$ on $(\cD_T,\cB_t)$ such that 
\[\mu(\pi_\Lambda^{-1}(A_0\times\cdots\times A_n))=
\Pi_T\circ\pi_t^{-1}
(C_t^\Lambda(A_0\times\cdots\times A_n))
=
\rho_T\circ\pi_t^{-1}(C_t^\Lambda(A_0\times\cdots\times A_n)).
\]
$\Pi_T\circ\pi_t^{-1}$ and 
$\rho_T\circ\pi_t^{-1}$ are
probability measures on 
$(\cD_t,\cB_t)$, and 
$
\Pi_T\circ\pi_t^{-1}
(C_t^\Lambda(A_0\times\cdots\times A_n))
=\Pi_T\circ\pi_t^{-1}
(\pi_\Lambda^{-1}(A_0\times\cdots\times A_n))
=
\rho_T\circ\pi_t^{-1}(C_t^\Lambda(A_0\times\cdots\times A_n))
=
\rho_T\circ\pi_t^{-1}(\pi_\Lambda^{-1}(A_0\times\cdots\times A_n))$. 
Since the extension is unique, 
$\Pi_T\circ\pi_t^{-1}=\mu=\rho_T\circ\pi_t^{-1}$ on $(\cD_t, \cB_t)$ follows. 
\qed

{\it Proof of Proposition \ref{hir}}\\
By $s\text{-}\lim_{T\to\infty} \one_T=\grr$ 
we obtain that 
$s\text{-}\lim_{T\to\infty}\one_{T-t}=\grr$ and
$\lim_{T\to\infty}\|\one_{T}\|=~1$. 
Then 
for each $\alpha\in\ZZ_2$, 
$(\one_{T-t}/\|\one_T\|)(\cdot,\s)\to\gr(\cdot,\s)$
as $T\to\infty$ in $L^2(\RR,\rd \mu)$. 
Let 
$\grr^T=
\frac{\one_{T-t}}{\|\one_T\|}$. 
Note that $\grr,\grr^T\in L^\infty(\ZZ_2\times \RR)$. 
Let $A\in \pi_t^{-1}(\cB_t)$. 
Then $\Pi_T(A)=\rho_T(A)$ by Lemma \ref{identity} 
and $\nu(A)={\Pi_\infty}(A)$ by Lemma \ref{A1}. 
We have 
\begin{align*}
&\Pi_T(A)-{\Pi_\infty}(A)=
\rho_T(A)-\nu(A)
\\
&=
e^{2Et}e^{2\triangle t}
\sum_{\alpha\in\ZZ_2} 
\EE_\rQ^\alpha
\left[
\one_A
\int_{\RR} 
\EE_{\bar\rP}^x
\left[
\left(
\grr(\hat q^\triangle_ {-t}) \grr(\hat q^\triangle_t)-
\grr^T(\hat q^\triangle_ {-t}) \grr^T(\hat q^\triangle_t)
\right) e^{-g\int_{-t}^t W(\hat q^\triangle_s)\rd s}
\right] \rd \mu(x)\right].
\end{align*}
Then 
\begin{align*}
&
\int_{\RR} 
\EE_{\bar\rP}^x
\left[
\left|
\grr(\hat q^\triangle_ {-t}) \grr(\hat q^\triangle_t)-
\grr^T(\hat q^\triangle_ {-t}) \grr^T(\hat q^\triangle_t)
\right|
e^{-g\int_{-t}^t W(\hat q^\triangle_s)\rd s}
\right] \rd \mu(x)\\
&
\leq
\int_{\RR} 
\EE_{\bar\rP}^x
\left[
\left|
\grr(\hat q^\triangle_ {-t})- \grr^T(\hat q^\triangle_{-t})|
\grr(\hat q^\triangle_ {t}) 
\right|e^{-g\int_{-t}^t W(\hat q^\triangle_s)\rd s}
\right] \rd \mu(x)\\
&+
\int_{\RR} \EE_{\bar\rP}^x
\left[
|\grr^T(\hat q^\triangle_ {-t})|
\left|
 \grr(\hat q^\triangle_t)-
\grr^T(\hat q^\triangle_ {t})
\right|
e^{-g\int_{-t}^t W(\hat q^\triangle_s)\rd s}\right] \rd \mu(x).
\end{align*}
We estimate 
$\int_{\RR} 
\EE_{\bar\rP}^x
\left[
\left|
\grr(\hat q^\triangle_ {-t})- \grr^T(\hat q^\triangle_{-t})|
\grr(\hat q^\triangle_ {t}) 
\right|e^{-g\int_{-t}^t W(\hat q^\triangle_s)\rd s}
\right]\rd \mu(x)$. 
By the shift invariance we have
\begin{align*}
&\int_{\RR} 
\EE_{\bar\rP}^x
\left[
\left|
(\grr(\hat q^\triangle_ {-t})- \grr^T(\hat q^\triangle_{-t}))
\grr(\hat q^\triangle_ {t}) 
\right|e^{-g\int_{-t}^t W(\hat q^\triangle_s)\rd s}
\right] \rd\mu(x)\\
&=
\int_{\RR} 
|\grr(\hat q^\triangle_ {0})- \grr^T(\hat q^\triangle_{0})|
\EE_{\bar\rP}^x
\left[
\left| 
\grr(\hat q^\triangle_ {2t}) 
\right|e^{-g\int_{0}^{2t} W(\hat q^\triangle_s)\rd s}
\right] \rd\mu(x).
\end{align*}
By the Schwarz inequality we also have
\begin{align*}
&\leq
\int_{\RR} 
|\grr(\hat q^\triangle_ {0})- \grr^T(\hat q^\triangle_{0})|
\left(
\EE_{\bar\rP}^x
\left[\left| \grr(\hat q^\triangle_ {2t}) \right|^2\right]\right)^{\han}
\left(
\EE_{\bar\rP}^x
\left[e^{-2g\int_{0}^{2t} W(\hat q^\triangle_s)\rd s}
\right]\right)^{\han}\\
&\leq
\left(
\int_{\RR} 
|\grr(\hat q^\triangle_ {0})- \grr^T(\hat q^\triangle_{0})|^2\rd\mu(x)\right)^{\han}
\!\!
\!\!
\left(
\int_{\RR} 
\EE_{\bar\rP}^x
\left[\left| \grr(\hat q^\triangle_ {2t}) \right|^2\right]
\rd\mu(x)\right)^{\han}
\!\!
\!\!
\left(
\EE_{\bar\rP}^x
\left[
e^{-2g\int_{0}^{2t} W(\hat q^\triangle_s)\rd s}
\right]\right)^{\han}.
\end{align*}
Since by Lemma \ref{c}, 
\begin{align*}
\EE_{\bar\rP}^x
\left[
e^{-2g\int_{0}^{2t} W(\hat q^\triangle_s)\rd s}
\right]\leq 
e^{|g|(1-e^{-2t})|x|}
e^{{g^2}\int_0^{(1-e^{-4t})/2}|2t-y|^2\rd y}, 
\end{align*}
we have 
\begin{align*}
&\int_{\RR} 
\EE_{\bar\rP}^x
\left[
\left|
(\grr(\hat q^\triangle_ {-t})- \grr^T(\hat q^\triangle_{-t}))
\grr(\hat q^\triangle_ {t}) 
\right|e^{-g\int_{-t}^t W(\hat q^\triangle_s)\rd s}
\right]\rd \mu(x)\\
&\leq
C\left(
\int_{\RR} 
|\grr(\hat q^\triangle_ {0})- \grr^T(\hat q^\triangle_{0})|^2\rd\mu(x)\right)^{\han}
\left(
\int_{\RR} 
\EE_{\bar\rP}^x
\left[\left| \grr(\hat q^\triangle_ {2t}) \right|^2\right]
e^{|g|(1-e^{-2t})|x|}
\rd\mu(x)\right)^{\han}\\
&\leq
C'\left(
\int_{\RR} 
|\grr(\hat q^\triangle_ {0})- \grr^T(\hat q^\triangle_{0})|^2\rd\mu(x)\right)^{\han}
\left(
\int_{\RR} 
e^{|g|(1-e^{-2t})|x|}
\rd\mu(x)\right)^{\han}.
\end{align*}
Here we used that $\grr\in L^\infty(\ZZ_2\times \RR)$. 
Since $\int_{\RR} 
|\grr(\hat q^\triangle_ {0})- \grr^T(\hat q^\triangle_{0})|^2\rd\mu(x)\to0$ as $T\to\infty$, 
\[
\int_{\RR} 
\EE_{\bar\rP}^x
\left[
\left|
\grr(\hat q^\triangle_ {-t})- \grr^T(\hat q^\triangle_{-t})|
\grr(\hat q^\triangle_ {t}) 
\right|e^{-g\int_{-t}^t W(\hat q^\triangle_s)\rd s}
\right] \rd\mu(x)\to0\]
as $T\to\infty$. Similarly we can also show that
\[
\int_{\RR} \EE_{\bar\rP}^x
\left[
|\grr^T(\hat q^\triangle_ {-t})|
\left|
 \grr(\hat q^\triangle_t)-
\grr^T(\hat q^\triangle_ {t})
\right|
e^{-g\int_{-t}^t W(\hat q^\triangle_s)\rd s}\right]\rd\mu(x) 
\to0\] as $T\to\infty$. Then the proof is complete. 
\qed

\noindent
{\bf Acknowledgments:} 
FH and TS acknowledge the kind hospitality of the Institute for Mathematical Sciences of 
National University of Singapore, where we stayed from 10/12/2023 to 16/12/2023 
and  began with investigating the quantum Rabi model. 
We also thank Professor Naotaka Kajino for useful discussion and Professor Itaru Sasaki for appearing   
\cite{HMS17} to us. TS was supported by JSPS KAKENHI Grant Numbers 
JP20K20884, JP22H05105, JP23H01077, and JP23K25774, 
and also supported in part by JP20H00119 and JP21H04432.
FH was also financially supported by 
JSPS KAKENHI Grant Numbers  JP20H01808, JP20K20886 and 23K20217.	

\bibliographystyle{plain}
{\bibliography{hiro8}}

\begin{thebibliography}{10}

\bibitem{BHLMS01}
V.~Betz, F.~Hiroshima, J.~L\H{o}rinczi, R.~A. Minlos, and H.~Spohn.
\newblock {Ground state properties of the Nelson Hamiltonian - A Gibbs
  measure-based approach}.
\newblock {\em Rev. Math. Phys.}, 14:173--198, 2002.

\bibitem{bil68}
P.~Billingsley.
\newblock {\em {Convergence of Probability Measures}}.
\newblock Wiley Interscience, 1968.

\bibitem{BZ21}
A.~Boutet~de Monvel and L.~Zielinski.
\newblock {Oscillatory behavior of large eigenvalues in quantum Rabi models}.
\newblock {\em Int. Math. Res. Not.}, 2021:5155--5213, 2021.

\bibitem{bra11}
D.~Braak.
\newblock {Integrability of Rabi model}.
\newblock {\em Phys. Rev. Lett.}, 107:100401,7pages, 2011.

\bibitem{bra13b}
D.~Braak.
\newblock {A generalized G-function for the quantum Rabi model}.
\newblock {\em Ann. Phys.}, 525:23--28, 2013.

\bibitem{bra13a}
D.~Braak.
\newblock {Continued fractions and the Rabi model}.
\newblock {\em J. Phys.A}, 46:175301,10pp, 2013.

\bibitem{cai22}
M.~L. Cai, Y.~K. Wu, Q.~X. Mei, W.~D. Zhao, Y.~Jiang, L.~Yao, L.~He, Z.~C.
  Zhou, and L.~M. Duan.
\newblock Observation of supersymmetry and its spontaneous breaking in a
  trapped ion quantum simulator.
\newblock {\em Nature Communications}, 13(1):3412, 2022.

\bibitem{ALS98}
G.~F. De~Angelis, G.~Jona-Lasinio, and V.~Sidoravicius.
\newblock {Berezin integrals and Poisson processes}.
\newblock {\em J. Phys. A}, 31:289--308, 1998.

\bibitem{ALS83}
G.~F. De~Angelis, G.~Jona-Lasinio, and M.~Sirugue.
\newblock {Probabilistic solution of Pauli type equations}.
\newblock {\em J. Phys. A}, 16:2433--2444, 1983.

\bibitem{ARS91}
G.~F. De~Angelis, A.~Rinaldi, and M.~Serva.
\newblock {Imaginary-time path integral for a relativistic spin-(1/2) particle
  in a magnetic field}.
\newblock {\em Europhys. Lett.}, 14:95--100, 1991.

\bibitem{EK86}
S.~N. Ethier and T.~G. Kurtz.
\newblock {\em {Markov Processes}}.
\newblock Wiley Interscience, 1986.

\bibitem{GV81}
B.~Gaveau and J.~Vauthier.
\newblock {Int{\'e}grales oscillantes stochastiques: l'{\'e}quation de Pauli}.
\newblock {\em J. Funct. Anal.}, 44:388--400, 1981.

\bibitem{GJ68}
J.~Glimm and A.~Jaffe.
\newblock {A $\la \phi^4$ quantum field theory without cutoffs. I}.
\newblock {\em Phys. Rev.}, 176:1945--1951, 1968.

\bibitem{GMM17}
B.~G{\"u}neysu, O.~Matte, and J.~S. M{\o}ller.
\newblock Stochastic differential equations for models of non-relativistic
  matter interacting with quantized radiation fields.
\newblock {\em Probability Theory and Related Fields}, 167(3):817--915, 2017.

\bibitem{hir15}
M.~Hirokawa.
\newblock {The Rabi model gives off a flavor of spontaneous SUSY breaking}.
\newblock {\em Quantum Stud.: Math. Found}, 2:379--388, 2015.

\bibitem{HH14}
M.~Hirokawa and F.~Hiroshima.
\newblock {Absence of energy level crossing for the ground state energy of the
  Rabi model}.
\newblock {\em Commun. Stochastic Anal.}, 8:551--560, 2014.

\bibitem{HHL14}
M.~Hirokawa, F.~Hiroshima, and J.~L\H{o}rinczi.
\newblock {Spin-boson model through a Poisson driven stochastic process}.
\newblock {\em Math. Zeitschrift}, 277:1165--1198, 2014.

\bibitem{HMS17}
M.~Hirokawa, J.S. M{\o}ller, and I.~Sasaki.
\newblock A mathematical analysis of dressed photon in ground state of
  generalized quantum {R}abi model using pair theory.
\newblock {\em J. Phys. A}, 50(18):184003, 20, 2017.

\bibitem{hir14}
F.~Hiroshima.
\newblock {Functional integral approach to semi-relativistic Pauli-Fierz
  models}.
\newblock {\em Adv. Math.}, 259:784--840, 2014.

\bibitem{HIL12}
F.~Hiroshima, T.~Ichinose, and J.~L\H{o}rinczi.
\newblock {Path integral representation for Schr\"odinger operators with
  Bernstein functions of the Laplacian}.
\newblock {\em Rev. Math. Phys.}, 24:1250013,40pages, 2012.

\bibitem{HM21}
F.~Hiroshima and O.~Matte.
\newblock {Ground states and their associated Gibbs measures in the
  renormalized nelson model}.
\newblock {\em Rev. Math.Phys.}, 33:2250002 (84 pages), 2021.

\bibitem{IW05}
T.~Ichinose and M.~Wakayama.
\newblock {Special values of the spectral zeta function of the non-commutative
  harmonic oscillator and confluent Heun equations}.
\newblock {\em Kyushu.J.Math.}, 59:39--100, 2005.

\bibitem{IW05b}
T.~Ichinose and M.~Wakayama.
\newblock {Zeta functions for the spectrum of the noncommutative harmonic
  oscillators}.
\newblock {\em Commun. Math.Phys}, 258:697--739, 2005.

\bibitem{JC63}
E.T. Jaynes and F.W. Cummings.
\newblock {On the process of space quantization}.
\newblock {\em Proceedings of the IEEE}, 51:89--109, 1963.

\bibitem{kat52}
T.~Kato.
\newblock {Notes on some inequalities for linear operators}.
\newblock {\em Math. Ann.}, 125:208--212, 1952.

\bibitem{KRW21}
K.~Kimoto, C.~Reyes-Bustos, and M.~Wakayama.
\newblock {Determinant expressions of constraint polynomials and the spectrum
  of the asymmetric quantum Rabi model}.
\newblock {\em Int. Math. Res. Not.}, 2021:9458--9544, 2021.

\bibitem{KW23}
K.~Kimoto and M.~Wakayama.
\newblock {Ap\'ery-like numbers for non-commutative harmonic oscillators and
  automorphic integrals}.
\newblock {\em Ann. Inst. Henri Poincar\'e D}, 10:205--275, 2023.

\bibitem{kus85}
M.~Ku\'s.
\newblock {On the spectrum of a two-level system}.
\newblock {\em J. Math.Phys.}, 26:2792--2795, 1985.

\bibitem{LB15}
Z.M. Li and M.T. Batchelor.
\newblock {Algebraic equations for the exceptional eigenspectrum of the
  generalized Rabi model}.
\newblock {\em J. Phys. A: Math. Theor.}, 48:454005,13pp, 2015.

\bibitem{LHB20}
J.~L{\H o}rinczi, F.~Hiroshima, and V.~Betz.
\newblock {\em {Feynman-Kac type theorems and its applications, volume 1 (2nd
  ed)}}.
\newblock De Gruyter, 2020.

\bibitem{MPS14}
A.~J. Maciejewski, M.~Przybylska, and T.~Stachowiak.
\newblock {Full spectrum of the Rabi model}.
\newblock {\em Phys. Lett. A}, 378:16--20, 2014.

\bibitem{och08}
H.~Ochiai.
\newblock {A special value of the spectral zeta function of the non-commutative
  harmonic oscillators}.
\newblock {\em Ramanujan Journal}, 15:31--36, 2008.

\bibitem{rab36}
I.~I. Rabi.
\newblock {On the process of space quantization}.
\newblock {\em Phys. Rev.}, 49:324--328, 1936.

\bibitem{rey23}
C.~Reyes-Bustos.
\newblock {The heat kernel of the asymmetric quantum Rabi model}.
\newblock {\em J. Phys. A}, 56:425302, 26 pp., 2023.

\bibitem{RBW21}
C.~Reyes-Bustos, D.~Braak, and M.~Wakayama.
\newblock {Remarks on the hidden symmetry of the asymmetric quantum Rabi
  model}.
\newblock {\em J. Phys. A}, 54:285202, 20 pp., 2021.

\bibitem{RW21}
C.~Reyes-Bustos and M.~Wakayama.
\newblock {Heat kernel for the quantum Rabi model: II. Propagators and spectral
  determinants.}
\newblock {\em J. Phys. A}, 54:115202, 30 pp., 2021.

\bibitem{RW22a}
C.~Reyes-Bustos and M.~Wakayama.
\newblock {Degeneracy and hidden symmetry for the asymmetric quantum Rabi model
  with integral bias}.
\newblock {\em Commun. Number Theory Phys.}, 16:615--672, 2022.

\bibitem{RW22b}
C.~Reyes-Bustos and M.~Wakayama.
\newblock {Heat kernel for the quantum Rabi model}.
\newblock {\em Adv. Theor. Math. Phys}, 26:1347--1447, 2022.

\bibitem{RW23}
C.~Reyes-Bustos and M.~Wakayama.
\newblock {Zeta limits for the spectrum of quantum Rabi models}.
\newblock {\em arXiv:2304.08943}, preprint, 2023.

\bibitem{sim74}
B.~Simon.
\newblock {\em {The $P (\phi)_2$ Euclidean (Quantum) Field Theory}}.
\newblock Princeton University Press, 1974.

\bibitem{spo89}
H.~Spohn.
\newblock {Ground state(s) of the spin-boson Hamiltonian}.
\newblock {\em Commun. Math. Phys.}, 123:277--304, 1989.

\bibitem{sug18}
S.~Sugiyama.
\newblock {Spectral zeta functions for the quantum rabi models}.
\newblock {\em Nagoya Math. J.}, 229:52--98, 2018.

\bibitem{WY14}
M.~Wakayama and T.~Yamasaki.
\newblock {The quantum Rabi model and Lie algebra representations of $sl_2$}.
\newblock {\em J. Phys. A}, 47:335203, 17 pp., 2014.

\end{thebibliography}

\end{document}